\documentclass[apj]{emulateapj}
\usepackage[colorlinks,linkcolor={blue},citecolor={blue},urlcolor={red}]{hyperref}
\usepackage{epsfig,graphicx,natbib,amsmath,amsfonts,amssymb}

\newcommand{\re}{\Reff}                 
\newcommand{\msolar}{M$_\odot$}
\newcommand{\mstar}{M$_\ast$}

\newcommand{\dindex}{D$_n$(4000)}           
\newcommand{\nuvr}{NUV$-r$}
\newcommand{\hd}{H$\delta$}                  
\newcommand{\hda}{\hd$_A$}
\newcommand{\ewhda}{EW(\hda)}               
\newcommand{\ha}{H$\alpha$}
\newcommand{\hae}{\ha}                    
\newcommand{\ewhae}{EW(\hae)}
\newcommand{\lgewhae}{$\log_{10}$\ewhae}

\newcommand{\hb}{H$\beta$}

\newcommand{\mustar}{$\mu_\ast$}

\newcommand{\Reff}{{$R_{\rm e}$}}

\newcommand{\N}[1]{N$_{#1}$}

\newcommand{\myemail}{\email{leech@shao.ac.cn}}

\slugcomment{Accepted to ApJ}
\shorttitle{Gradients in recent star formation histories}
\shortauthors{Li et al.}

\begin{document}

\title{P-MaNGA: Gradients  in  Recent  Star
  Formation Histories as Diagnostics for Galaxy Growth and Death}

\author{
Cheng  Li\altaffilmark{1},  
Enci Wang\altaffilmark{1},  
Lin Lin\altaffilmark{1},
Matthew  A. Bershady\altaffilmark{2},  
Kevin Bundy\altaffilmark{3},
Christy A. Tremonti\altaffilmark{2}, 
Ting Xiao\altaffilmark{1}, 
Renbin Yan\altaffilmark{4},
Dmitry Bizyaev\altaffilmark{5},
Michael  Blanton\altaffilmark{6}, 
Sabrina Cales\altaffilmark{7},
Brian  Cherinka\altaffilmark{8}, 
Edmond Cheung\altaffilmark{3,9},
Niv Drory\altaffilmark{9},      
Eric Emsellem\altaffilmark{10,11},
Hai Fu\altaffilmark{12},  
Joseph Gelfand\altaffilmark{13,6},  
David R. Law\altaffilmark{8},
Lihwai Lin\altaffilmark{14},
Nick MacDonald\altaffilmark{15}, 
Claudia Maraston\altaffilmark{16},
Karen L. Masters\altaffilmark{16},
Michael R. Merrifield\altaffilmark{17}, 
Kaike Pan\altaffilmark{5},
S.\,F. S\'anchez\altaffilmark{18},
Donald P. Schneider\altaffilmark{19,20},
Daniel Thomas\altaffilmark{16}, 
David Wake\altaffilmark{2,21},
Lixin Wang\altaffilmark{1},
Anne-Marie Weijmans\altaffilmark{22},
David Wilkinson\altaffilmark{16}, 
Peter Yoachim\altaffilmark{23},
Kai Zhang\altaffilmark{1,4}, 
Tiantian Zheng\altaffilmark{13}
} \myemail

\altaffiltext{1}{Partner    Group   of   Max-Planck    Institute   for
  Astrophysics,  Shanghai Astronomical  Observatory,  Nandan Road  80,
  Shanghai  200030, China}

\altaffiltext{2}{Department  of Astronomy, University   of
  Wisconsin-Madison,   Madison,   WI 53706,   USA}

\altaffiltext{3}{Kavli  Institute for the  Physics and Mathematics of
  the Universe (Kavli IPMU, WPI), Todai Institutes for Advanced Study,
  the    University    of     Tokyo,    Kashiwa 277-8583,    Japan}

\altaffiltext{4}{Department  of Physics  and Astronomy,  University of
  Kentucky,  Lexington, KY  40506, USA} 

\altaffiltext{5}{Apache Point Observatory and New Mexico  State
  University, P.O.  Box 59, Sunspot, NM,  88349-0059,  USA} 

\altaffiltext{6}{Center  for  Cosmology  and Particle  Physics,
  Department  of  Physics, New  York University,  4 Washington  Place,
  New   York,  NY  10003, USA}  

\altaffiltext{7}{Yale Center for Astronomy and Astrophysics, 
  Physics Department, Yale University, P.O. Box 208120, New Haven, 
  CT 06520-8120, USA}

\altaffiltext{8}{Dunlap Institute for Astronomy and  Astrophysics,
  University of Toronto, 50 St.    George Street,    Toronto, Ontario
  M5S   3H4,   Canada} 

\altaffiltext{9}{Department of Astronomy and Astrophysics, University
  of California, Santa  Cruz, CA 95064, USA} 

\altaffiltext{10}{European Southern  Observatory,
  Karl-Schwarzschild-Str.   2,  85748 Garching, Germany}

\altaffiltext{11}{Universit\'e Lyon 1, Observatoire de Lyon, Centre de 
  Recherche Astrophysique de Lyon and Ecole Normale Sup\'erieure de Lyon, 
  9 avenue Charles Andr\'e, F-69230 Saint-Genis Laval, France}

\altaffiltext{12}{Department  of  Physics  and  Astronomy, University
  of Iowa, Iowa  City, IA 52242, USA} 

\altaffiltext{13}{NYU Abu Dhabi, PO Box 129188, Abu  Dhabi, UAE} 

\altaffiltext{14}{Institute  of Astronomy  and  Astrophysics, Academia
  Sinica,   Taipei   106,   Taiwan}  

\altaffiltext{15}{Department   of Astronomy, Box 351580, University
  of Washington, Seattle, WA 98195, USA}  

\altaffiltext{16}{Institute   of  Cosmology  and  Gravitation,
  University    of    Portsmouth,    Portsmouth,    PO1    3FX,    UK}

\altaffiltext{17}{School  of  Physics  and  Astronomy,  University  of
  Nottingham,    University   Park,    Nottingham    NG7   2RD,    UK}

\altaffiltext{18}{Instituto  de  Astronom\'\i  a,Universidad  Nacional
  Auton\'oma   de   Mexico,   A.P.   70-264,   04510,   M\'exico,D.F.}

\altaffiltext{19}{Department   of  Astronomy  and   Astrophysics,  The
  Pennsylvania  State  University,  University  Park, PA  16802,  USA}

\altaffiltext{20}{Institute  for  Gravitation   and  the  Cosmos,  The
  Pennsylvania  State  University,  University  Park, PA  16802,  USA}

\altaffiltext{21}{Department of Physical Sciences, 
The Open University, Milton Keynes, MK7 6AA, UK}

\altaffiltext{22}{School of  Physics and Astronomy,  University of St.
  Andrews,    North    Haugh,    St.     Andrews   KY16    9SS,    UK}

\altaffiltext{23}{Department  of Astronomy, University  of Washington,
  Box 351580, Seattle WA, 98195, USA}

\begin{abstract}
  We present an analysis of the data produced by the MaNGA prototype
  run (P-MaNGA), aiming to test how the radial gradients in recent
  star formation histories, as indicated by the 4000\AA-break
  (\dindex), \hd\ absorption (\ewhda) and \ha\ emission (\ewhae)
  indices, can be useful for understanding disk growth and star
  formation cessation in local galaxies.  We classify 12 galaxies
  observed on two P-MaNGA plates as either centrally quiescent (CQ) or
  centrally star-forming (CSF), according to whether \dindex\ measured
  in the central spaxel of each datacube exceeds 1.6. 
  For each spaxel we generate both 2D maps and radial profiles
  of \dindex, \ewhda\ and \ewhae.  We find that CSF galaxies generally
  show very weak or no radial variation in these diagnostics.  In
  contrast, CQ galaxies present significant radial gradients, in the
  sense that \dindex\ decreases, while both \ewhda\ and
  \ewhae\ increase from the galactic center outward.  The outer
  regions of the galaxies show greater scatter on diagrams relating
  the three parameters than their central parts.  In particular, the
  clear separation between centrally-measured quiescent and
  star-forming galaxies in these diagnostic planes is largely filled
  in by the outer parts of galaxies whose global colors place them in
  the green valley, supporting the idea that the green valley
  represents a transition between blue-cloud and red-sequence phases,
  at least in our small sample. These results are consistent with a
  picture in which the cessation of star formation propagates from the
  center of a galaxy outwards as it moves to the red sequence.
\end{abstract}


\section{Introduction}
\label{sec:introduction}

Our understanding of galaxies in the local universe has improved
dramatically over the past decade, in significant part due to large
optical spectroscopic surveys such  as the Two-degree Field Galaxy
Redshift Survey \citep[2dFGRS;][]{Colless-01} and the Sloan Digital
Sky Survey \citep[SDSS;][]{York-00}. One of the major findings of the
SDSS is the discovery of `galaxy bimodality', through which local
galaxies are divided into two distinct populations, termed `red
sequence' and `blue cloud', according to their rest-frame colors or
specific star formation rate \citep[e.g.][]{Strateva-01, Blanton-03a,
  Kauffmann-03b, Baldry-04}. Red-sequence galaxies typically present
bulge-dominated morphology, red optical/UV color, and little gas and
star formation, while blue-cloud galaxies are usually gas-rich with
blue colors, ongoing star formation and disky morphology.  The smaller
number of galaxies found lying in the `green valley' between these two
main populations in color space are widely believed to be caught in a
transition phase from the blue cloud to the red sequence
\citep[e.g.][]{Bell-04b, Faber-07, Martin-07, Schawinski-07b,
  Schiminovich-07, Wyder-07, Mendez-11, Goncalves-12}, although this
picture has recently been disputed by some authors
\citep[e.g.][]{Schawinski-14}.  Narrow-field deep surveys reveal that
this color bimodality persists out to redshift of at least $2.5$
\citep[e.g.][]{Bell-04b, Bundy-06, Cirasuolo-07, Cooper-08, Faber-07,
  Martin-07, Cowie-Barger-08, Brammer-09, Williams-09, Muzzin-12,
  Huang-13a}, though the prevalence of the red population has
significantly increased since a redshift of unity \citep{Bell-04b,
  Bundy-06, Faber-07}. The cessation of star formation is thus an
important process which has been contributing to galaxy evolution over
the past $\sim8\,{\rm Gyr}$.

A complete picture of the way in which the star formation in
galaxies gets shut down remains elusive.  However, recent studies of
the scaling relations of galaxy properties and their dependence on
local environment have clearly established that, in addition to
stellar mass, both internal structural properties and external
environment are key indicators, or may even be drivers, of the star
formation cessation processes in galaxies
\citep[e.g.][]{Kauffmann-06,Bell-08,Franx-08,Peng-10b,Thomas-10,
  Bell-12,Cheung-12,Fabello-12,Li-12c,Fang-13,Mendel-13,Zhang-13}.
For instance, when studying the central galaxies in groups or
clusters, the presence of a prominent bulge-like structure is found to
be a necessary (but not sufficient) condition for stopping star
formation \citep{Bell-08,Bell-12,Cheung-12,Fang-13}. In addition,
studies of the relationship between galaxy morphology and color have
revealed a significant population of red-sequence galaxies with 
disc-dominated spirals at both low-z
\citep[e.g.][]{Wolf-Gray-Meisenheimer-05,Wolf-09,
Bamford-09,Masters-10a} and $z\sim1-2$ \citep[e.g.][]{Bundy-10},
which are preferentially found in galaxies with large bulges
\citep{Bundy-10, Masters-10a}.
These findings support the `morphological quenching' mechanism proposed 
by \citet{Martig-09}, although studies of cold gas in massive galaxies
indicate that a reduction in gas content is also required
\citep[e.g.][]{Fabello-11}.  For central galaxies in massive dark
matter halos above a critical mass of $\sim10^{12}M_\odot$,
``radio-mode'' AGN feedback and shock-heating may effectively reduce
the gas cooling efficiency, thus also playing an important role in
preventing further star formation
\citep{Silk-77,Rees-Ostriker-77,Blumenthal-84, Birnboim-Dekel-03,
  Keres-05, Dekel-Birnboim-06, Cattaneo-06}. Powerful AGN feedback can
also happen in the so-called ``quasar mode'', which is predicted  to
be triggered by major mergers of gas-rich galaxies with comparable
mass \citep[e.g.][]{DiMatteo-Springel-Hernquist-05, Hopkins-06}, and
is commonly adopted as  one of the quenching processes in
semi-analytic models. 

For satellite galaxies, star formation quenching seems to be driven
primarily by external processes occurring within their host
group/cluster, such as gas stripping by ram-pressure
\citep{Gunn-Gott-72, Abadi-Moore-Bower-99} and tidal interactions
\citep{Toomre-Toomre-72, Moore-96}. The `smoking gun' of observations
showing H{\sc i} gas being removed by ram-pressure stripping have been
obtained for spirals in nearby clusters of galaxies
\citep[e.g.][]{Vogt-04, Chung-09, Merluzzi-13}. Studies of color
profiles  and surface brightness profiles for satellite galaxies in
SDSS group systems prefer more gentle processes such as ``starvation''
or ``strangulation'' \citep[e.g.][]{Weinmann-09}. On the other hand, a
recent study of the specific star formation rate--stellar mass
relation for galaxy groups/clusters at higher redshift ($0.2<z<0.8$,
\citealt{Lin-14}) suggests that galaxy mergers play a primary role in
quenching satellites in galaxy groups, while strangulation is a
process more important  in cluster-scale environment.  There have also
been statistical studies of the cold gas content and star formation in
galaxies residing in different environments, using either direct
atomic gas mass measurements from large blind surveys of H{\sc i}
emission such as the H{\sc i} Parkes All Sky Survey
\citep[HIPASS;][]{Zwaan-05} and the Arecibo Legacy Fast ALFA
\citep[ALFALFA;][]{Giovanelli-05} survey, or indirect estimates of gas
content from the optical photometry in surveys like SDSS
\citep[e.g.][]{Li-12c,Zhang-13}. These studies have revealed that
gas-related quenching depends not only on the stellar mass of the
satellite galaxies, but also on their surface mass density.

A full understanding of star formation cessation in galaxies thus
requires spatially-resolved measurements of stellar and gaseous
components to be obtained for a large sample of galaxies covering wide
ranges in mass and color, probing a range of environmental conditions.
Previous studies for this purpose have been mainly based
    on multi-wavelength broadband photometry, for both nearby galaxies
    \citep[e.g.][]{deJong-96, Bell-deJong-00, Taylor-05,
      Munoz-Mateos-07,Zibetti-Charlot-Rix-09,Roche-Bernardi-Hyde-10,
      Suh-10,Tortora-10a,
      Gonzalez-Perez-Castander-Kauffmann-11,Tortora-11,Kauffmann-15}
    and those at high redshifts
    \citep[e.g.][]{Abraham-99,Azzollini-Beckman-Trujillo-09,
      Szomoru-12,Wuyts-12,Szomoru-13,Wuyts-13}, while some authors
    also made use of long-slit spectroscopy \citep[e.g.][]{Moran-12,
      Huang-13b}.  Significant improvements have been made in the past
    decade, thanks to many integral field unit (IFU) surveys
    which have obtained spatially resolved spectroscopy for samples of
    galaxies. However, to-date,  the samples have been relatively small
($\lesssim 100$ galaxies): SAURON \citep{Bacon-01}, DiskMass
\citep{Bershady-10}, ATLAS$^{\mbox{3D}}$ \citep{Cappellari-11b}.  

The ongoing CALIFA survey \citep[Calar Alto Large Integral Field
  Area][]{Sanchez-12} is observing a sample of 600 galaxies, making a
big step forward in 2D spectroscopic studies of nearby galaxies. For
example, this survey has,  for the first time, quantified the
spatially-resolved history of stellar mass assembly for galaxies
beyond the Local Group, demonstrating how massive galaxies grow their
stellar mass inside-out \citep{Perez-13}. Such an inside-out picture
of galaxy formation is supported by further measurements of gradients
in oxygen abundance of H{\sc ii} regions \citep{Sanchez-14}, in
stellar metallicity of face-on spirals \citep{Sanchez-Blazquez-14},
and in stellar age and local surface mass density of galaxies of
different morphologies and masses \citep{GonzalezDelgado-14a}, all
from the CALIFA survey. However, the sample is still relatively
modest, limiting the range of environments and galaxy properties that
can be explored. The more recently launched SAMI survey
\citep{Croom2012} takes a further step with a goal of obtaining IFU
observations for 3400 galaxies by 2016. In the meantime, the
KMOS$^{3D}$ Survey is pushing forward the IFU observations for high
redshift galaxies, observing a sample of 600 galaxies at $0.7<z<2.7$
using KMOS at the Very Large Telescope \citep{Wisnioski-15}.  

As a next step at low-z, the study of these phenomena is a key goal of
the upcoming MaNGA  \citep[Mapping Nearby Galaxies at Apache Point
  Observatory,][]{Bundy-15} survey.  As one of the major programs of
the fourth-generation of SDSS, MaNGA will obtain two-dimensional,
integral-field spectroscopy (IFS) for 10,000 galaxies in the nearby
redshift range $0.01<z<0.15$, optimally selected for uniform coverage
and resolution using the single (central) spectra from SDSS.  Each
galaxy will be spectroscopically observed with an IFU to obtain
multiple high signal-to-noise ratio (S/N) spectra across the full
optical wavelength range from 3600 to 10300\AA. MaNGA will thus
provide 2D maps of stellar populations and recent star formation
histories for a large number of quiescent and star-forming galaxies,
thus enabling extensive exploration of the various processes by which
star formation ceases in the local universe.

In this paper, to assess the potential use of the MaNGA survey in
studying the cessation of star formation, and to derive some initial
results in this area, we present an analysis of the MaNGA prototype
(P-MaNGA) data for a set of 12 galaxies, which were observed in the
pilot run in 2012 December and 2013 January.  These observations were
made through time granted by the Sloan Digital Sky
Survey-III \citep[SDSS-III,][]{Eisenstein-11}.  The prototype data are
described in more detail in \citet{Bundy-15}, which also
provides a general overview of the MaNGA project.

For each of the P-MaNGA galaxies and for each spectrum in the
associated data cube, we have performed full spectral fitting using
the methodology developed in \citet{Li-05}, in order to separate out
emission line components from the underlying absorption-line
stellar spectrum. From these decomposed spectra, we have obtained full
2D maps of three key parameters---\dindex\ (the depth of the
4000\AA\ break), \ewhda\ (the equivalent width of the H$\delta$
absorption line) and \hae\ (the equivalent width of the H$\alpha$
emission line)---which are known to be sensitive indicators of stellar
populations of different ages \citep{Kauffmann-03c,Kauffmann-14}. We
then investigate any radial variations in the recent star formation
histories of the galaxies by plotting the values of these parameters
for different subregions of each galaxy, to see how they change with
position.  As we will show, the P-MaNGA sample, though small in size,
reveals intriguing systematic trends in the three diagnostic parameters
across these galaxies, demonstrating the potential for using MaNGA 
(and similar IFU surveys) to understand disk growth and star formation 
cessation in local galaxies. 

Throughout this paper we assume a $\Lambda$ cold dark matter cosmology
model with $\Omega_m=0.27$, $\Omega_\Lambda=0.73$ and $h=0.7$.  A
\citet{Chabrier-03} stellar initial mass function (IMF) is adopted.

\section{Data}
\label{sec:data}

\subsection{P-MaNGA Observations}
\label{sec:p-manga}

\begin{deluxetable*}{llccccccccc}
\tablecolumns{13} \tablewidth{0pc}
\tablecaption{Prototype         Run         Galaxies         Observed}
\tabletypesize{\footnotesize}
\tablehead{       \colhead{Bundle\_Field}       &      \colhead{P-MaNGA}
  &\colhead{mangaID}   &\colhead{IAU Name}   &\colhead{$z$}
  &\colhead{$\log$}         &\colhead{$M_i$}        &\colhead{$(g-r)$}
  &\colhead{\Reff}   &\colhead{type${\dagger}$}   &   \colhead{$R_{\rm
      IFU}$}   \\    
\colhead{}   &  \colhead{Name}   &   \colhead{}
  &  & \colhead{} & \colhead{$M_*$}&
  \colhead{mag}  &   \colhead{mag}  &  \colhead{''}   &  \colhead{}  &
  \colhead{(\Reff)} \\
}

\startdata

\cutinhead{Field 9: PlateID =  6650, 3.0 hr, seeing 1\farcs7} 

ma003\_9 & p9-127A & 8-188794  & J093457.30+214220.9 & 0.013 &  10.7 &
-21.3 & 0.70 & 23.7 & $\star$ & 0.7 \\   
ma008\_9 & p9-127B  & 8-131835 & J093506.31+213739.5 &  0.013 & 9.1 &
-18.5 & 0.51 & 6.8 & 1 & 2.4 \\ 
ma002\_9 & p9-61A  & 8-188807  & J093712.30+214005.0 & 0.019 & 10.1 &
-20.5 & 1.30 & 9.3 & $\star$ & 1.2 \\  
ma005\_9 & p9-19B & 8-131893   & J093109.60+224447.4 & 0.051  & 10.6 &
-22.2  & 0.86  & 4.0 &  1 & 1.6 \\  
ma007\_9 &  p9-19D & 8-131577  & J093109.07+205500.5 & 0.034 & 10.3 &
-21.2 & 0.79 & 3.2 & 1 & 2.0 \\ 
ma001\_9 & p9-19E & 8-131821   & J094030.22+211513.7 & 0.024 & 9.7 &
-20.0 & 0.73 & 2.9 & 1 & 2.2 \\ 

\cutinhead{Field 4: PlateID =  6652, 2.0 hr, seeing 1\farcs3} 

ma003\_4 & p4-127A  & 8-109661 & J105555.26+365141.4  & 0.022 & 10.7
& -22.1 & 0.84 & 10.3 & $\star$ &1.6 \\

ma008\_4 & p4-127B  & 8-109682 & J105259.05+373648.2 &  0.042 & 11.0 &
-22.7 & 0.77 & 13.9 & $\star$  & 1.2 \\

ma002\_4 & p4-61A & 8-113576   & J105746.61+361657.8 & 0.030 & 9.7 &
-20.0 & 0.71 & 3.4 & $\star$ & 3.4 \\ 

ma004\_4 & p4-19A & 8-113557   & J110012.10+362313.8 & 0.027 & 9.4 &
-19.5  & 0.49 & 2.5 & 1 & 2.5 \\

ma005\_4  & p4-19B  & 8-113506 & J104958.69+362454.0 & 0.023 & 9.5
& -19.5 & 0.53 & 4.6 & 1 &  1.4 \\ 

ma006\_4  & p4-19C & 8-109657  & J105605.68+365736.1 & 0.022 & 9.5 &
-19.4 & 0.72 & 4.8 &1 & 1.3 

\enddata
 \label{tab:sample}
 \tablecomments{The P-MaNGA Name is composed of a ``p'' for ``prototype''
followed by the field ID, a hyphen, and then the shorthand ID for the
bundle used. This shorthand includes a number corresponding to N$_{\rm IFU}$
for that bundle. The ($g-r$) magnitude is extinction-corrected. \\
$^{\dagger}$ Target type 1 indicates the galaxy would
   be selected in the MaNGA Survey's Primary Sample. These galaxies
   often have spatial coverage to larger than 1.5\Reff\ due to the 
   different bundle size distribution used in P-MaNGA.  A ``$\star$''
   indicates a galaxy that was chosen by hand for the prototype run and
 does not pass the nominal selection criteria. 
 }
\end{deluxetable*}

In the present work, we use data obtained using the MaNGA engineering
prototype instrument (hereafter P-MaNGA) in 2013 January.  P-MaNGA was
designed to explore a variety of instrument design options, observing
strategies, and data processing algorithms. These P-MaNGA data offer
valuable insights into MaNGA's potential, but differ substantially
from the MaNGA survey data in several ways.  First, P-MaNGA used just
one of the two BOSS spectrographs, with only 560 total fibers
and 8 IFUs with three sizes: 19 fibers (\N{19}),
61 fibers (\N{61}), and 127 fibers (\N{127}).  The P-MaNGA IFU
complement was dramatically different from the MaNGA survey
instrument, with $5 \times $\N{19} (instead of just two), $1 \times
$\N{61} (instead of 4), $2 \times $\N{127} (instead of five), and no
37-fiber or 91-fiber IFUs 
\citep[see][for detailed description of MaNGA instrumentation]{Drory-15}.  
Finally, we also note that the masses and
sizes of the P-MaNGA target galaxies were not selected to be
representative of the full MaNGA sample.

P-MaNGA observations were obtained of three galaxy fields (Field 9,
11, 4) using SDSS plates 6650, 6651, and 6652, each containing 6 
galaxies.  Some of the P-MaNGA targets were drawn from
early versions of the full MaNGA sample (Wake et al., in prep),
but in many cases P-MaNGA targets were chosen for specific reasons.
In each of the three plates, one \N{127} IFU was allocated for comparison purposes to a
galaxy observed by the CALIFA survey \citep{sanchez12}, even if the
galaxy would not otherwise satisfy the MaNGA selection cuts.
Additionally, the non-optimal IFU complement of the P-MaNGA instrument
required some targets to be selected manually.  For each plate,
observations were obtained in sets of three 20-minute exposures, which
were dithered by approximately a fiber radius along the vertices of an
equilateral triangle to provide uniform coverage across each IFU.
These three plates were observed to varying depths and in varying
conditions, as required by the P-MaNGA engineering tasks.  Although
plate 6650 (Field 9) was observed to a depth comparable to what will
be regularly achieved during MaNGA operations, plates 6651 (Field 11)
and 6652 (Field 4) are both significantly shallower than MaNGA survey
data, and plate 6651 was intentionally observed at high airmass
($\sim1.5$) resulting in particularly poor image quality. In this
paper we use data obtained with plates 6650 and 6652 (see Table \ref{tab:sample}), 
while excluding plate 6651.

The raw data were reduced using a prototype of the MaNGA Data
Reduction Pipeline (DRP), which is described in detail by \citet{Law-15}.
In brief, individual fiber flux and inverse variance
spectra were extracted using a row-by-row algorithm, wavelength
calibration was obtained from a series of Neon-Mercury-Cadmium arc
lines, and the spectra were flatfielded using internal quartz
calibration lamps.  Sky-subtraction of the IFU fiber spectra was
performed by constructing a cubic basis spline model of the sky
background flux as seen by the 41 individual fibers placed on blank
regions of sky, and subtracting the resulting composite spectrum
shifted to the wavelength solution of each IFU fiber.

Flux calibration of the P-MaNGA data is performed by fitting Kurucz
model stellar spectra to the spectra of calibration standard stars
covered with single fibers at each of the three dither positions.  The
flux calibration vectors derived from these single-fiber spectra were
found to vary by $\sim$ 10\% from exposure to exposure, depending on
the amount of light lost from the fiber due to atmospheric seeing and
astrometric misalignments.  While this uncertainty is acceptable for
the present science purposes, the flux calibration uncertainty of the
single fibers ultimately drove the design decision of the MaNGA survey
to instead use 7-fiber IFU ``mini-bundles'' for each calibration
standard star, which results in a few percent photometric uncertainty
\citep[see][]{Yan-15}.

Flux-calibrated spectra from the blue and red cameras were combined
across the dichroic break using an inverse-variance weighted basis
spline function.  Astrometric solutions were derived for each
individual fiber spectrum that incorporate information about the IFU
bundle metrology (i.e., fiber location within an IFU), dithering, and
atmospheric chromatic differential refraction, among other effects.
Fiber spectra from all exposures for a given galaxy were then combined
into a single datacube (and corresponding inverse variance array)
using these astrometric solutions and a nearest-neighbor sampling
algorithm similar to that used by the CALIFA survey.  For the P-MaNGA
datacubes, a spaxel size of 0\farcs5 was chosen.  The typical
effective spatial resolution in the reconstructed datacubes can be
described by a Gaussian with a full width at half maximum ${\rm FWHM}
\approx 2$\farcs5.  When binning the datacubes, we scale the resulting
error vectors to account, at least approximately, for wavelength and
spatial covariance in the P-MaNGA error cubes.

\subsection{SDSS-I perspective of the P-MaNGA galaxies}
\label{sec:sdss_data}

\begin{figure*}
  \begin{center}
    \epsfig{figure=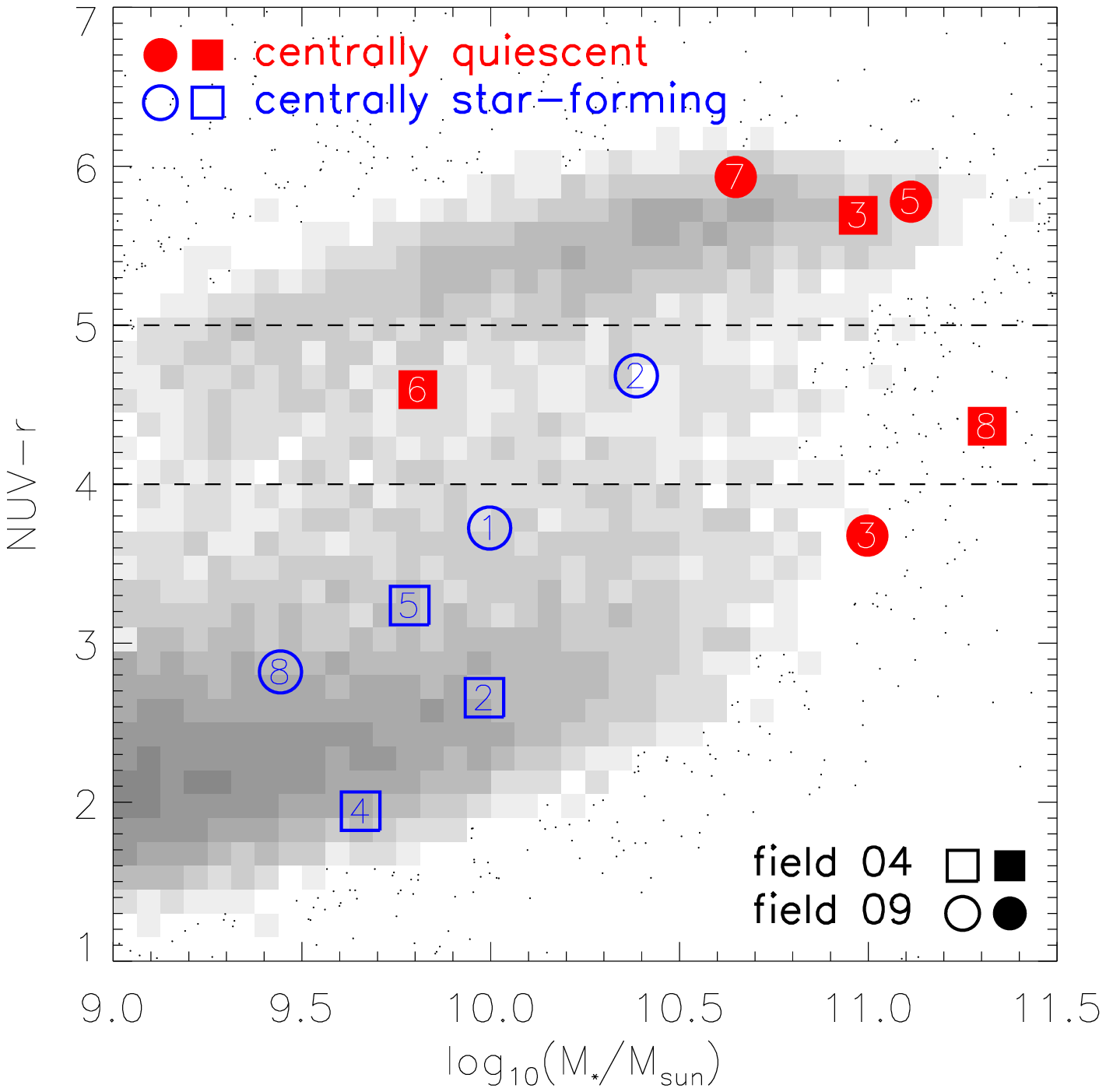,clip=true,width=0.4\textwidth}
    \epsfig{figure=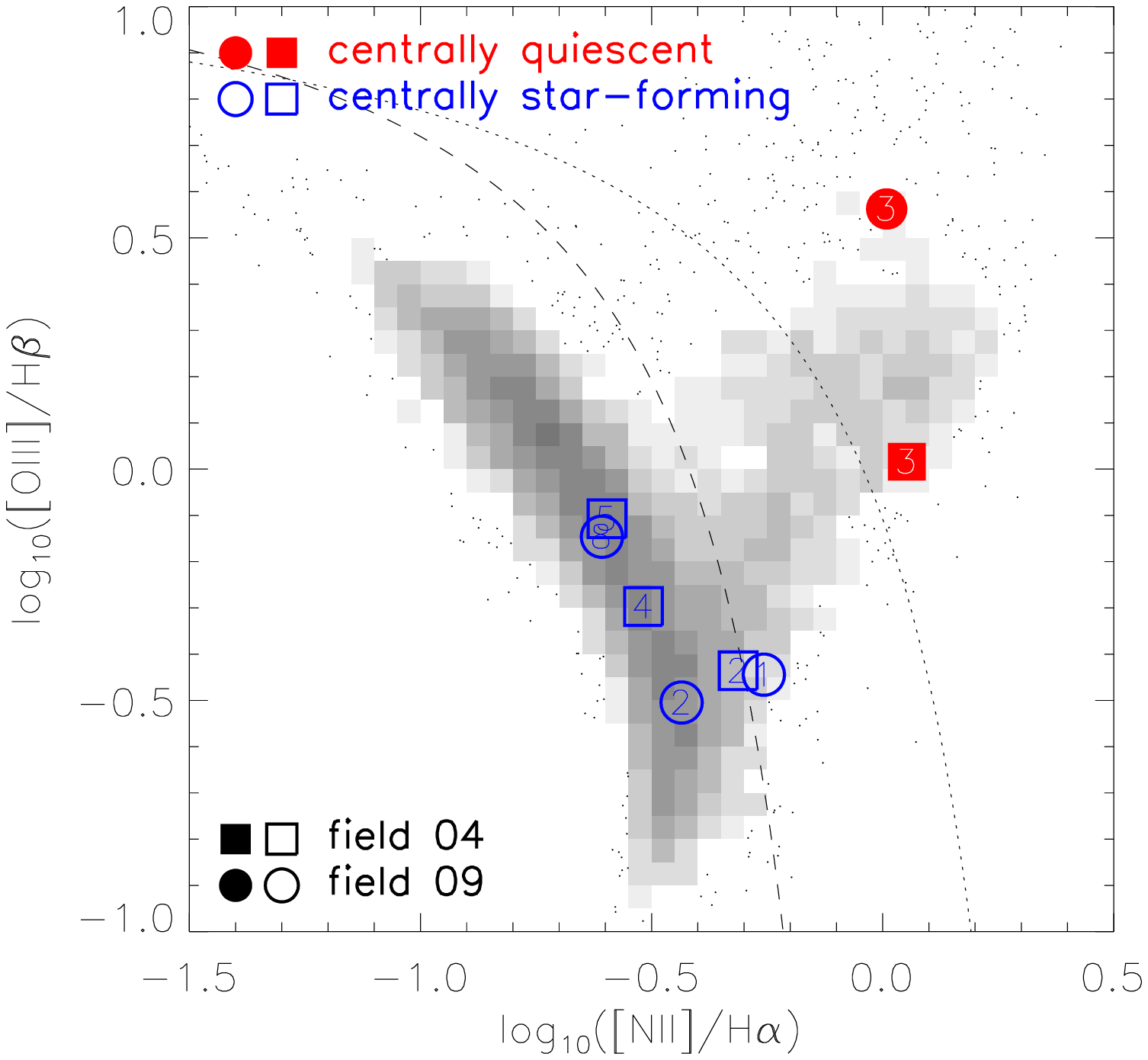,clip=true,width=0.4\textwidth}
  \end{center}
  \caption{P-MaNGA galaxies are plotted as large red or blue symbols on the
    stellar mass versus  \nuvr\ color plane (left panel)  and the BPT
    \citep{Baldwin-Phillips-Terlevich-81}  diagram (right  panel). The
    different fields of the  P-MaNGA observations are distinguished by
    the different  symbols as indicated,  while the number  within the
    symbols indicates  the IFU Bundle  for each galaxy.   Galaxies are
    divided into  two classes: centrally quiescent  (red solid symbols)
    and star-forming  (blue open  symbols) galaxies, defined  as those
    with   \dindex$>1.6$   and    \dindex$\le1.6$   in   the   central
    spaxel of the P-MaNGA datacube. Distributions  of a
    volume-limited  sample selected  from the  SDSS are  shown  as
    grey scale for comparison. The dashed horizontal lines in the left-hand
    panel are for $NUV-r$=4 and 5, which are commonly used in the literature
    to divide galaxies into blue-cloud, green-valley and red-sequence.
    The dashed and dotted lines in the right-hand panel are the AGN
    classification line of \citet{Kauffmann-03c} and \citet{Kewley-06}.}
  \label{fig:sample_properties}
\end{figure*}

\begin{figure*}
  \begin{center}
    \epsfig{figure=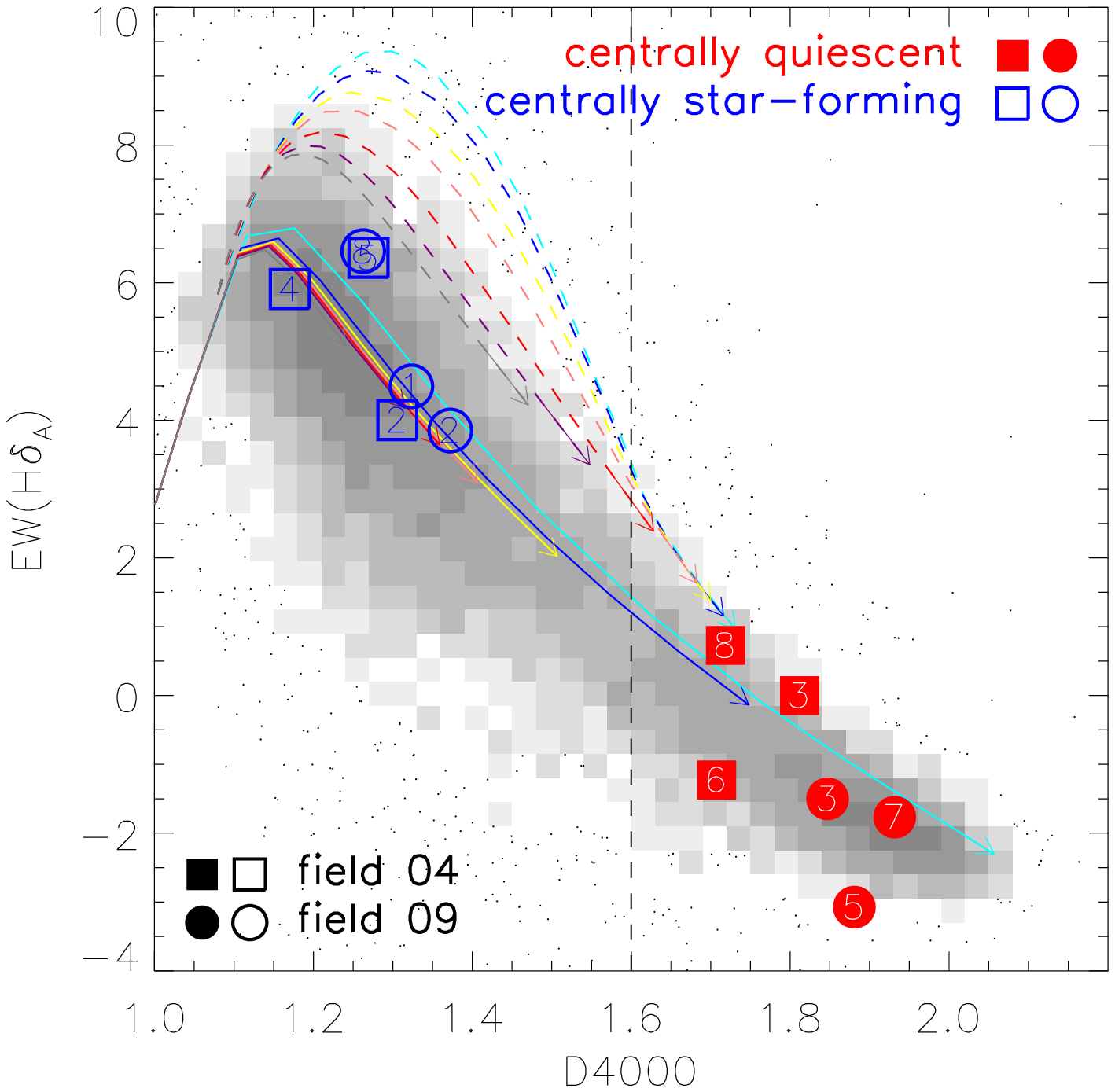,clip=true,width=0.4\textwidth}
    \epsfig{figure=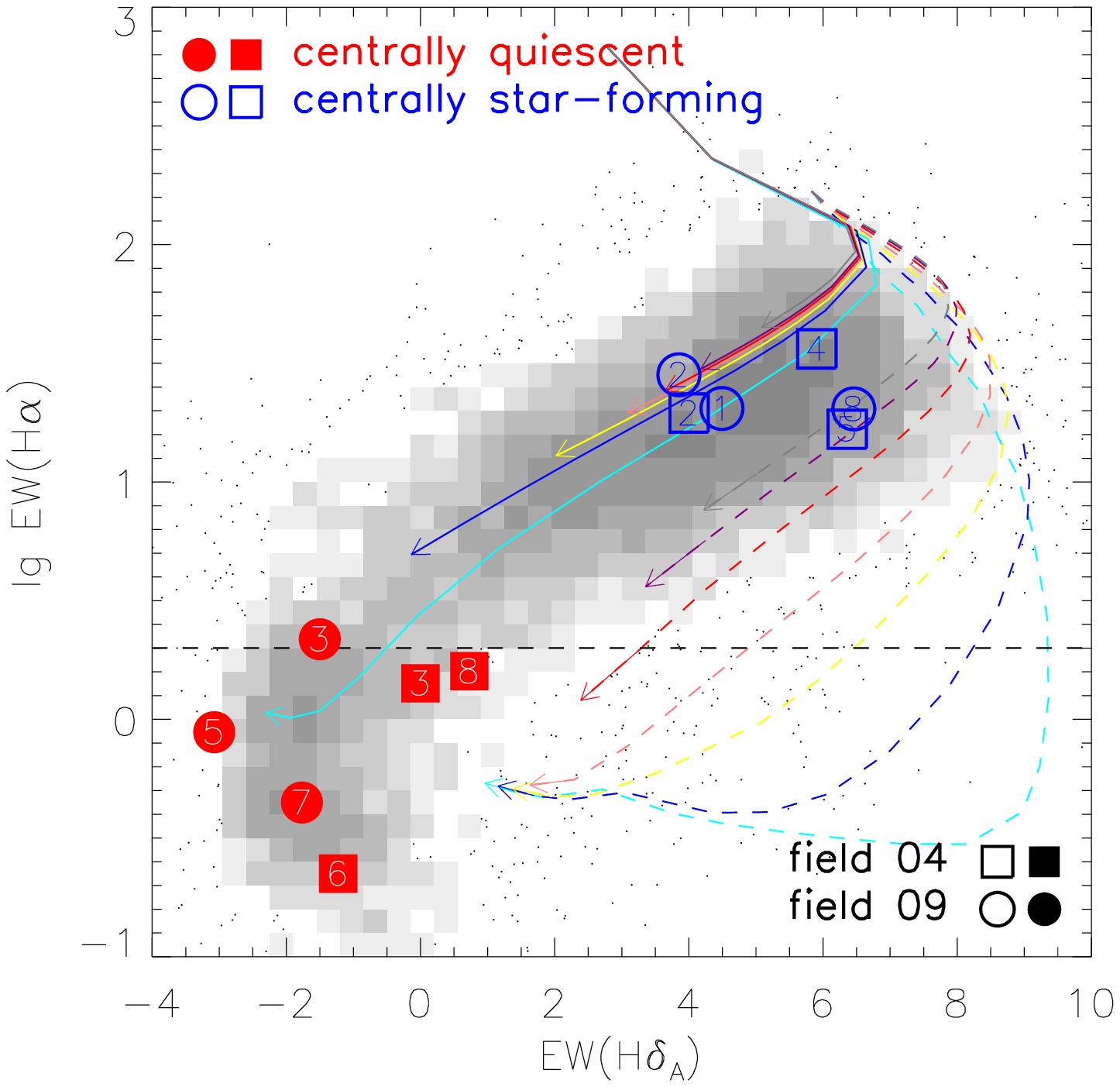,clip=true,width=0.4\textwidth}
  \end{center}
  \caption{P-MaNGA  galaxies are  plotted as  large red or blue  symbols on
    planes   of   4000\AA-break    vs.    \hda\   (left   panel)   and
    \hda\  vs. \lgewhae\  (right panel).   Shown in  solid  and dashed
    lines are  solar metallicity  models of continuous  star formation
    histories and star formation bursts calculated using
    the      stellar       population      synthesis      code      of
    \citet{Bruzual-Charlot-03}.   Distributions  of  a  volume-limited
    sample  selected from  the SDSS  are shown  as grey scale for
    comparison.   Symbols/colors  are  the  same as  in  the  previous
    figure. The dashed vertical line in the left-hand panel indicates
    a constant \dindex=1.6, while the dashed horizontal line in the 
    right-hand panel is for \ewhae=2\AA; these are the typical values 
    adopted in the literature to divide a galaxy into quiescent and 
    star-forming populations.}
  \label{fig:sdss_d4k_hda_hae}
\end{figure*}

All our galaxies have photometry in the SDSS-I five bands, as well as archival
spectroscopy from the SDSS-I 3\farcs0-diameter fiber. 
Figures~\ref{fig:sample_properties} and~\ref{fig:sdss_d4k_hda_hae} show
the properties of this sample based on the SDSS data. With
these two figures we aim to show how our galaxies compare to the larger
SDSS-I sample from which they are drawn.

Figure~\ref{fig:sample_properties} presents the galaxies on the plane
of stellar mass versus \nuvr\ color and the BPT diagram
\citep{Baldwin-Phillips-Terlevich-81}.  We use estimates of stellar
mass and \nuvr\ from the New York University Value-Added Galaxy
Catalog
\citep[NYU-VAGC\footnote{http://sdss.physics.nyu.edu/vagc/};][]{Blanton-05b}
and NASA Sloan Atlas (NSA)\footnote{http://www.nsatlas.org}.  NSA is a
catalog of images and parameters of local galaxies based on data from
SDSS \citep{York-00}, GALEX \citep{Martin-05} and
2MASS \citep{Skrutskie-06}; for details see
\citet{Blanton-05a,Blanton-05b,Blanton-11}.  The stellar mass for each
galaxy is estimated based on its spectroscopically-measured redshift
and the five-band Petrosian magnitudes from SDSS photometric data, as
described in detail in \citet{Blanton-Roweis-07}.  The \nuvr\ color is
defined by the integrated light in the NUV-band from GALEX and the
Petrosian magnitude in the $r$-band from SDSS, both corrected for Galactic
extinction. Measurements of the emission line ratios for the BPT diagram
are taken from the MPA/JHU SDSS database
\footnote{http://www.mpa-garching.mpg.de/SDSS/DR7/}
\citep{Brinchmann-04}.

Figure~\ref{fig:sdss_d4k_hda_hae} displays the galaxies in the
\dindex\ versus \ewhda\ and \ewhda\ versus \lgewhae\ planes, where
\dindex\ is the 4000\AA\ break in the optical spectrum as defined by
\citet{Balogh-99}, \ewhda\ the Lick/IDS index of the H$\delta$
absorption line defined by \citet{Worthey-Ottaviani-97}, and \lgewhae\
the logarithm of H$\alpha$ emission line equivalent width.  We use
measurements of these parameters from the MPA/JHU
database, which were obtained from the SDSS
3\farcs0-diameter spectra, thus probing the central $1$ -- $2\,{\rm
  kpc}$ for the P-MaNGA galaxies.

In both these figures, we use squares and circles for galaxies in
Fields 4 and 9 respectively, and label each galaxy by the BundleID used
in those fields.  We divide these galaxies into two subsets according to
the 4000\AA\ break in the central spaxel of their datacube (see 
\S~\ref{sec:fitting}): {\em centrally quiescent}
(CQ) with \dindex$>1.6$ and {\em centrally star-forming} (CSF) with
\dindex$\le1.6$.  The two subsets are highlighted in the figures with
red solid and blue open symbols respectively.  We will use the same
symbols and colors for subsequent figures unless otherwise stated.
For comparison, we have selected a volume-limited sample of 21,328
galaxies with stellar mass above $10^9$\msolar\ and redshift in the
range $0.01<z<0.03$ from the NSA.  Distributions of this sample are
plotted in gray-scale maps in Figures~\ref{fig:sample_properties}
and~\ref{fig:sdss_d4k_hda_hae}.

As can be seen from Figure~\ref{fig:sample_properties}, the P-MaNGA
sample (though small in size) spans a wide range in both the
color--mass and BPT diagrams, similar to that of the general
population.  In the literature, galaxies are usually classified into
three types according to their \nuvr\ color: red-sequence (\nuvr\
$>5$), green-valley ($4\le$\ \nuvr\ $\le5$) and blue-cloud (\nuvr\
$<4$). Accordingly, half of the sample galaxies fall in the
blue-cloud region, indicative of ongoing star formation and relatively
rich cold gas content.  For all the galaxies except {\tt ma002\_9} and
{\tt ma003\_9}, the
color-based classification and the \dindex-based classification are
consistent with each other, in the sense that the blue-cloud galaxies
are centrally star forming, while the objects in the green valley and
red sequence combine to form the centrally-quiescent galaxy class.
This result indicates that green-valley and red-sequence galaxies
share similar properties in their central regions, and that their
different global color can be mainly attributed to the different
stellar population and gas content in their outer parts.

One of the two exceptions, {\tt ma003\_9}, is
a strong Seyfert-type AGN according to its location in the BPT
diagram, and is an outlier in the color--mass diagram due to its
relatively blue color at its stellar mass.  As shown below, this
galaxy presents a red core and a blue outer disk in the optical image,
such that its classification would be more reasonable if the
measurement of \nuvr\ color were confined to its central region.  The
strong contradiction between the central and global classification in
this case highlights the importance of spatially-resolved
measurements.  The other exception, {\tt ma002\_9}, is a dusty, 
inclined spiral.  It is a centrally star-forming
galaxy according to its small value of \dindex, but is classified as a
green-valley galaxy based on the relatively red \nuvr\ color, which in
this case is likely caused by dust.

The \dindex\ and \hda\ indices are known to be indicators of the
recent star formation history of galaxies, with \dindex\ being a good
proxy for the luminosity-weighted stellar age, and \hda\ a sensitive
tracer of the star formation that occurred 0.1 -- $1\,{\rm Gyr}$ ago
\citep[][hereafter BC03]{Bruzual-Charlot-03}.  As shown in
\citet{Kauffmann-03c}, the location of a galaxy in the plane of
\dindex\ and \hda\ provides a powerful diagnostic of whether the
galaxy has experienced a continuous star formation history or bursts
over the past 1 -- $2\,{\rm Gyr}$.  This property is illustrated in
the left-hand panel of Figure~\ref{fig:sdss_d4k_hda_hae}, where we
present the \dindex\ and \hda\ indices for the BC03 models of solar
metallicity that follow exponentially declining star formation
histories (${\rm SFR}\propto\exp(-t/\tau)$), either continuous 
star formation decline with long e-folding times ($\tau>5\times10^8$yr, 
solid lines), or bursts of star formation with fairly short e-folding times
($\tau<5\times10^8$yr, dashed lines).  Different
colors indicate the adopted values of the characteristic timescale
$\tau$.  The right-hand panel presents the same sets of models on the
\hda\ index versus \lgewhae\ plane.  The \ha\ luminosity is computed 
from the output of the BC03 models.  We convert Lyman continuum photons 
to \ha\ photons following \citet[][see equations B2-B4 in their appendix]
{Hunter-Elmegreen-04}.  We take the recombination coefficients and 
\ha/\hb\ ratios from \citet{Hummer-Storey-87}. For the LMC, solar, and 
super-solar metallicity models we assume nebular temperatures of 
$T_e=$15,000 K, 10,000 K, 5000 K respectively. The P-MaNGA galaxies occupy
roughly the same region as the general population in both panels, with
the CSF galaxies located in the low-\dindex\ and high-\hda\ region and
the CQ galaxies in the opposite corner.  It is, however, notable that
the P-MaNGA sample lacks galaxies of intermediate values in these
indices, and consequently the current sample is somewhat biased to a
more bimodal distribution on both planes compared to the full
population of SDSS galaxies.

In the rest of this paper we will use the \dindex--\ewhda\
and \ewhda--\lgewhae\ planes to investigate the maps and radial profiles
of recent star formation histories of the P-MaNGA galaxies. In this case we 
use relative measurements of the three indices as indicators of recent
star formation history, but we do not present quantitative measurements 
of mean stellar age, which depends on the sensitivity of these indices
to stellar metallicity, element abundance and dust attenuation 
\citep[e.g.][]{Worthey-Ottaviani-97, Bruzual-Charlot-03, Thomas-04, 
Thomas-Davies-06, Sanchez-12}. In a parallel
paper \citep{Wilkinson-15} we attempt to simultaneously obtain
the stellar age, metallicity and dust content for the P-MaNGA galaxies,
by fitting stellar population models to the full spectrum. In addition,
in the present paper we assume that \ha\ emission is just a signature of 
recent star formation. However, \ewhae\ lower than 6\AA\ was found to be
more compatible with light from post-AGB stars \citep{Papaderos-13,Sarzi-13}.
One should keep these caveats in mind when interpreting the observational 
results presented in this paper.

\subsection{Spectral fitting}
\label{sec:fitting}

\begin{figure*}
  \begin{center}
    \epsfig{figure=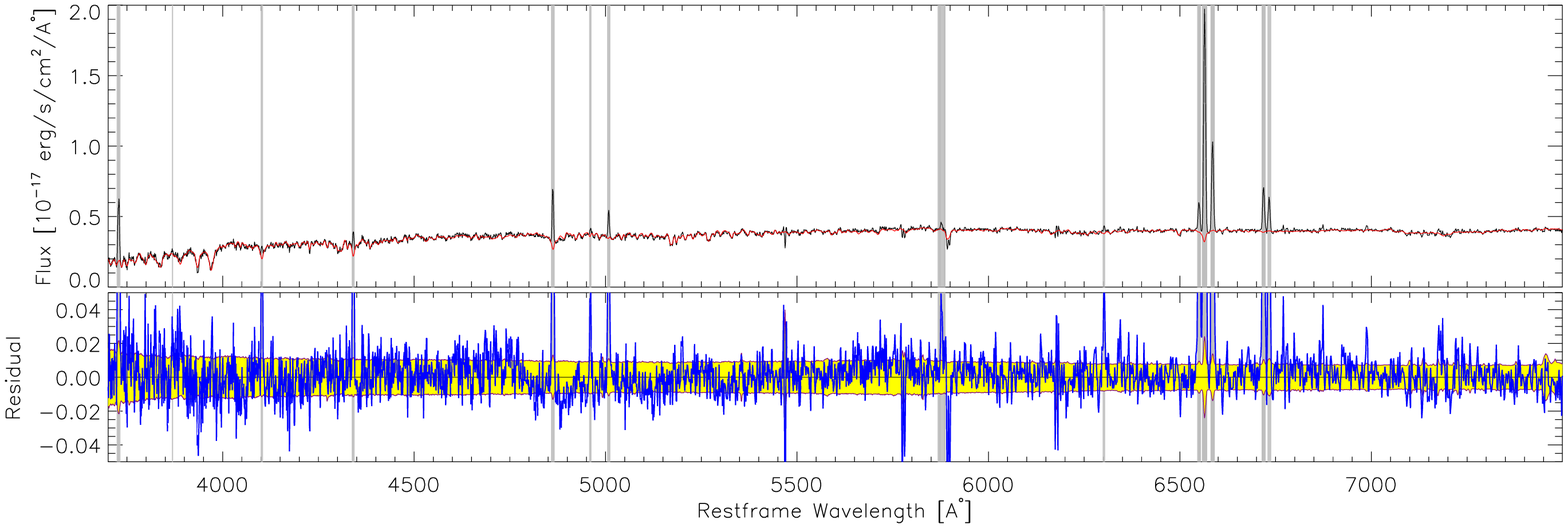,clip=true,width=0.9\textwidth}
    \epsfig{figure=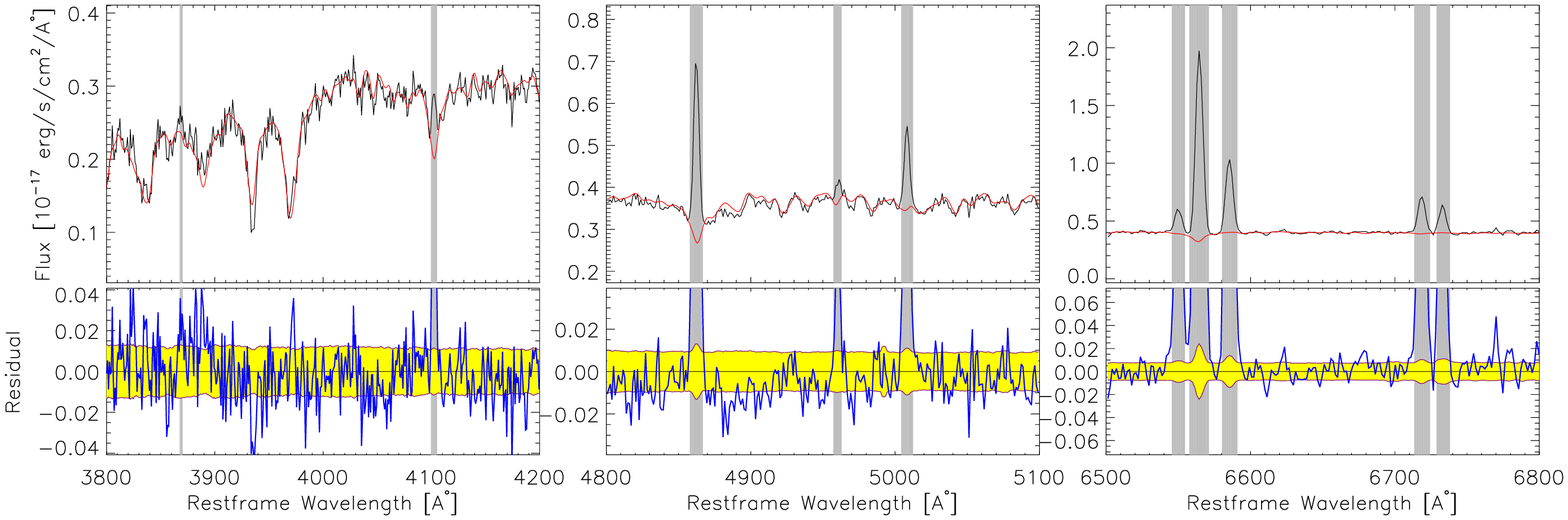,clip=true,width=0.9\textwidth}
    \epsfig{figure=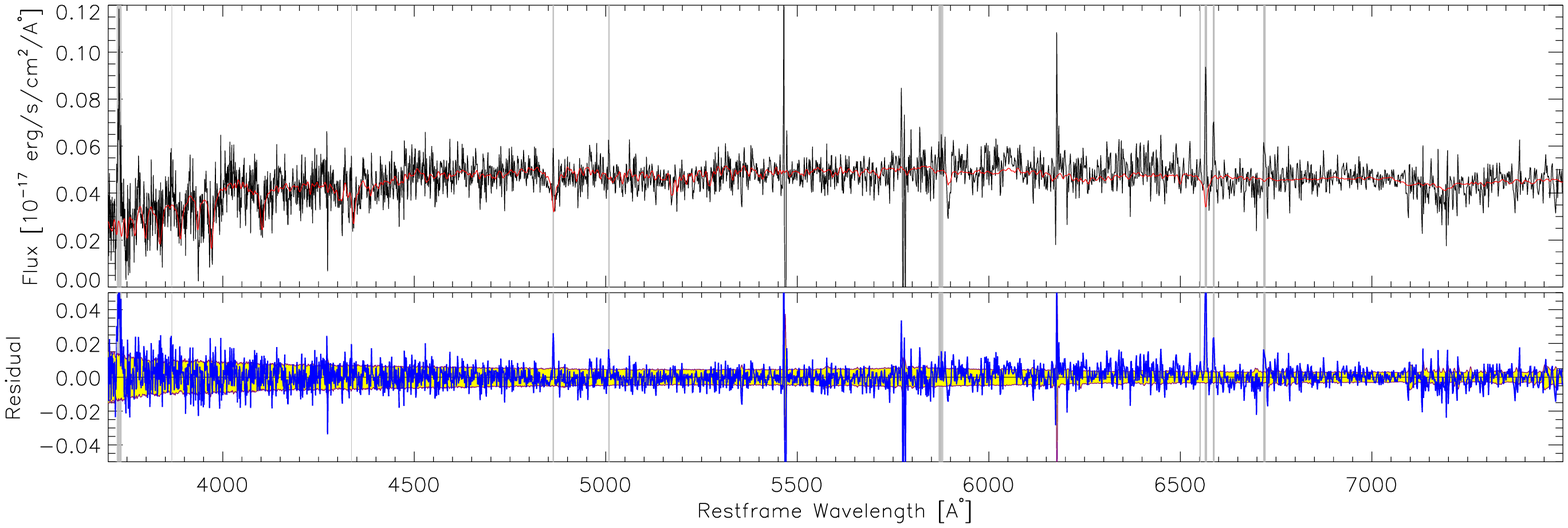,clip=true,width=0.9\textwidth}
    \epsfig{figure=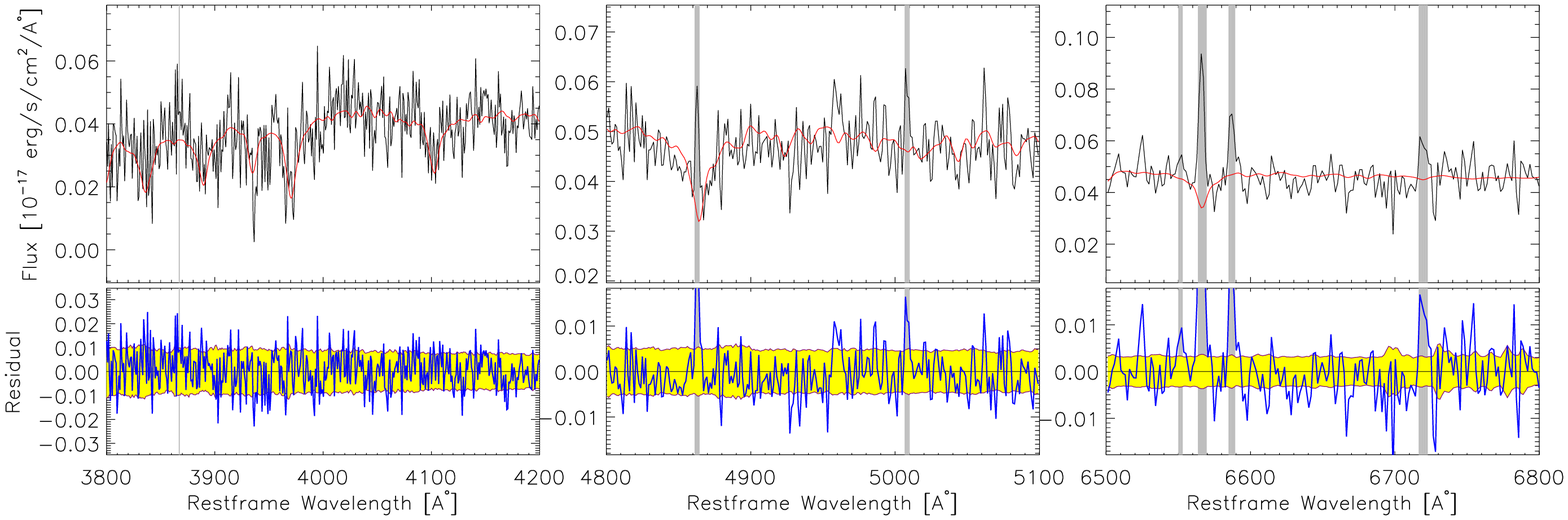,clip=true,width=0.9\textwidth}
  \end{center}
  \caption{Example  spectra  of the P-MaNGA galaxies. The top two
    rows of panels are for a high-S/N spectrum with $S/N\sim20$ at 5500\AA\ in
    the  continuum, while  the bottom two rows of panels are  for a spectrum with
    $S/N\sim8$ defined in the same  way. In each case the upper panel shows
    the whole wavelength range, while the lower three panels 
    show the three wavelength intervals with prominent emission lines.
    In  each  panel the  black  and red  lines  are  the observed  and
    best-fit spectra, while  the blue line in the  lower smaller panel
    is  the  difference between  the  two.   The vertical,  gray-shaded
    regions indicate the bandpasses  of the emission-line region which
    is masked during the spectral fitting.}
  \label{fig:spectral_fitting}
\end{figure*}

For each spectrum in the datacubes generated as described in
\S\ref{sec:p-manga}, we have performed a decomposition of the
emission-line component and the continuum plus absorption-line
component (hereafter called the `stellar component'), using both the
public spectral fitting code {\tt STARLIGHT} \citep{CidFernandes-04b}
and our own code described in \citet{Li-05}.  
We have carefully masked out the emission-line regions when doing the
spectral fitting.
Figure~\ref{fig:spectral_fitting} displays the results of this
procedure when applied to two representative spectra from the MaNGA
datacubes, one with a high signal-to-noise ratio ($S/N\sim20$) at
5500\AA\ in the continuum and one with a more modest $S/N\sim8$.

\begin{figure*}
  \begin{center}
    \epsfig{figure=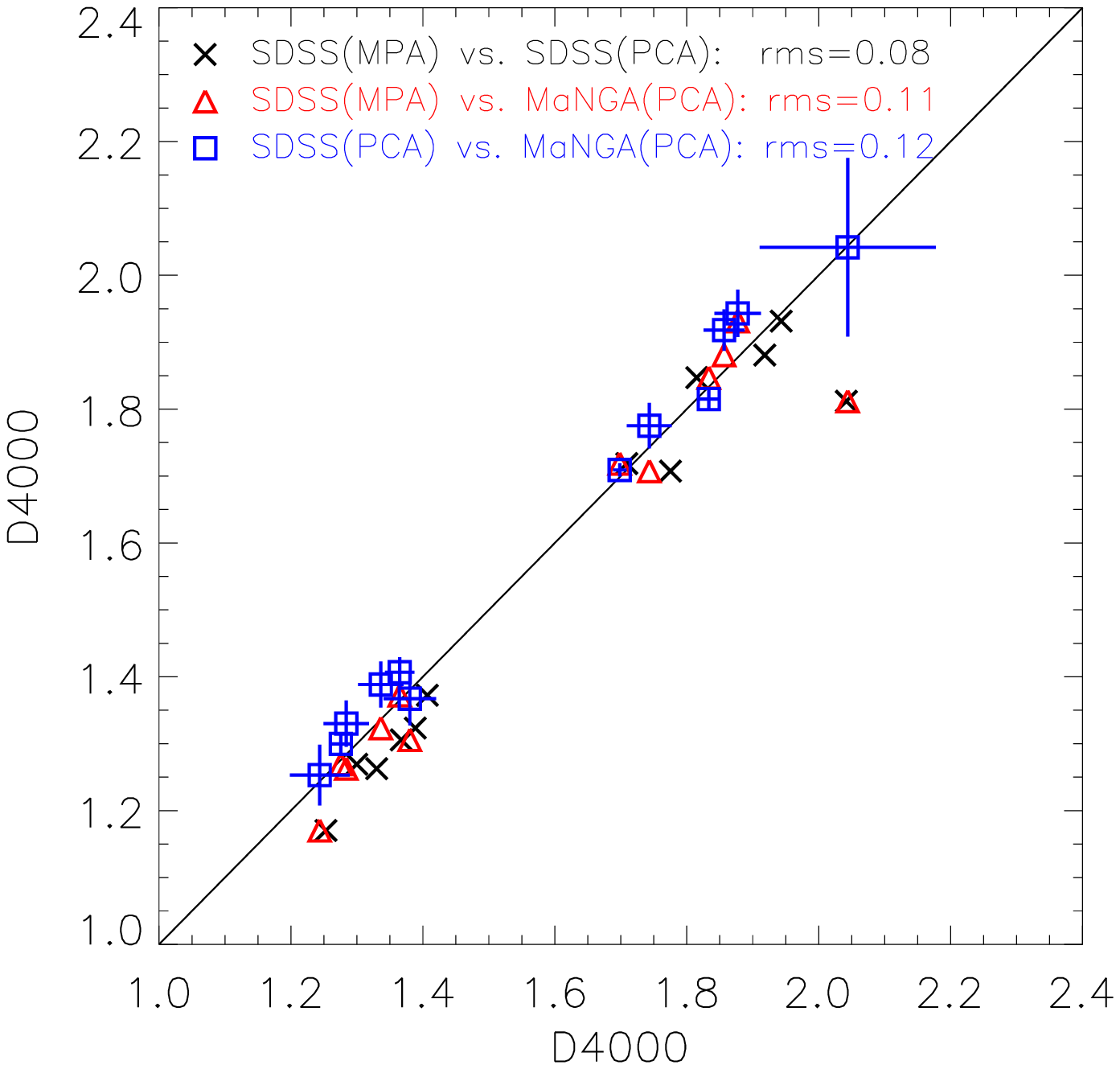,clip=true,width=0.3\textwidth}
    \epsfig{figure=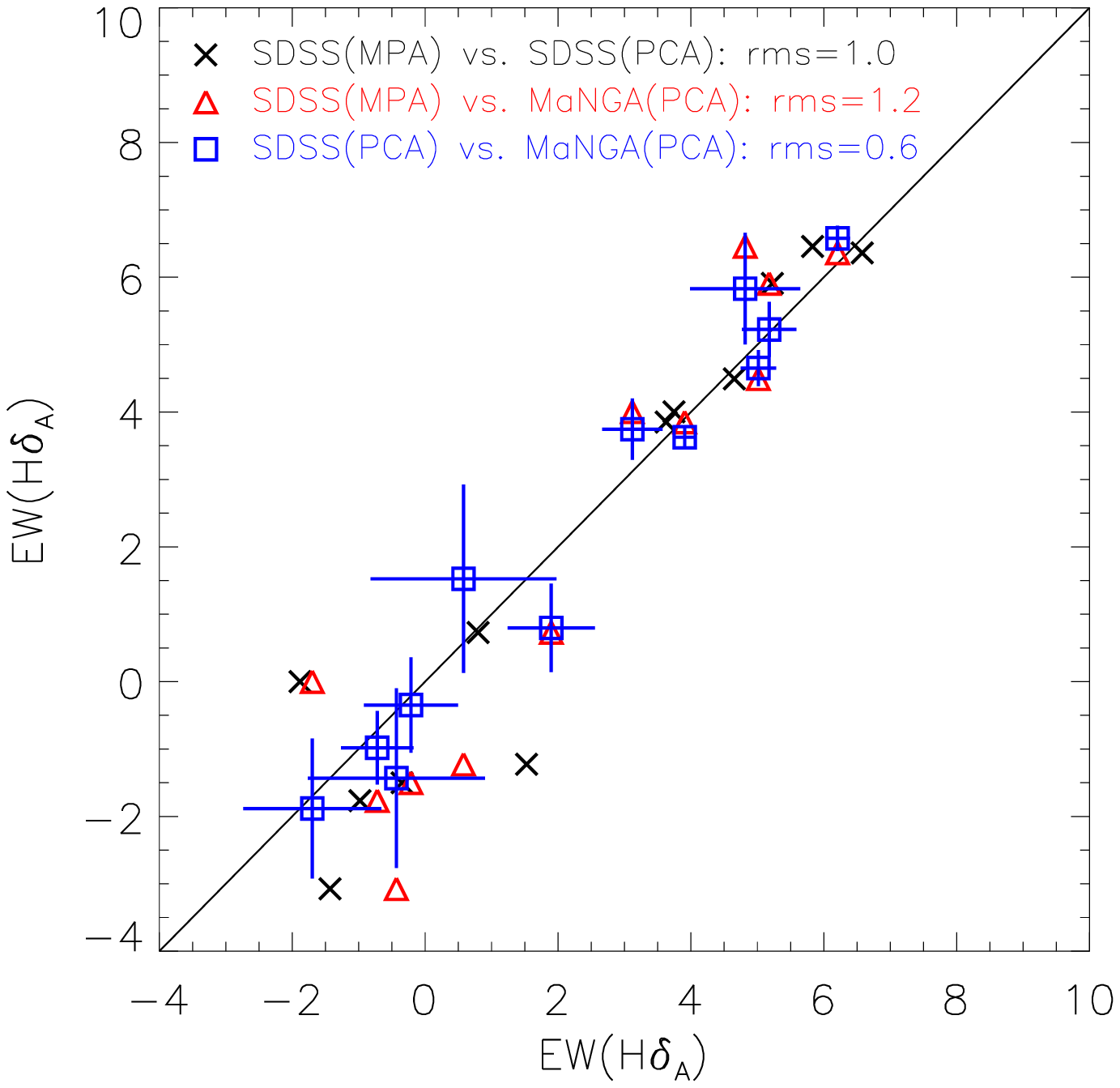,clip=true,width=0.3\textwidth}
    \epsfig{figure=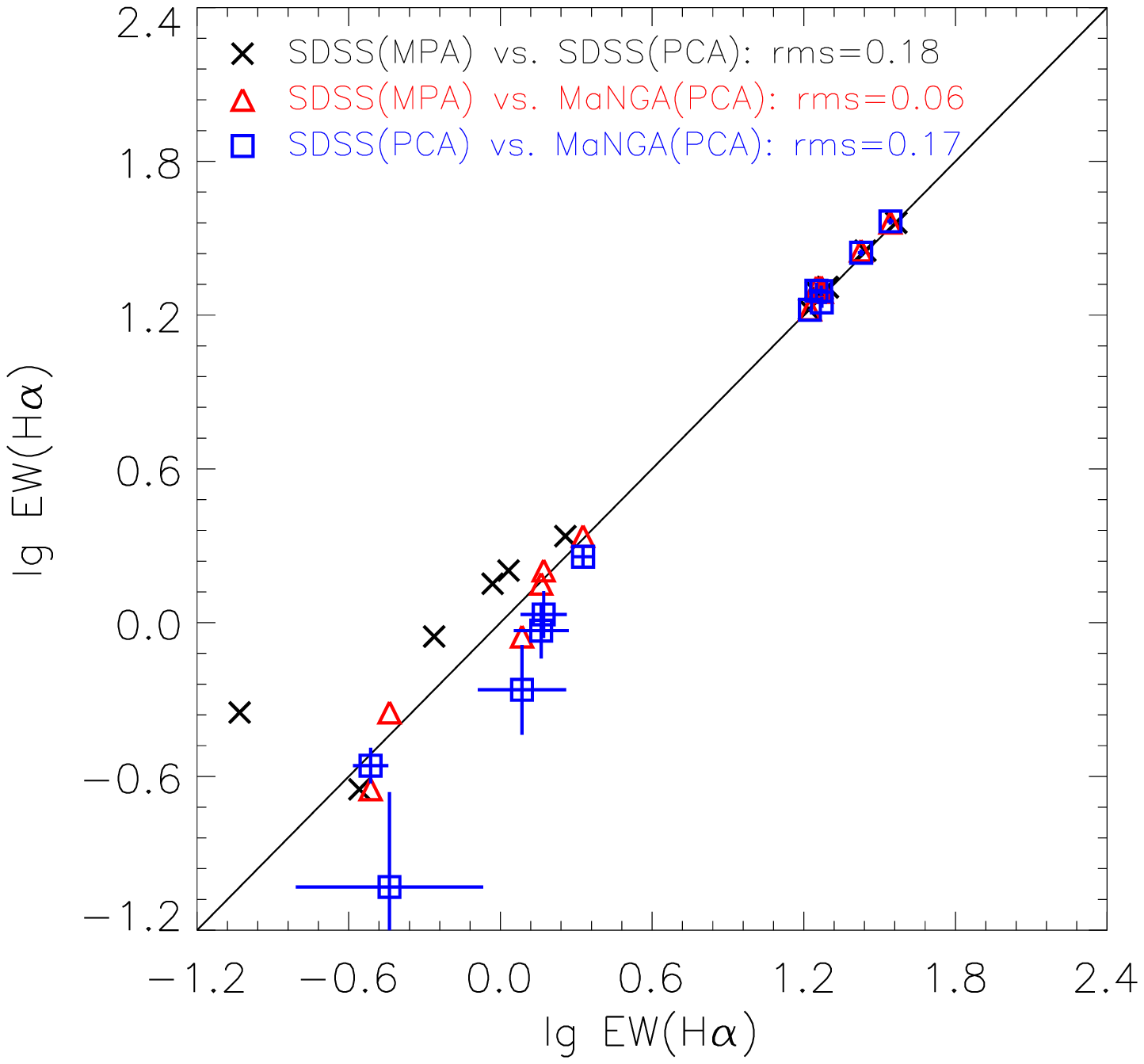,clip=true,width=0.3\textwidth}
  \end{center}
  \caption{Measurements  of  \dindex\  (left panel),  \ewhda\  (center
    panel) and \lgewhae\ (right  panel) for the P-MaNGA galaxies. Blue
    squares compare  the measurements obtained by applying  our code to
    the  central spaxel  of P-MaNGA  datacubes and the SDSS spectra,
    which  are compared  in red  triangles  and black  crosses to  the
    measurements taken from the MPA/JHU database. For clarity errors are
    shown only on the blue squares, which compare the measurements
    obtained from the SDSS and P-MaNGA spectra. For given parameter
    its error is estimated by the $1-\sigma$ scatter between the three measurements.}
  \label{fig:compare_d4k_hda_hae}
\end{figure*}

\begin{figure*}
  \begin{center}
    \epsfig{figure=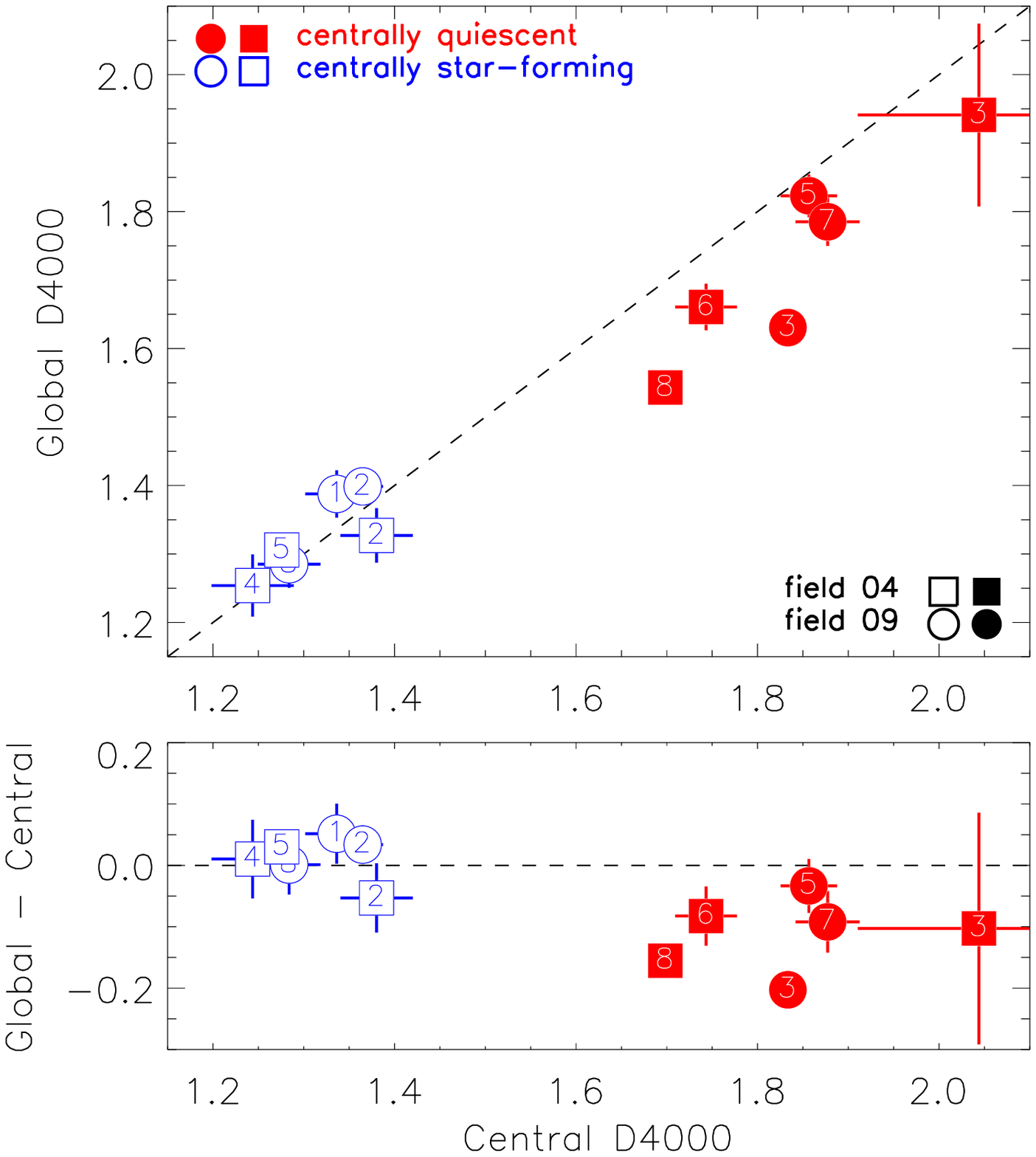,clip=true,width=0.3\textwidth}
    \epsfig{figure=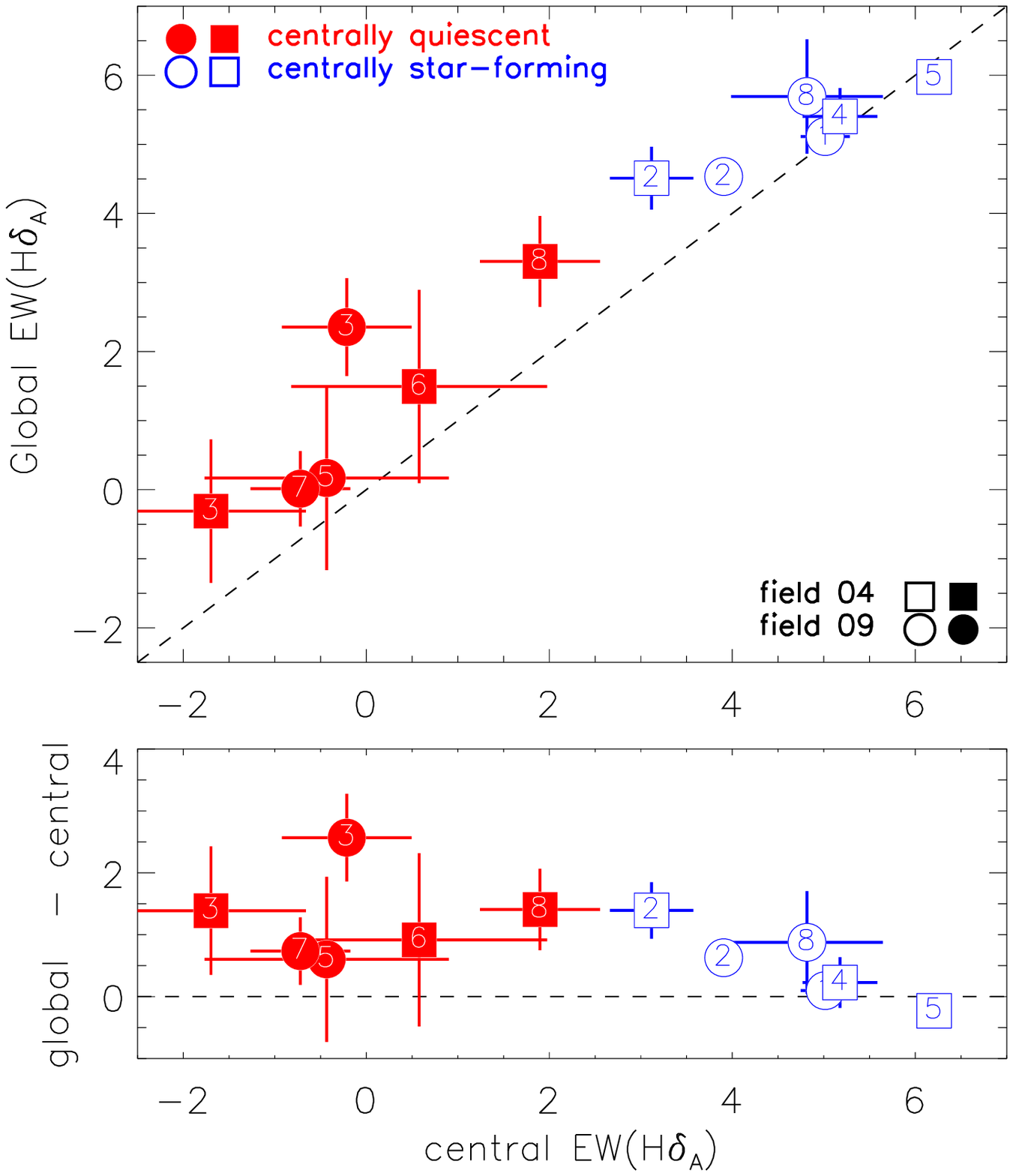,clip=true,width=0.3\textwidth}
    \epsfig{figure=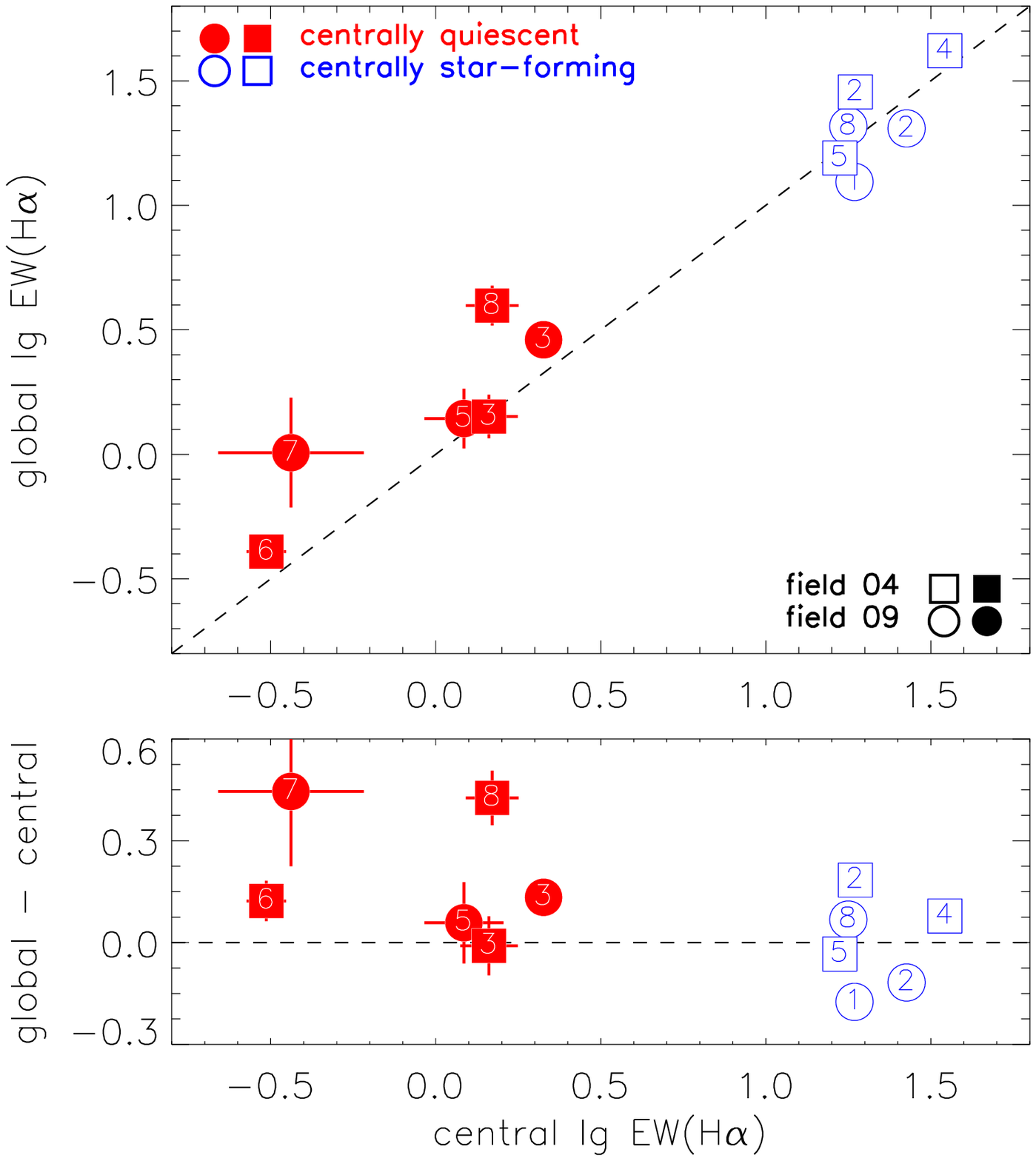,clip=true,width=0.3\textwidth}
  \end{center}
  \caption{Upper  panels  compare the global \dindex, \ewhda\  and
    \lgewhae\  of  the  P-MaNGA  galaxies  with  the  same  parameters
    measured from the central  spaxel of their datacubes. Lower panels
    show the difference between the global and central parameters as a
    function of  the central one.  The global  parameters are measured
    from the average spectrum for each galaxy obtained by stacking the
    entire  datacube,  with each  spaxel  being  weighted by  $S/N^2$.
    Symbols     and     colors      are     the     same     as     in
    Figure~\ref{fig:sample_properties}. For given parameter, the error 
    is given by the 1-$\sigma$ scatter between the three different measurements
    as shown in Figure~\ref{fig:compare_d4k_hda_hae}.}
  \label{fig:cen_vs_glb}
\end{figure*}

We measure the \dindex\ and \hda\ indices from each best-fit spectrum,
defining the indices in the same way as in the previous subsection. We
have also measured the flux and equivalent width of the emission lines
(both Balmer and forbidden) by fitting a Gaussian profile to these
lines in the stellar component-subtracted spectrum.  The emission-line
parameters from {\tt STARLIGHT} were compared to those obtained from
our code. We have also performed the same procedure of spectral
fitting and parameter measuring on Voronoi-binned versions of the
datacubes to assess the effects of spatial averaging before fitting.
All these analyses were found to return indistinguishable results, and
so in what follows we will only present the values we obtain when
applying our code to the unbinned datacubes.  

We have also applied our code to the archival SDSS spectra of the
centers of the P-MaNGA galaxies, and in
Figure~\ref{fig:compare_d4k_hda_hae} we compare the resulting
measurements of \dindex, \ewhda\ and \lgewhae\ to those obtained from
the central spaxel of the P-MaNGA datacubes (plotted as blue squares
in the figure).  In the same figure, we also compare our measurements,
from both SDSS (black crosses) and the central spaxel of P-MaNGA
datacubes (red triangles) with the measurements taken from the MPA/JHU
database.  Overall, the values from different observations and
pipelines agree well with each other.  We have visually examined the
best-fit spectrum for the few outliers, and find that our code
provides a reasonably good fit to both SDSS and P-MaNGA spectra even
in these cases.  The relatively large discrepancy in one of the
outliers (the blue square and black cross located in the left-bottom
corner in the right-hand panel) is caused by the bad data pixels over
the H$\alpha$ band in the SDSS spectrum.

\begin{figure*}
  \begin{center}
    \epsfig{figure=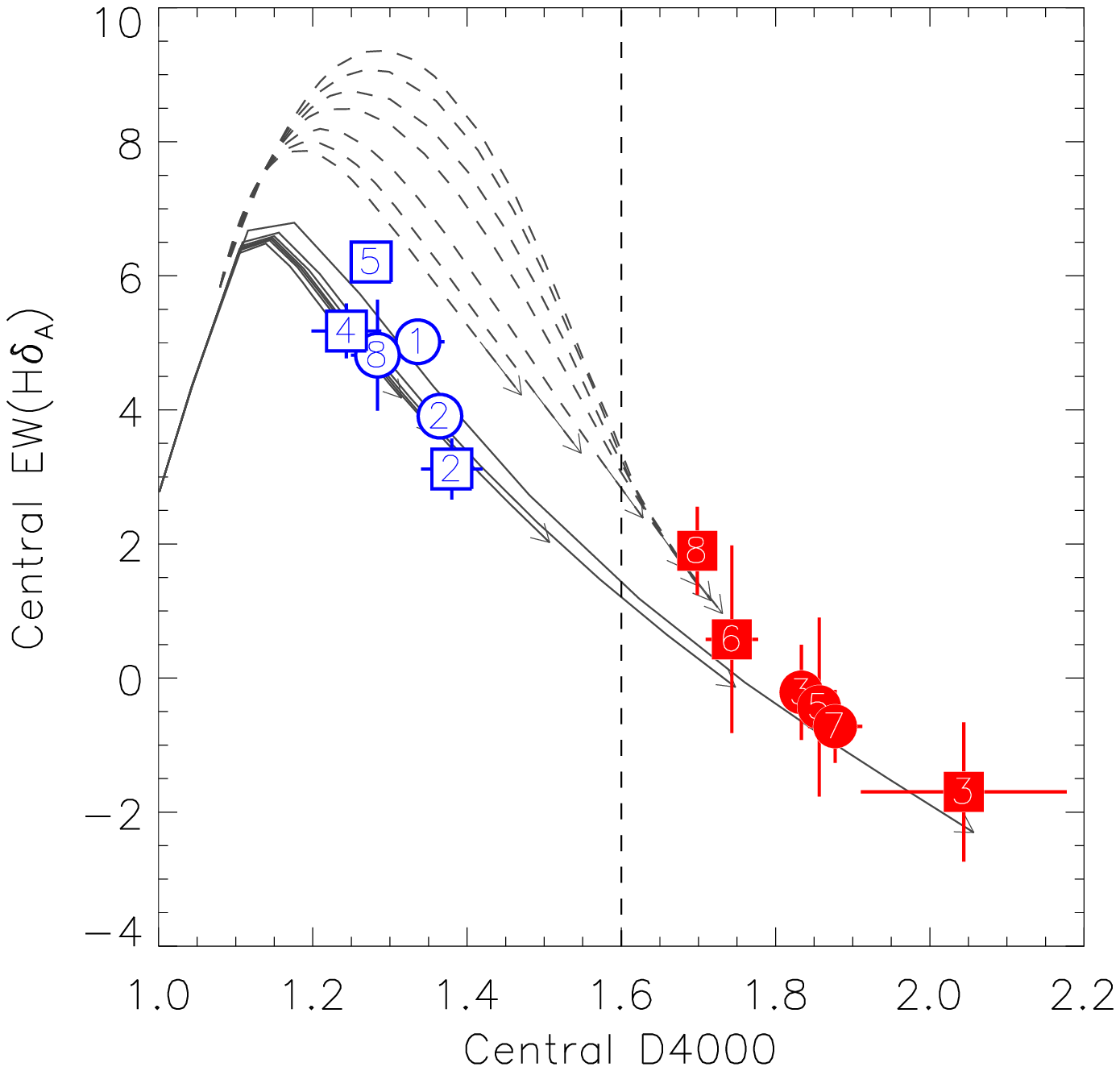,clip=true,width=0.35\textwidth}
    \epsfig{figure=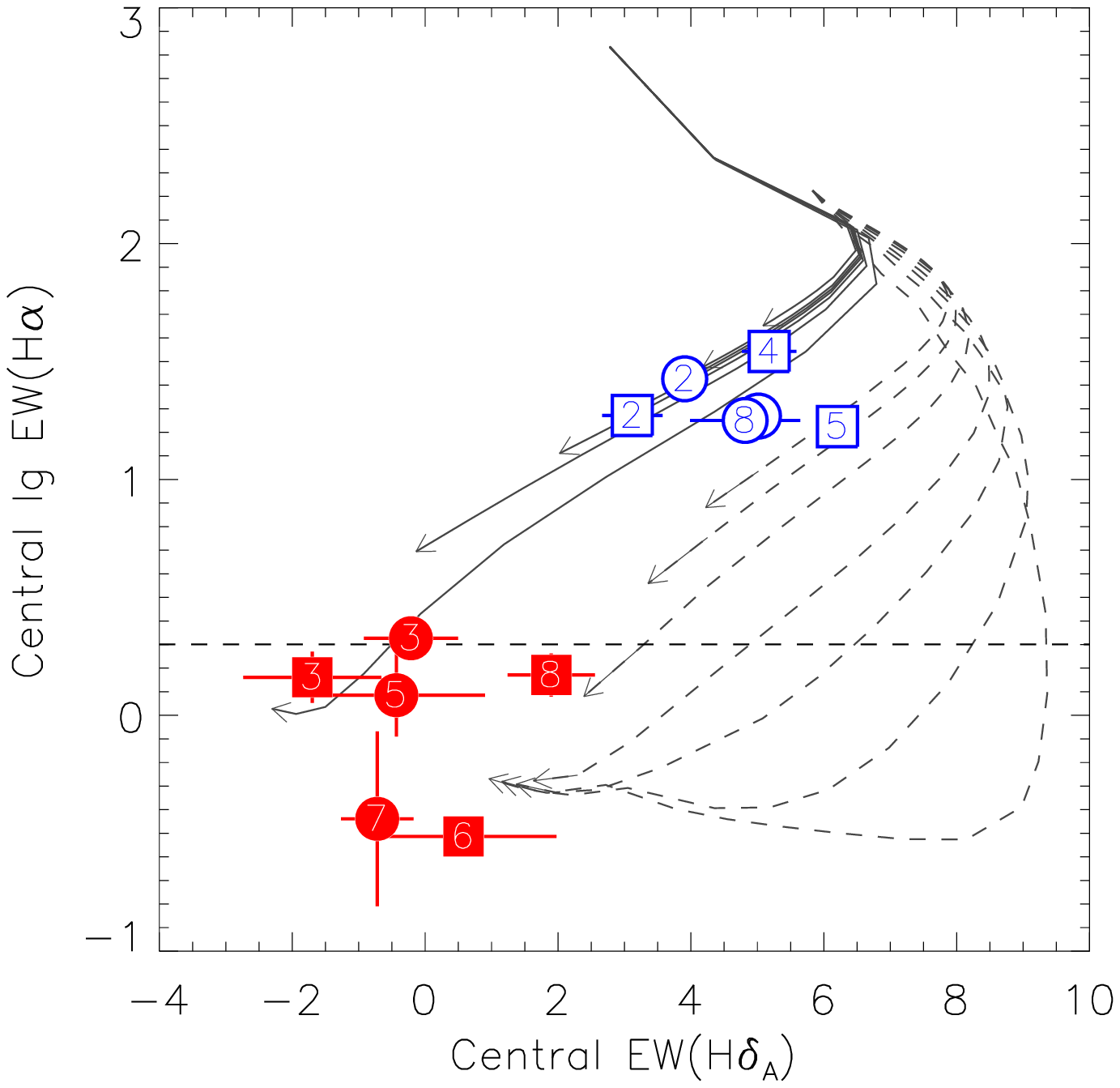,clip=true,width=0.35\textwidth}
  \end{center}
  \begin{center}
    \epsfig{figure=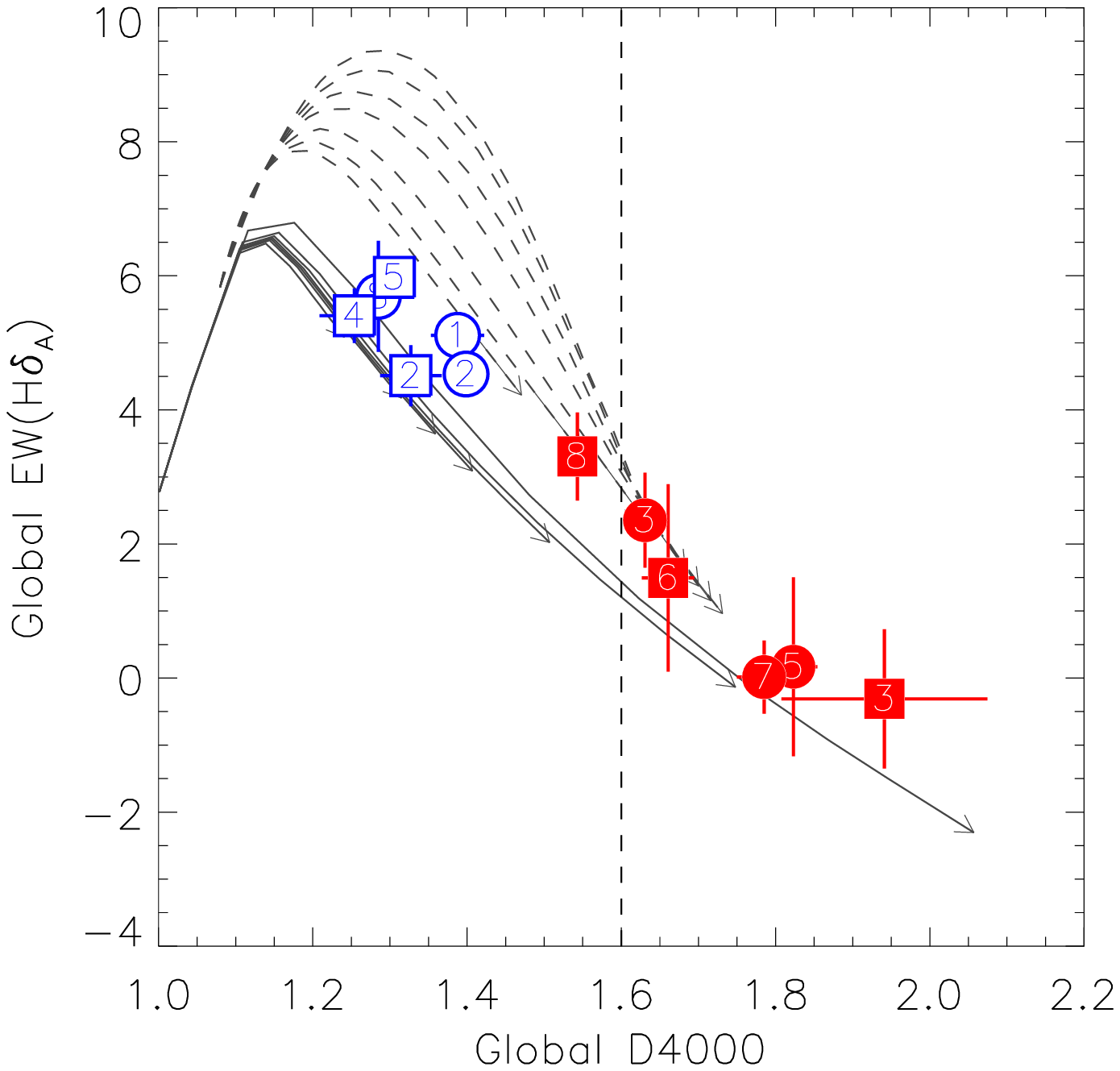,clip=true,width=0.35\textwidth}
    \epsfig{figure=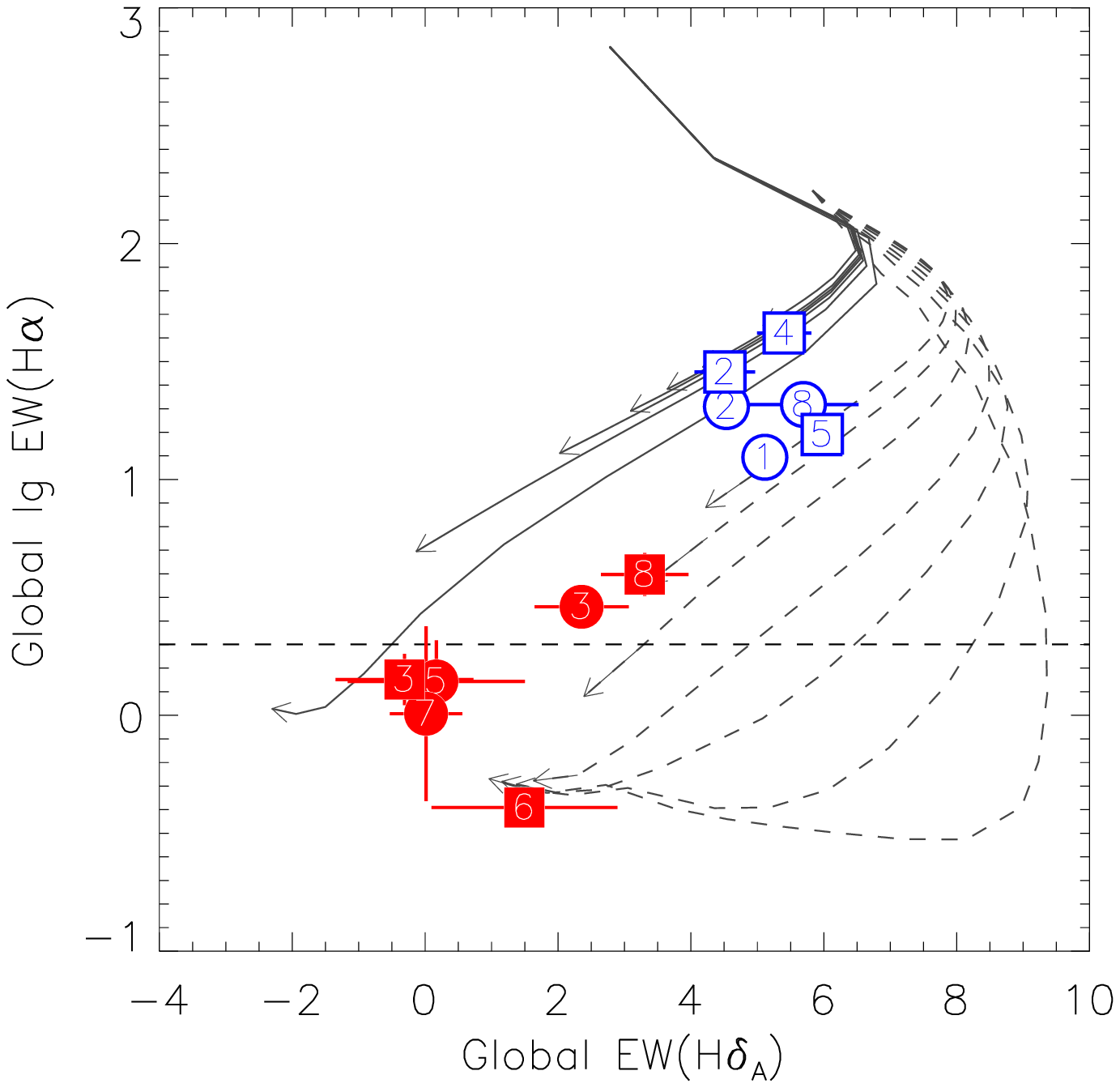,clip=true,width=0.35\textwidth}
  \end{center}
  \caption{Distribution of the P-MaNGA galaxies (symbols) on the plane
    of 4000\AA-break vs. \hda\ index (left-hand panels), and the plane
    of \hda\ index vs. \lgewhae, the equivalent width of \ha\ emission
    line in logarithm (right-hand panels). Results in the upper panels
    are for the  central spaxel of the P-MaNGA  datacubes.  Results in
    the lower panels are based  on the average spectrum, obtained from
    stacking the whole  datacube for given galaxy. Shown  in solid and
    dashed  lines  are solar  metallicity  models  of continuous  star
    formation  histories  and   star  formation  bursts
    calculated  using   the  stellar  population   synthesis  code  of
    \citet{Bruzual-Charlot-03}. Symbols for the MaNGA galaxies are the
    same as in the previous figure. For given parameter, the error
    is given by the 1-$\sigma$ scatter between the three different measurements
    as shown in Figure~\ref{fig:compare_d4k_hda_hae}.}
  \label{fig:d4k_hda_hae}
\end{figure*}

We note that, when fitting the spectra and measuring the
diagnostic parameters, we do not take into account the covariance
between wavelengths, nor the covariance between pixels. Methods for
tracking such covariances have been studied in depth \citep[e.g.][for
the SAMI survey]{Sharp-15}. As pointed out in \citet{Wilkinson-15},
when compared to P-MaNGA, MaNGA will provide more accurately estimated
spectral errors (thus accurate error propagation through to
the final datacubes), allowing construction of covariances between
pixels. We will apply a detailed treatment of covariance 
in MaNGA data in future work.

\begin{figure*}
  \begin{center}
    \epsfig{figure=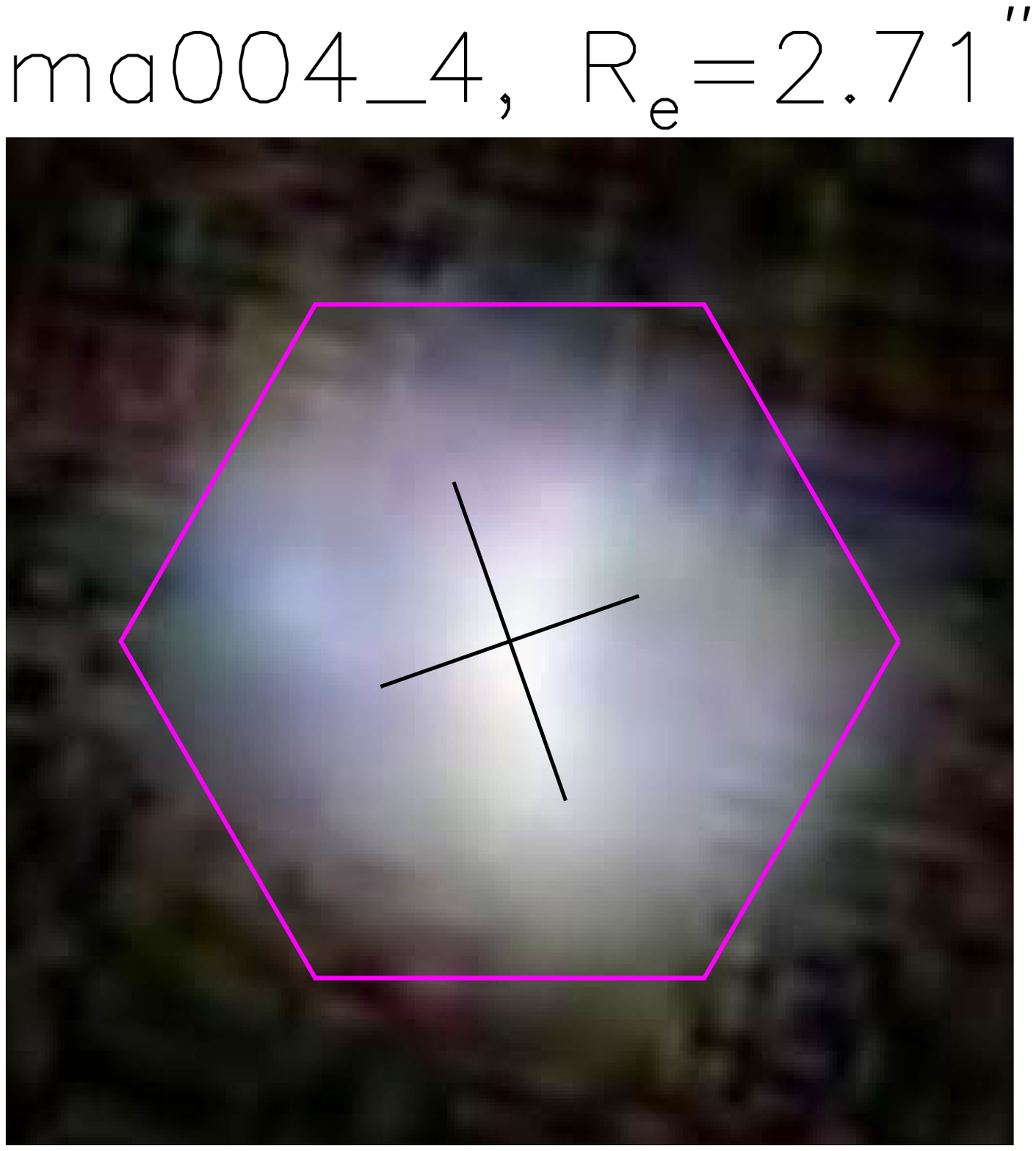,clip=true,height=0.140\textheight}
    \epsfig{figure=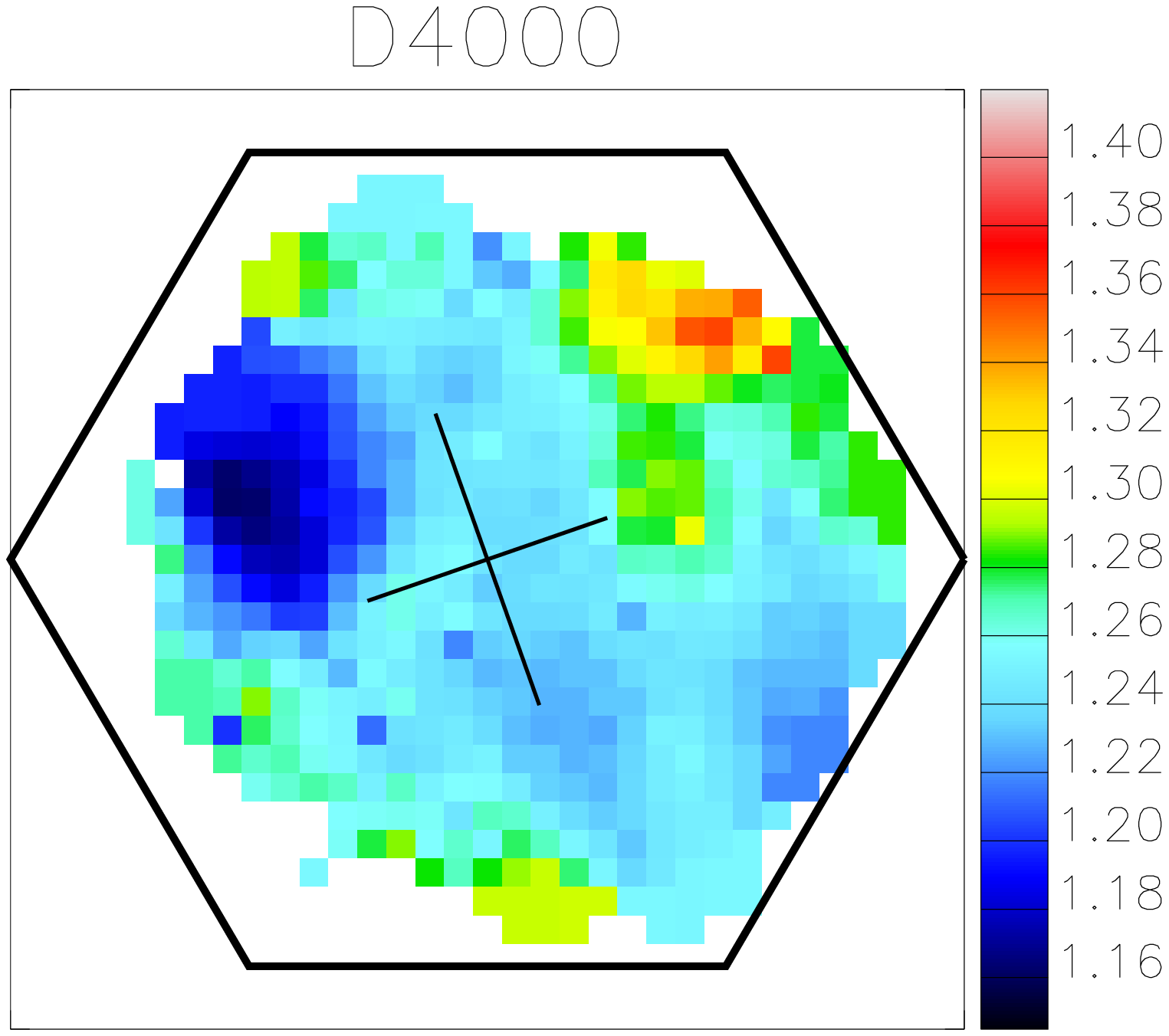,clip=true,height=0.140\textheight}
    \epsfig{figure=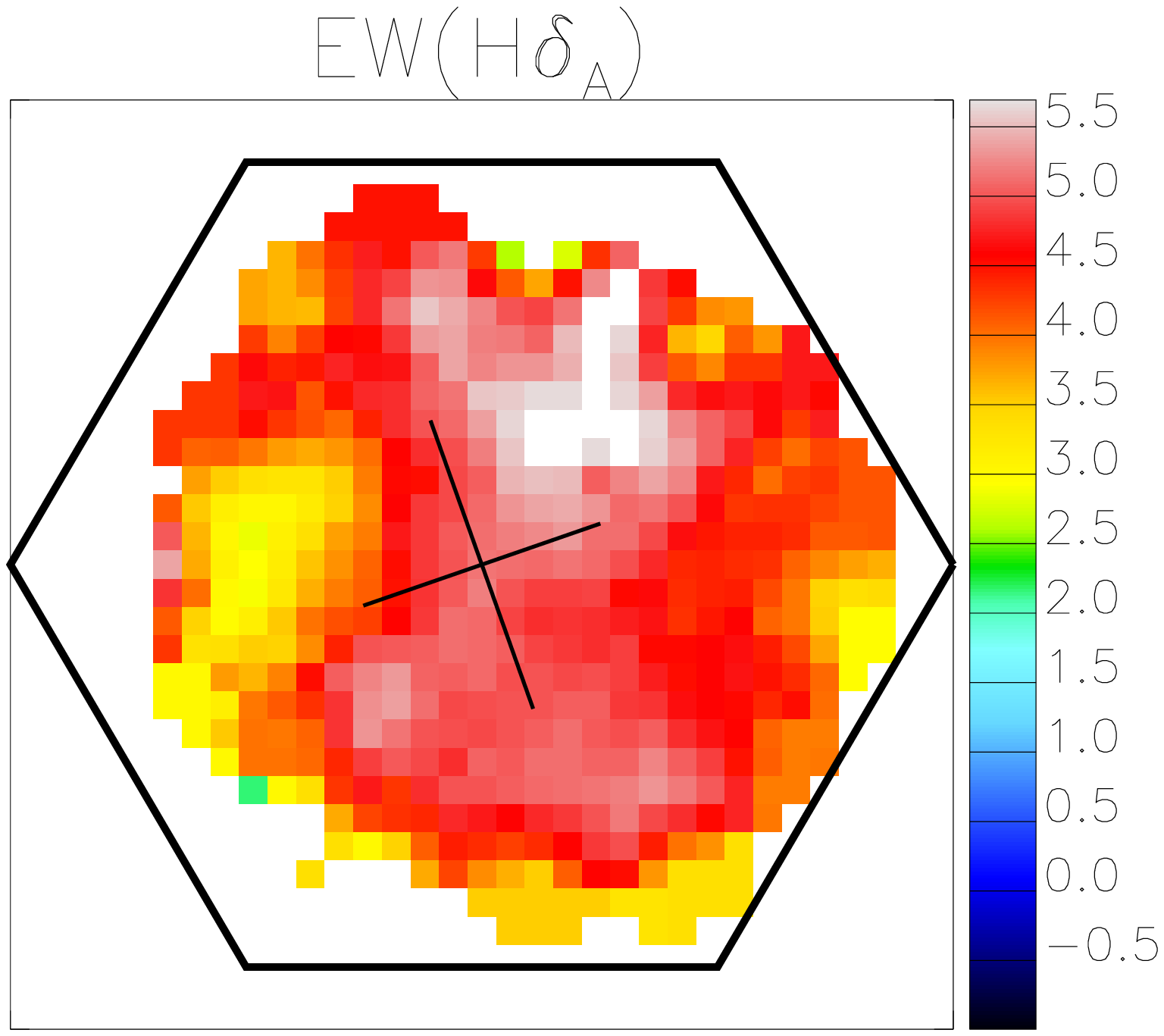,clip=true,height=0.140\textheight}
    \epsfig{figure=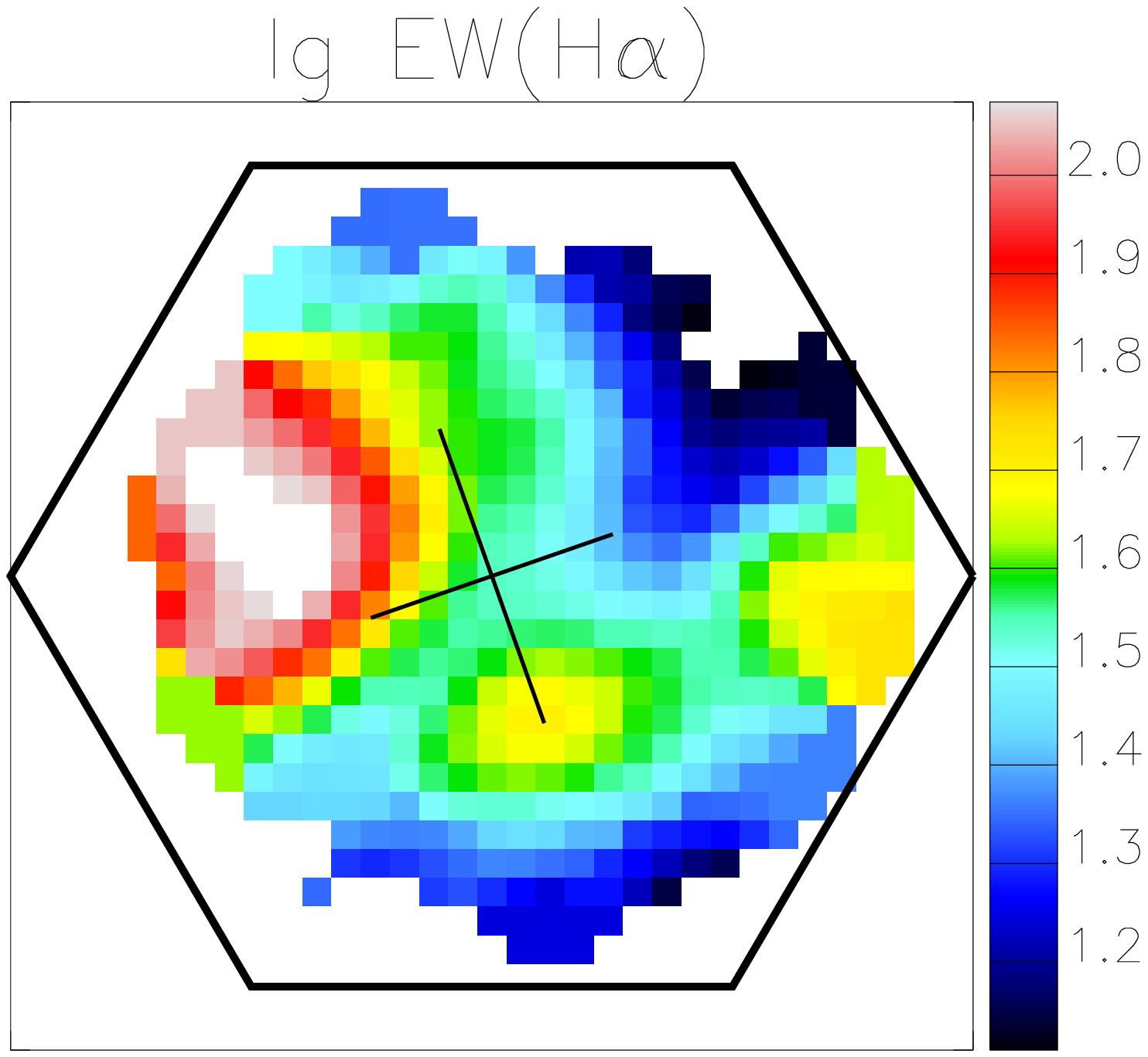,clip=true,height=0.140\textheight}
  \end{center}
  \begin{center}
    \epsfig{figure=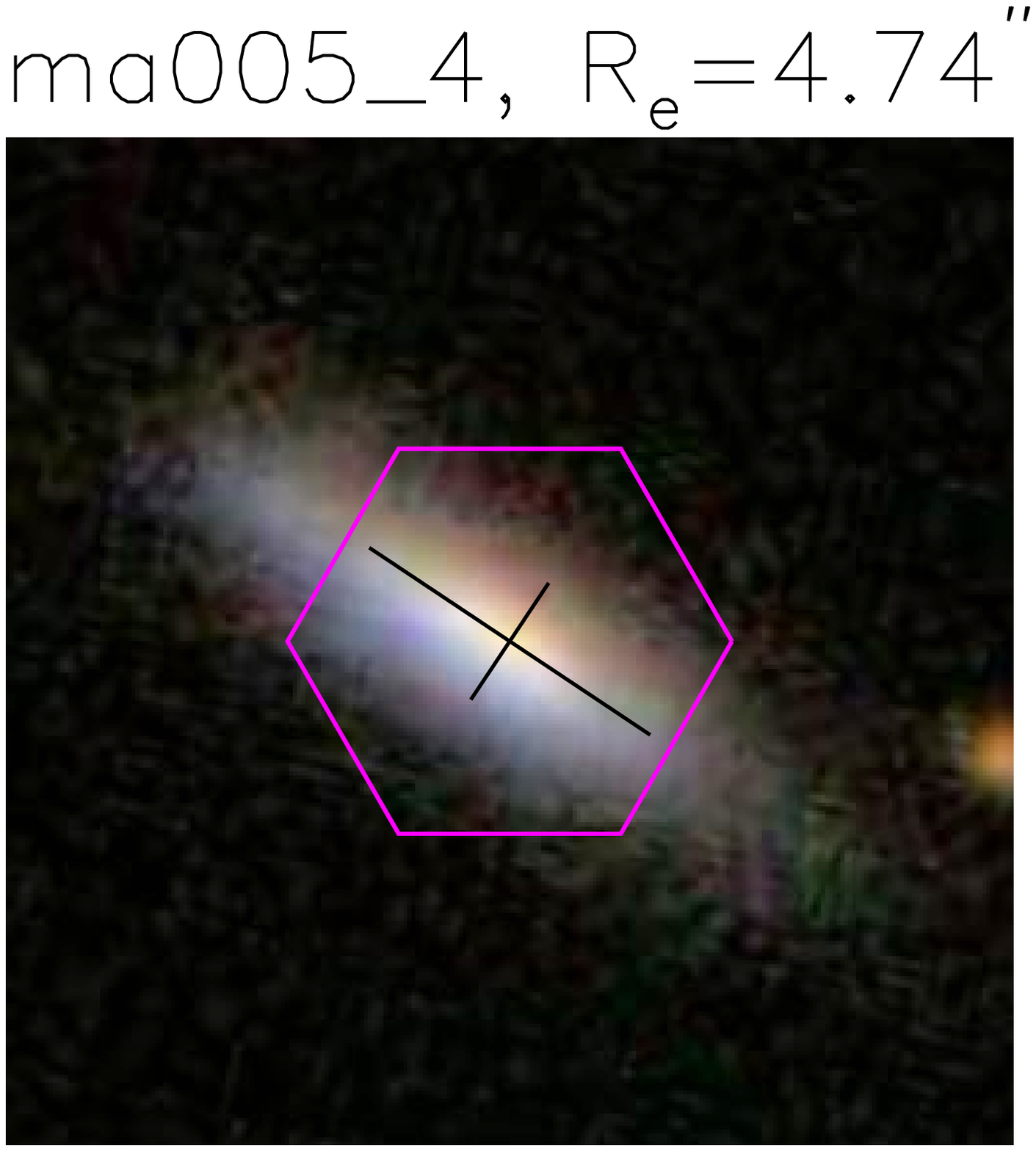,clip=true,height=0.140\textheight}
    \epsfig{figure=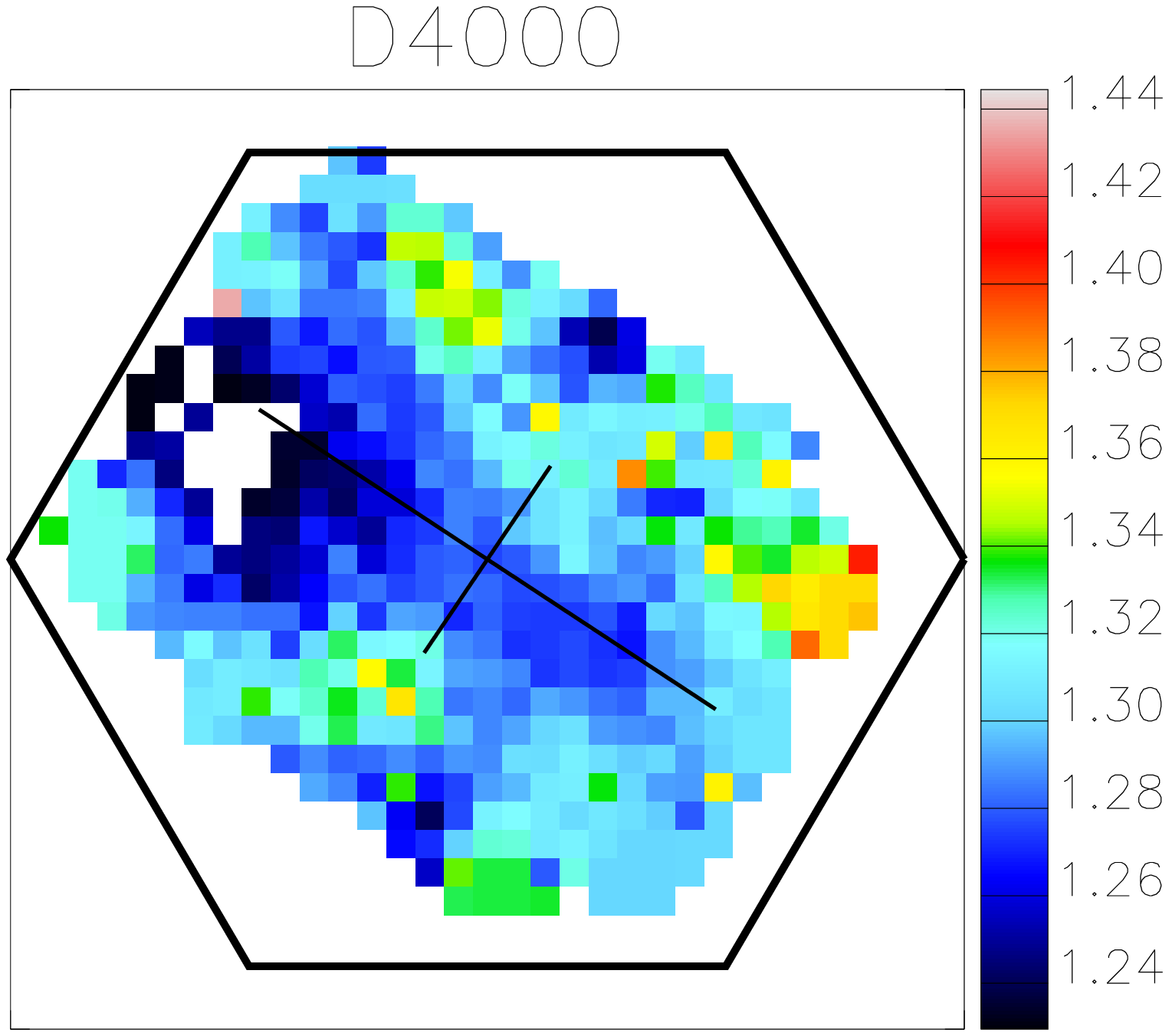,clip=true,height=0.140\textheight}
    \epsfig{figure=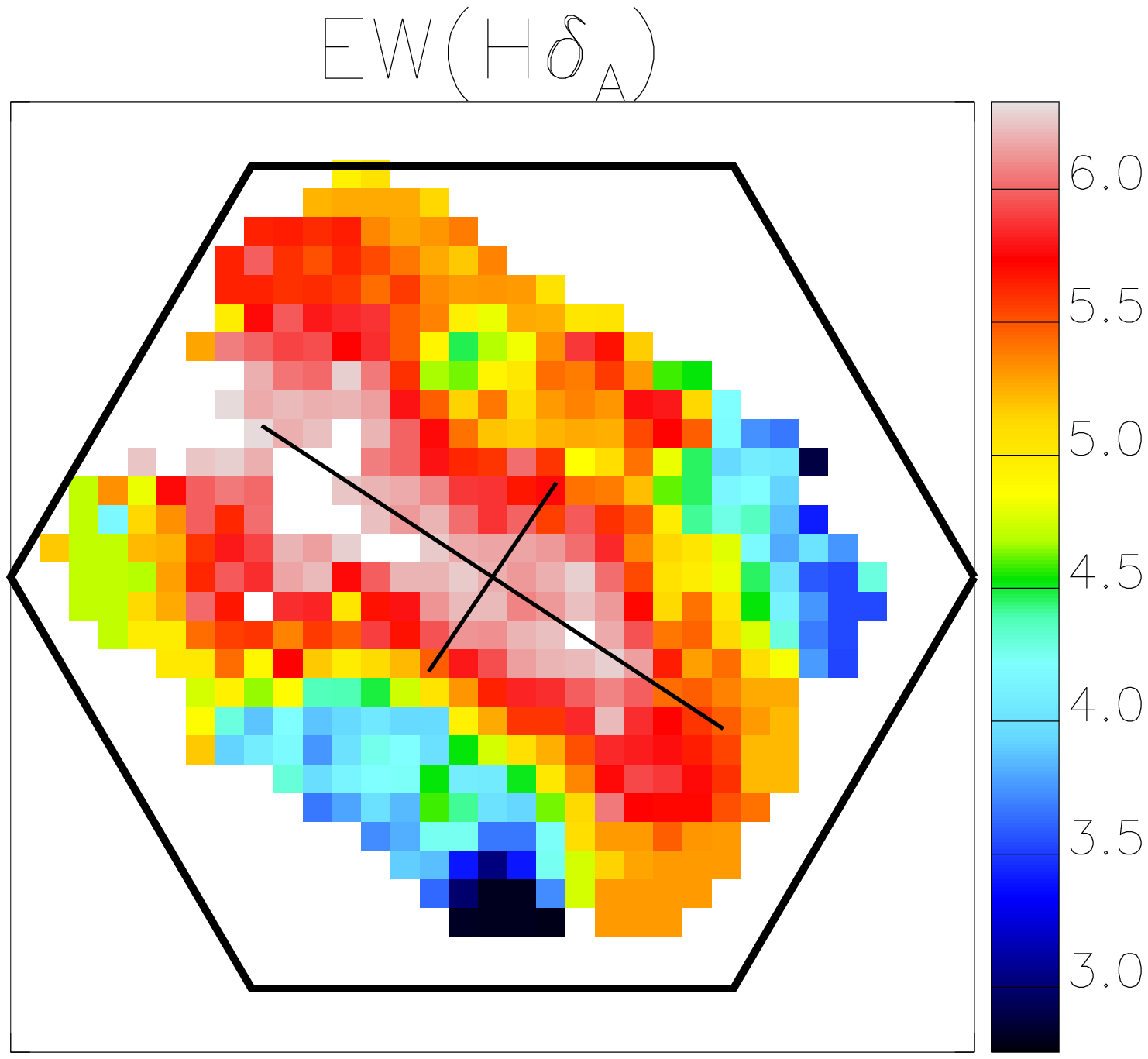,clip=true,height=0.140\textheight}
    \epsfig{figure=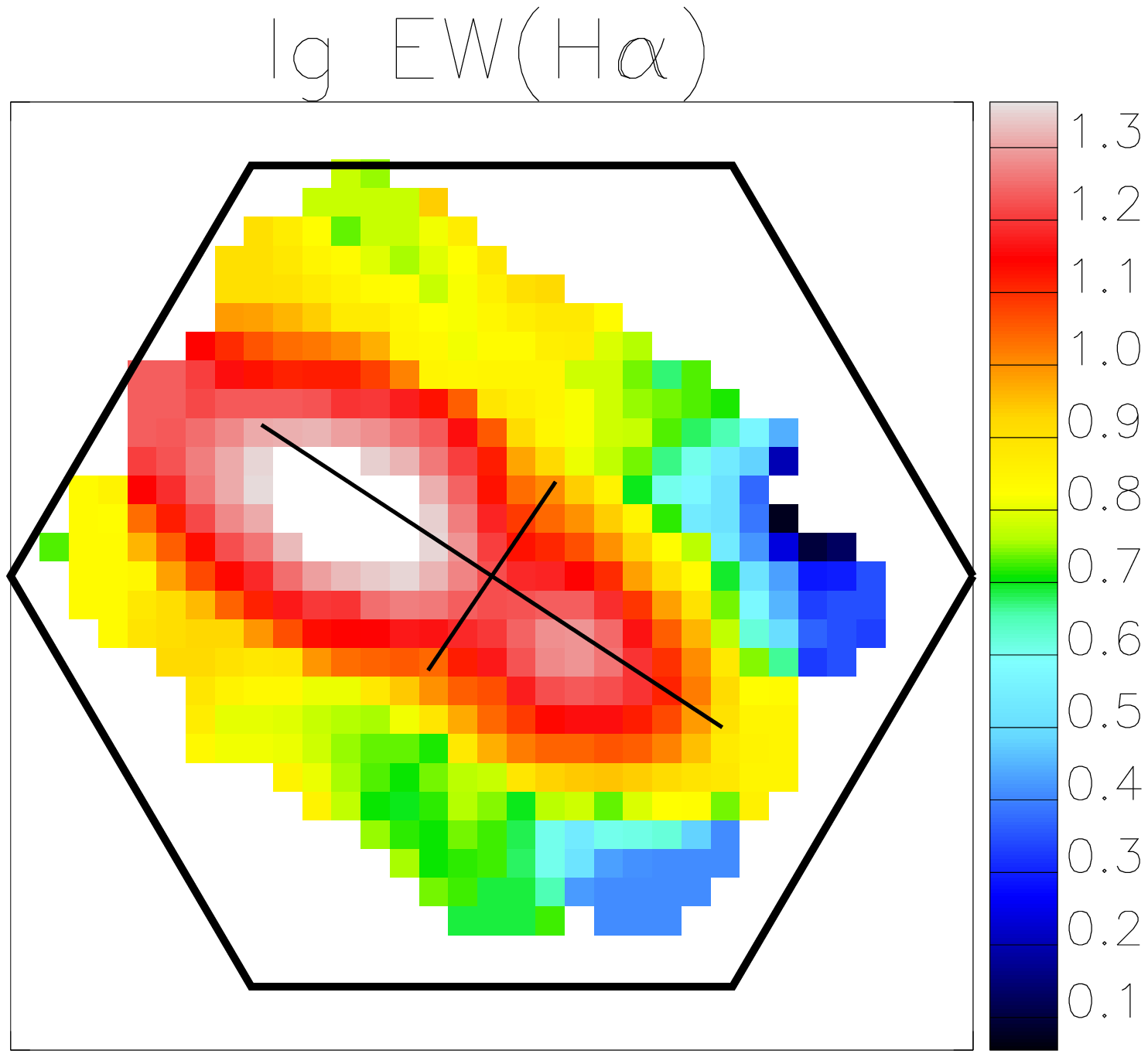,clip=true,height=0.140\textheight}
  \end{center}
  \begin{center}
    \epsfig{figure=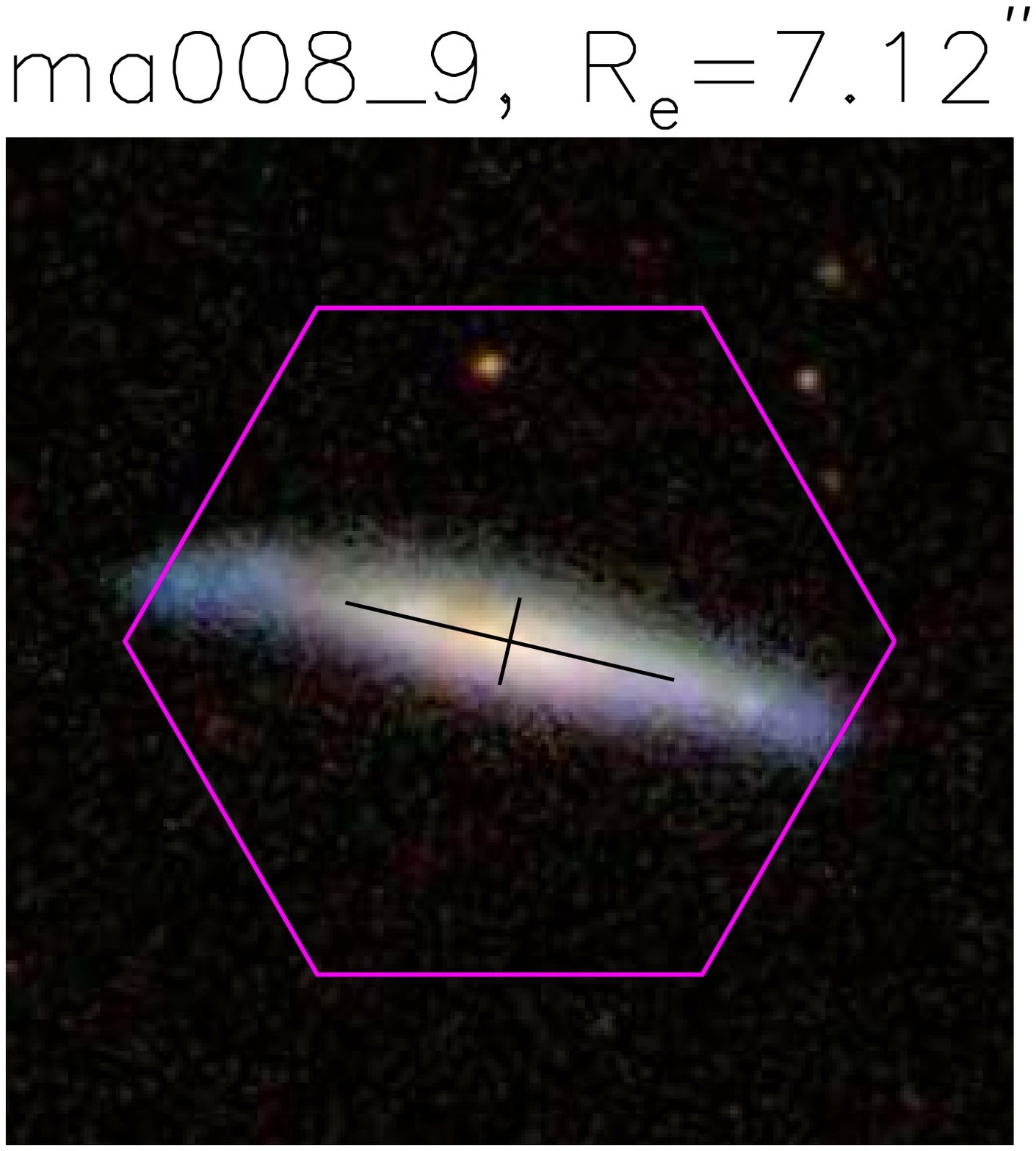,clip=true,height=0.140\textheight}
    \epsfig{figure=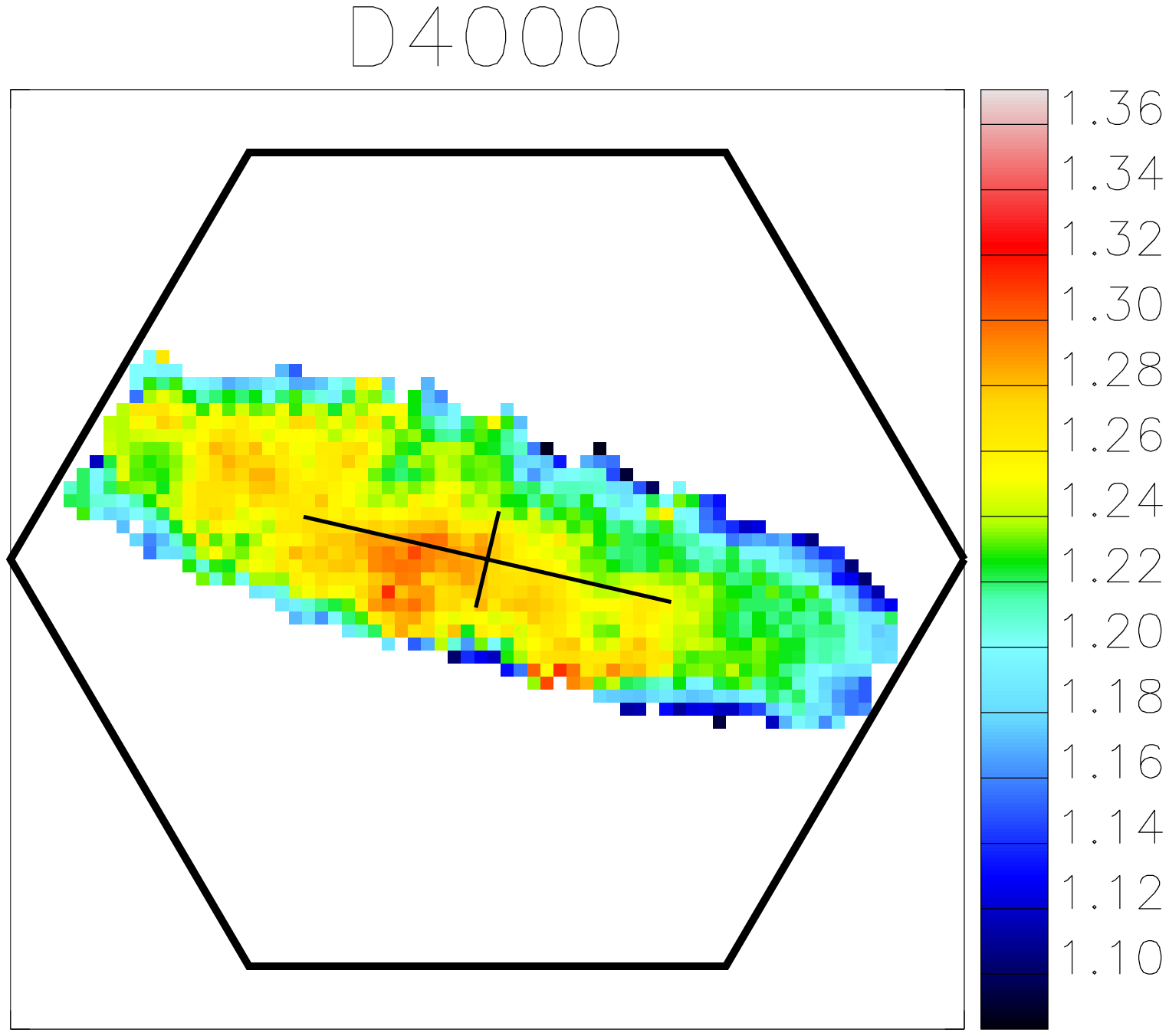,clip=true,height=0.140\textheight}
    \epsfig{figure=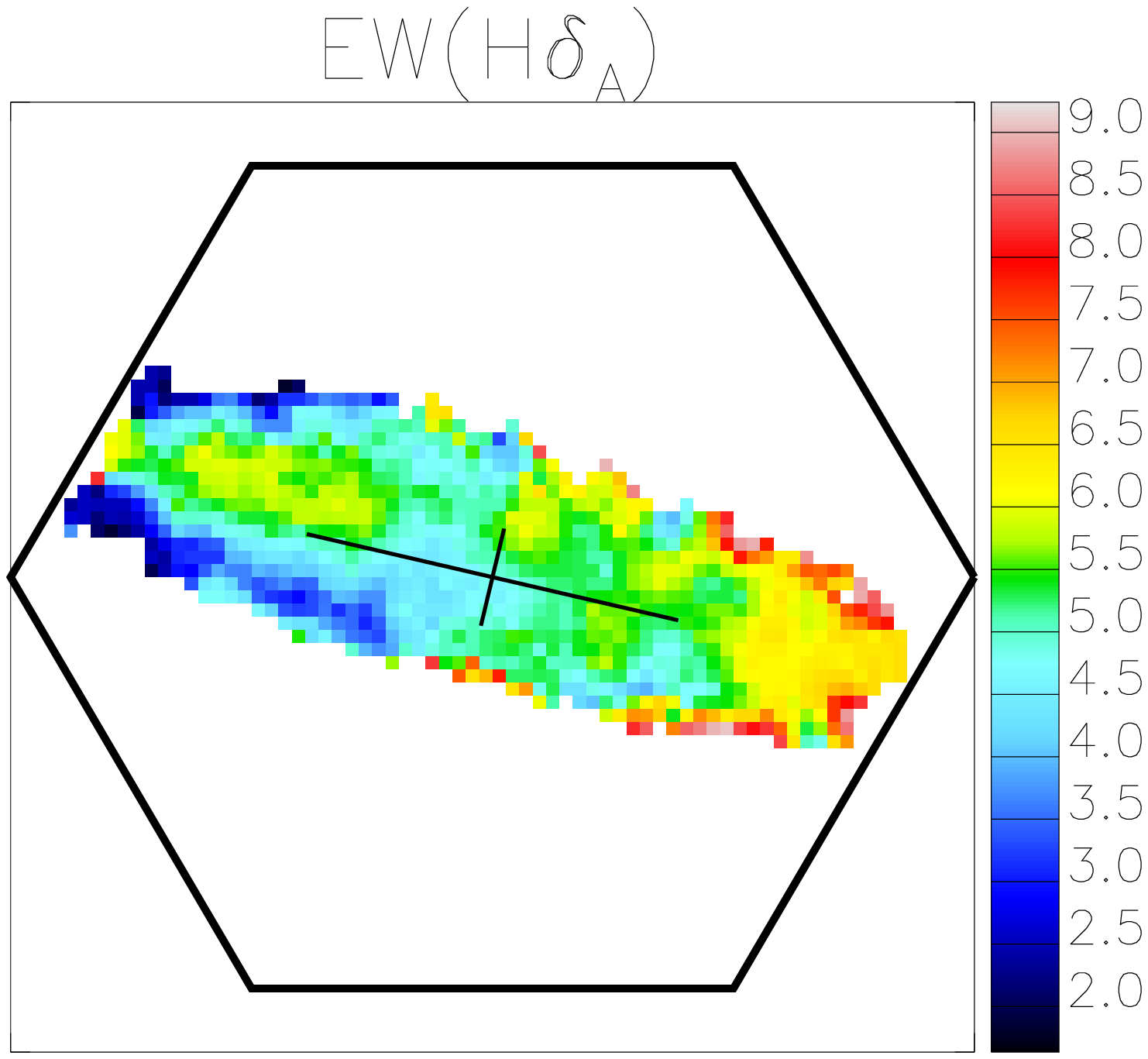,clip=true,height=0.140\textheight}
    \epsfig{figure=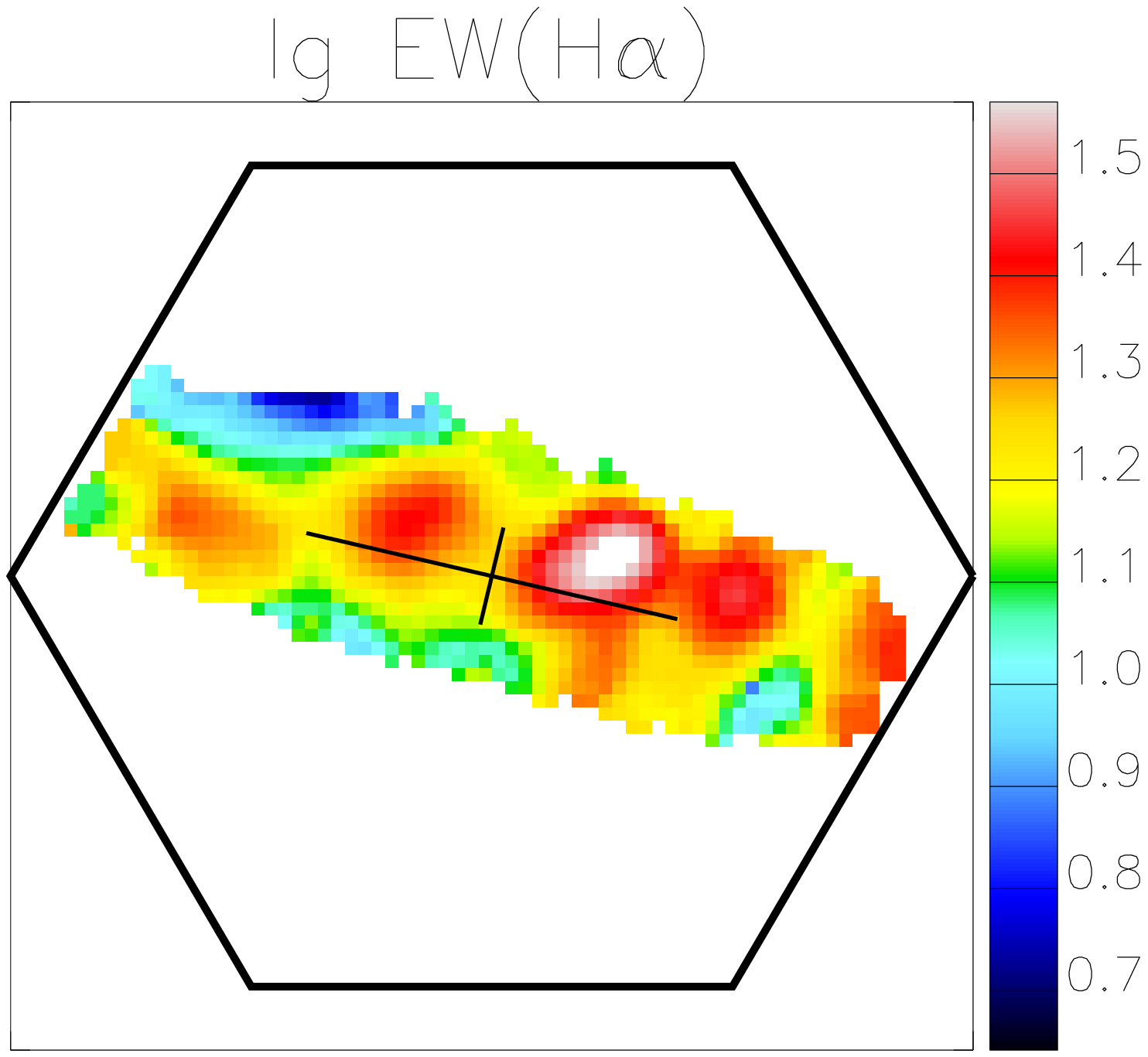,clip=true,height=0.140\textheight}
  \end{center}
  \begin{center}
    \epsfig{figure=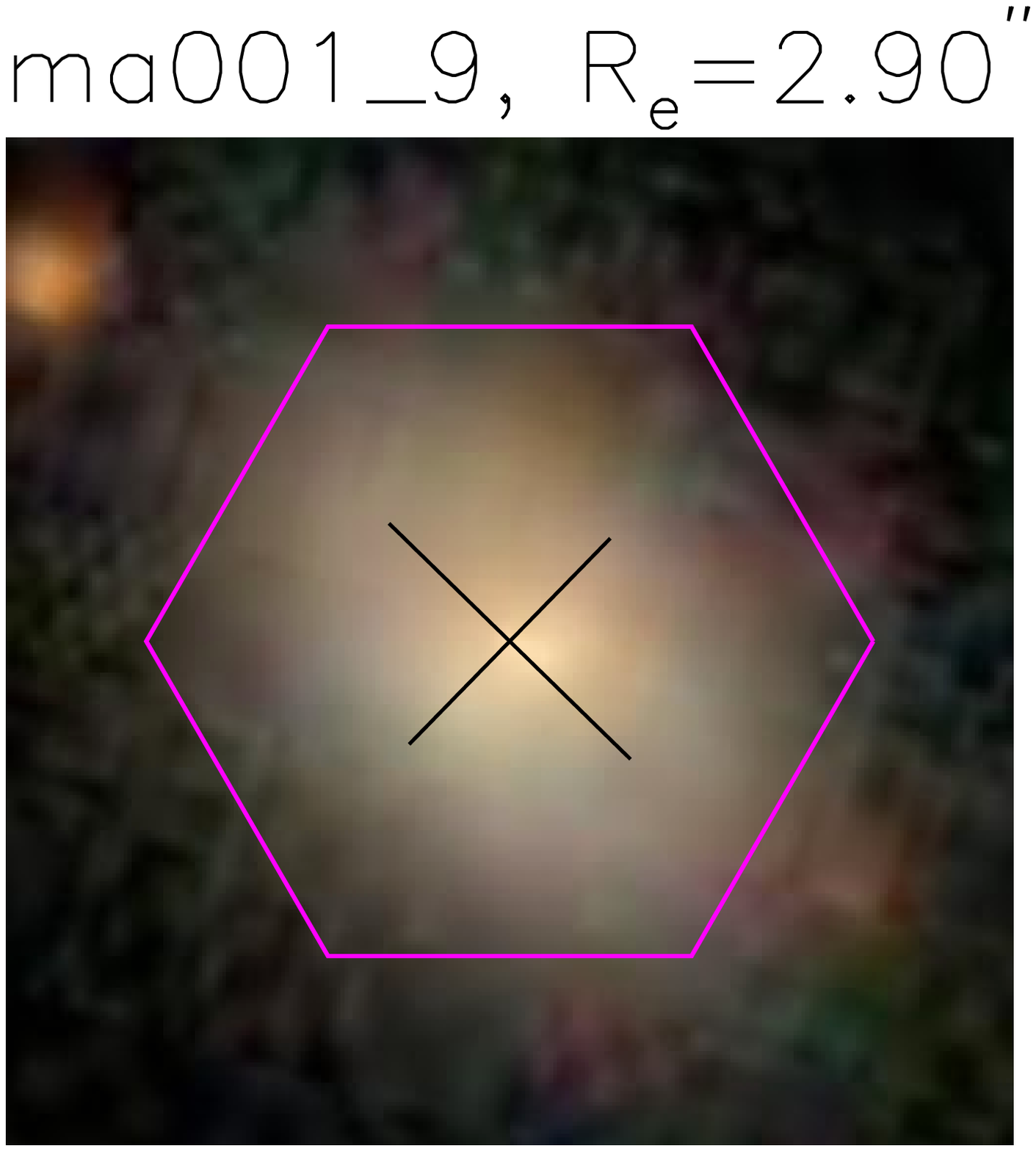,clip=true,height=0.140\textheight}
    \epsfig{figure=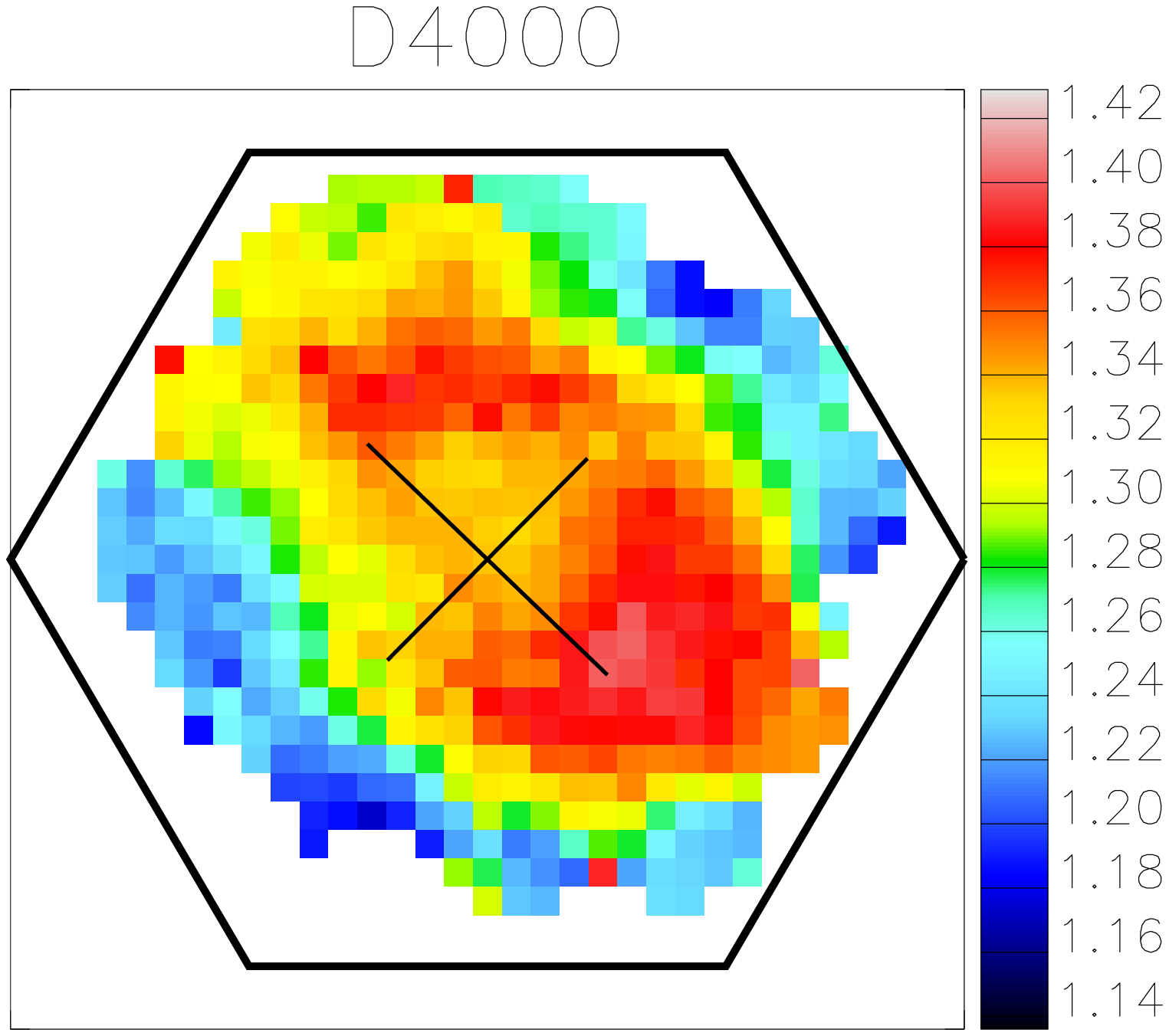,clip=true,height=0.140\textheight}
    \epsfig{figure=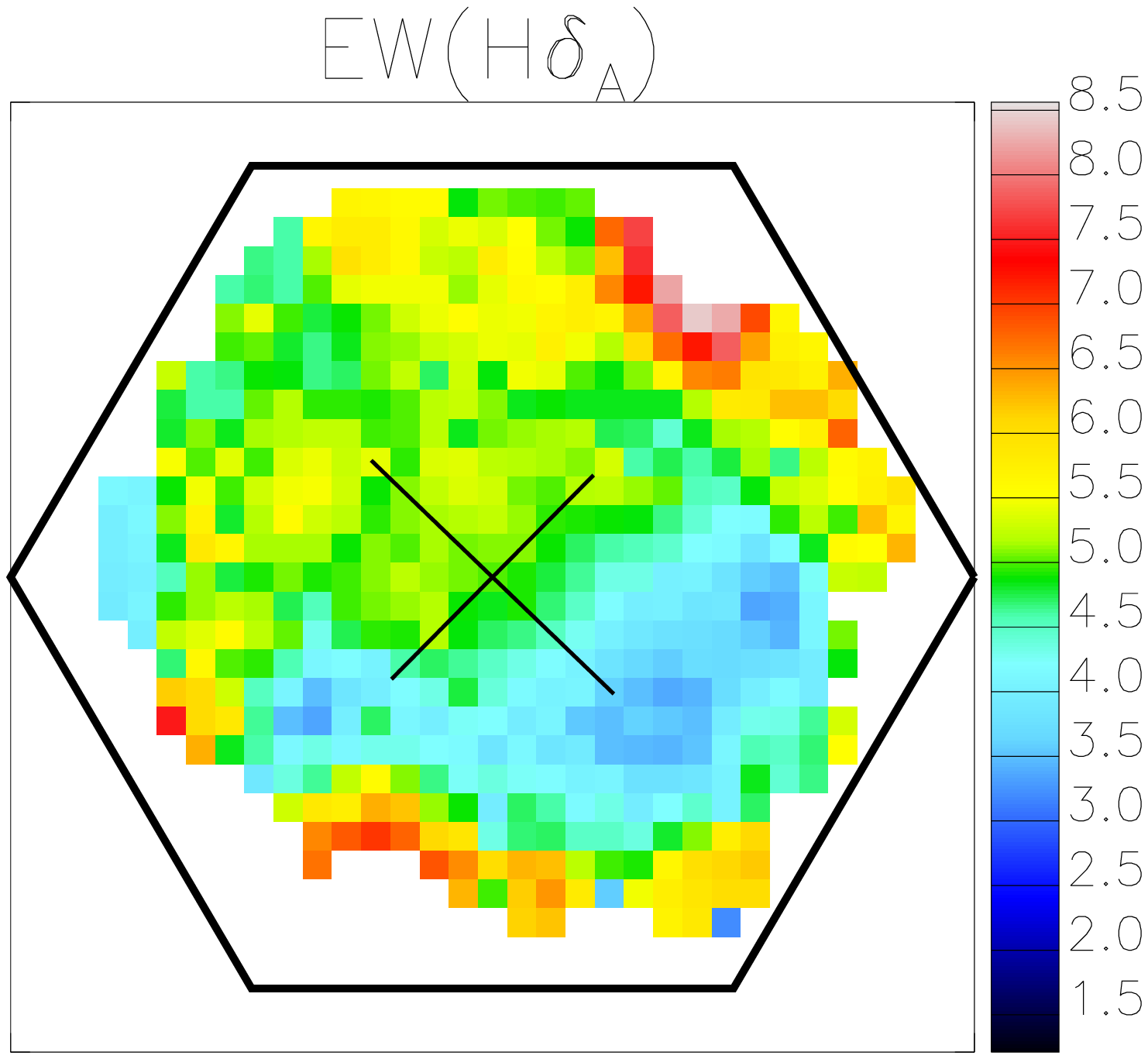,clip=true,height=0.140\textheight}
    \epsfig{figure=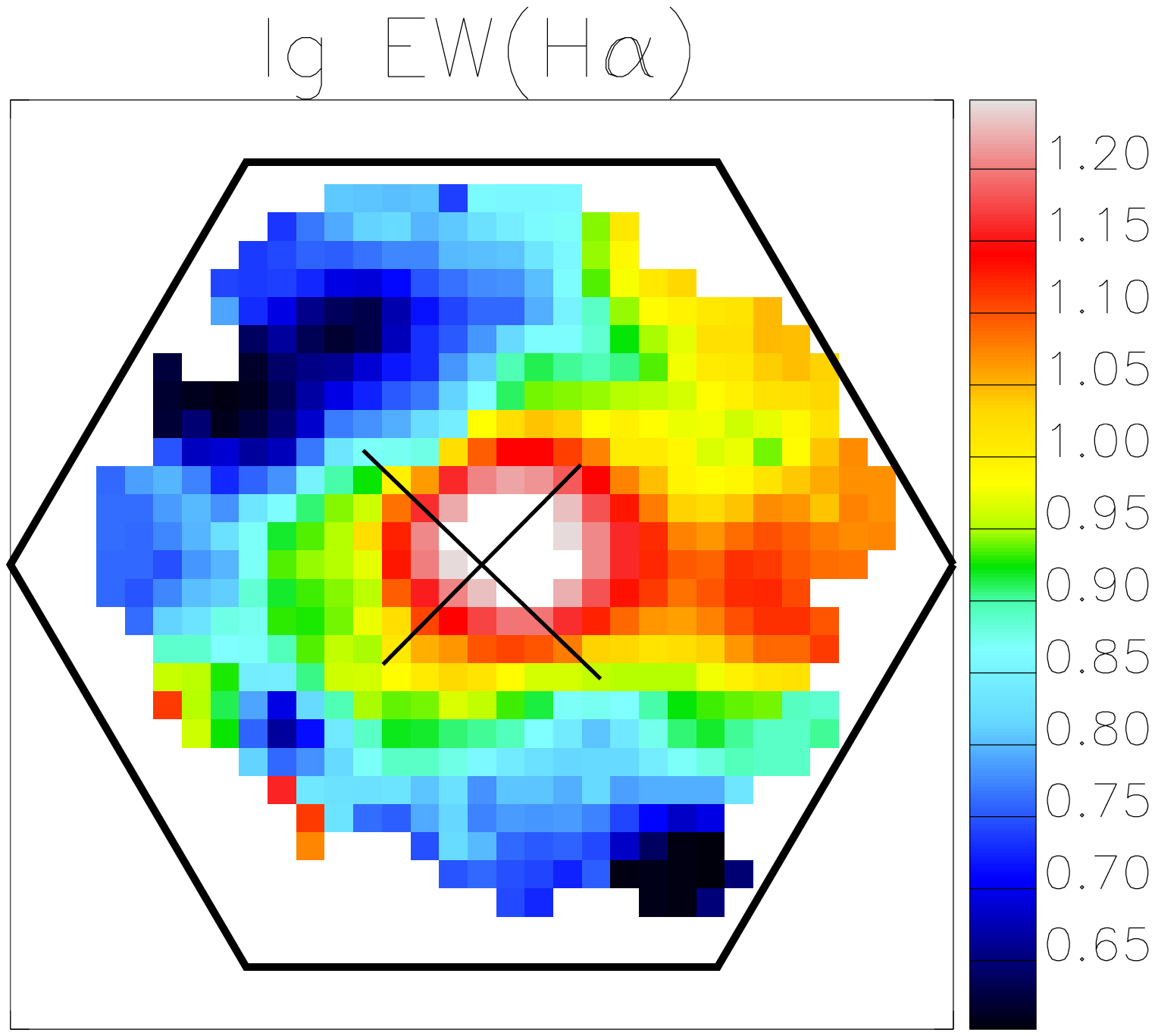,clip=true,height=0.140\textheight}
  \end{center}
  \begin{center}
    \epsfig{figure=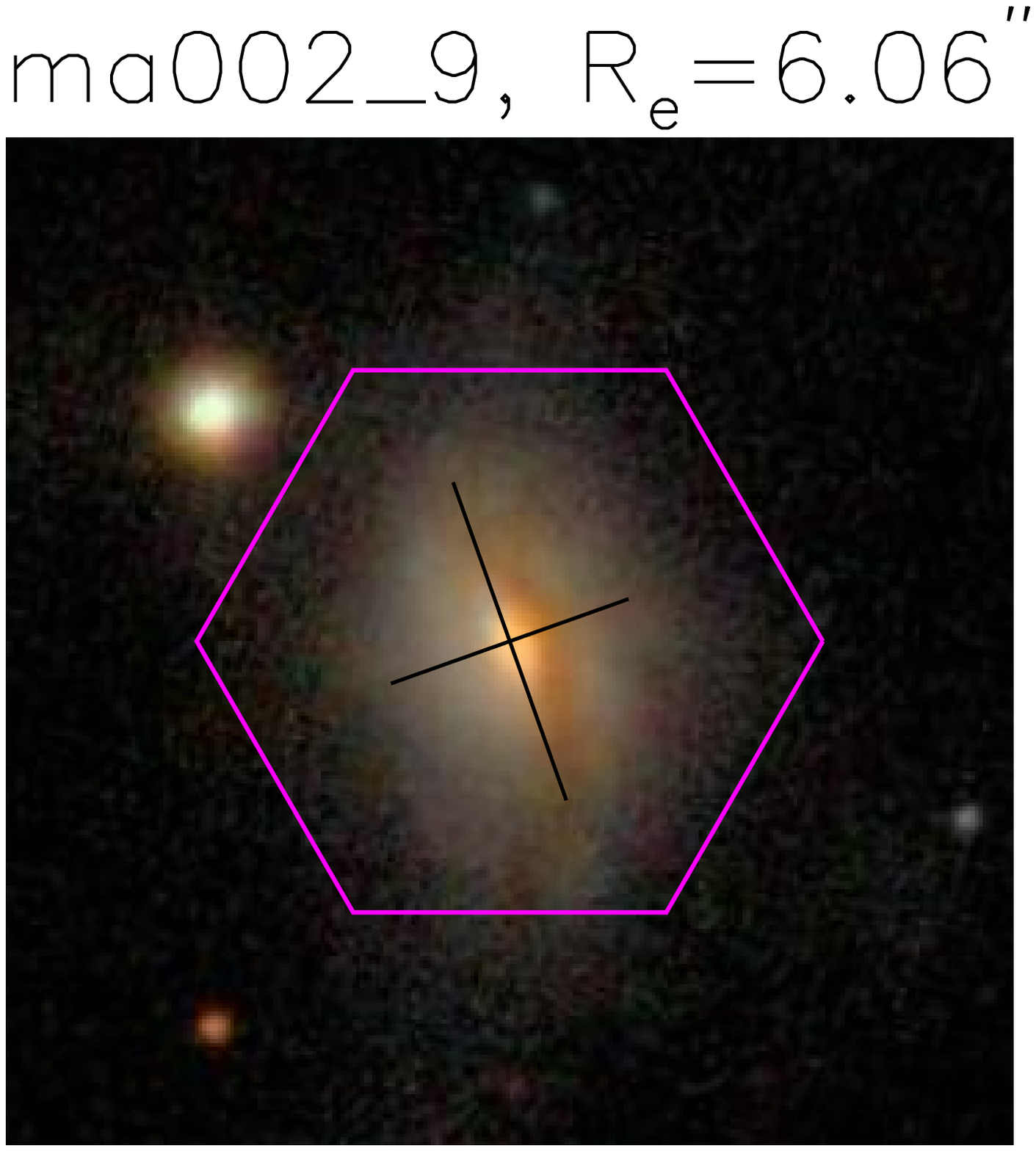,clip=true,height=0.140\textheight}
    \epsfig{figure=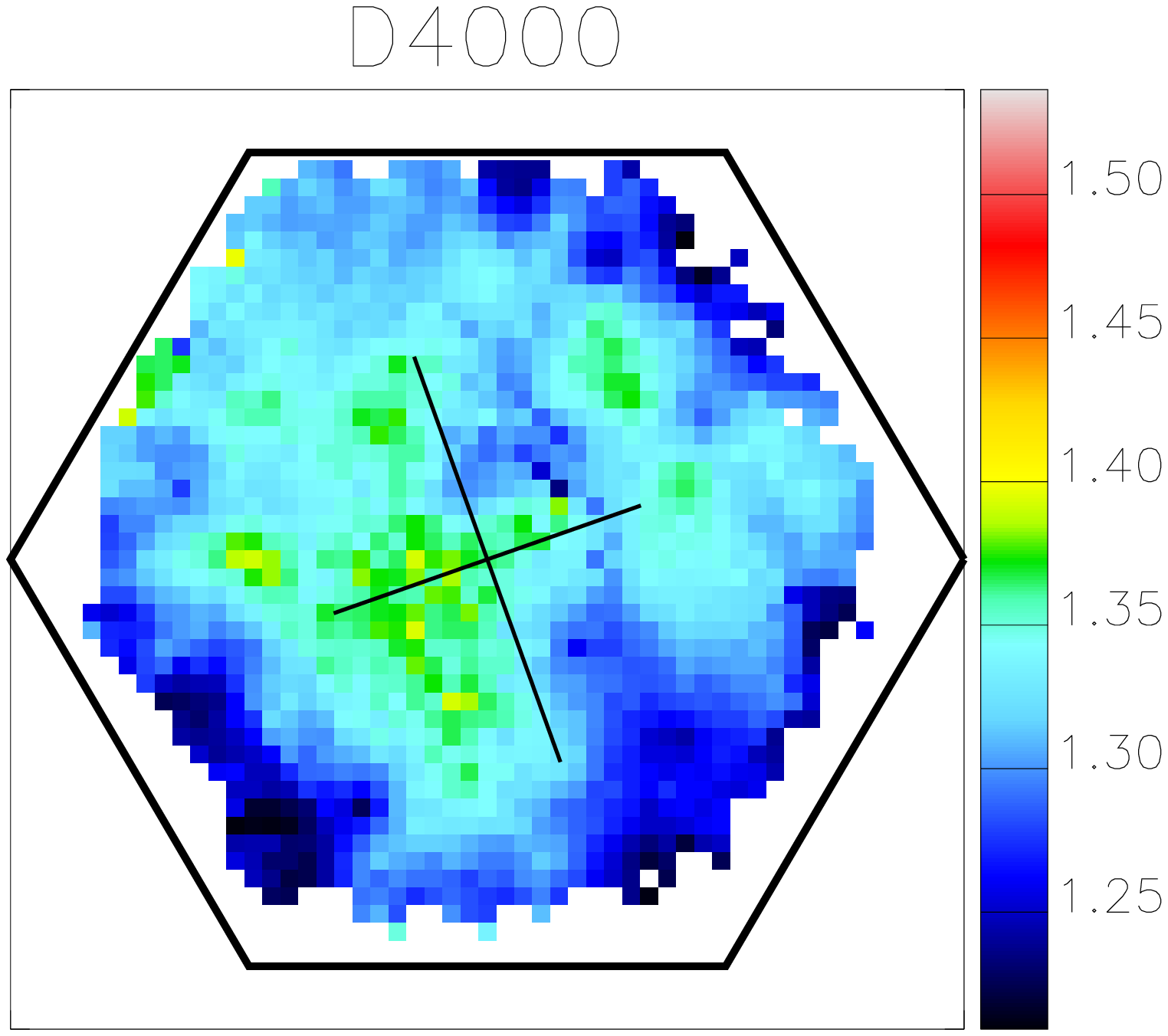,clip=true,height=0.140\textheight}
    \epsfig{figure=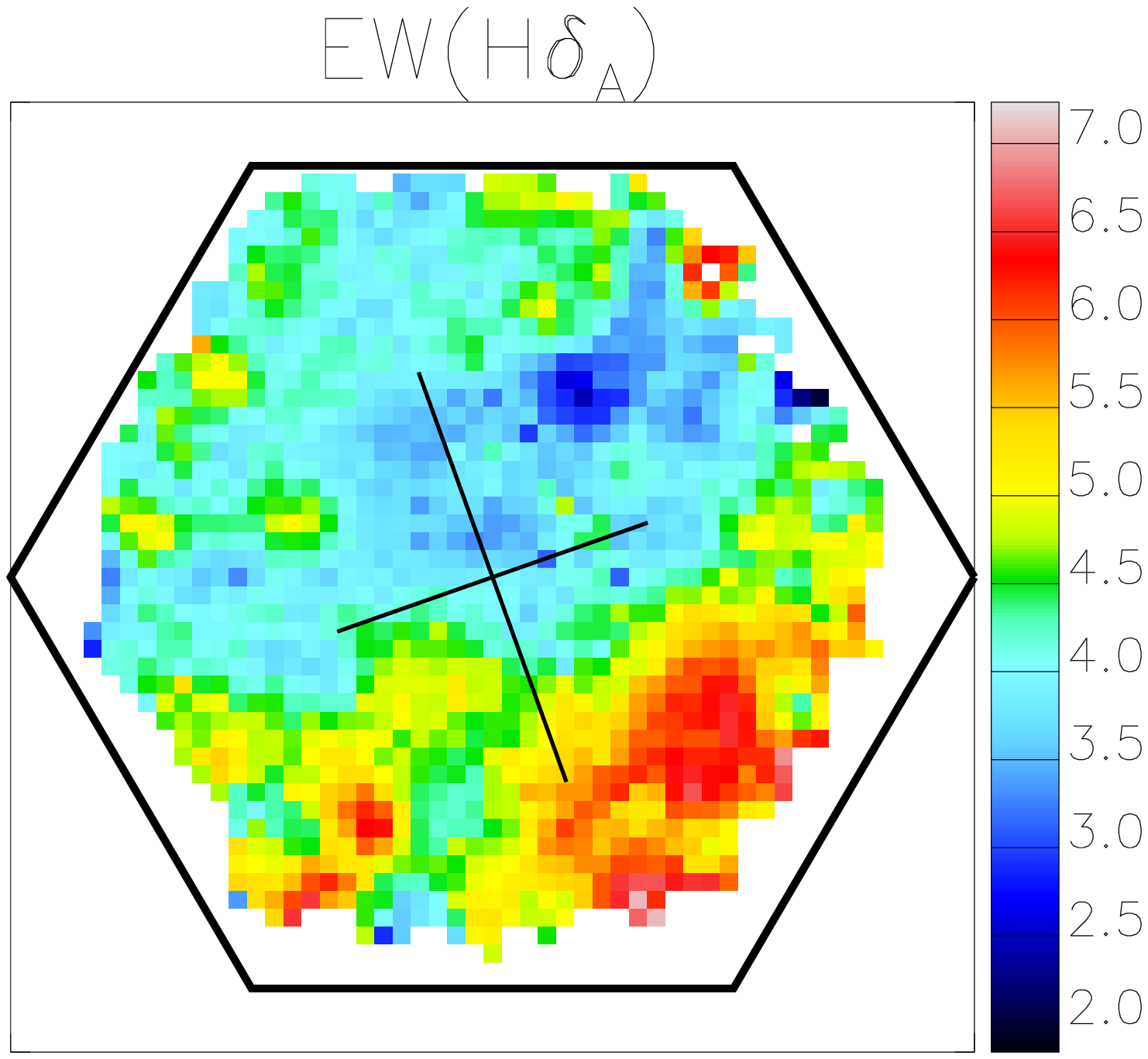,clip=true,height=0.140\textheight}
    \epsfig{figure=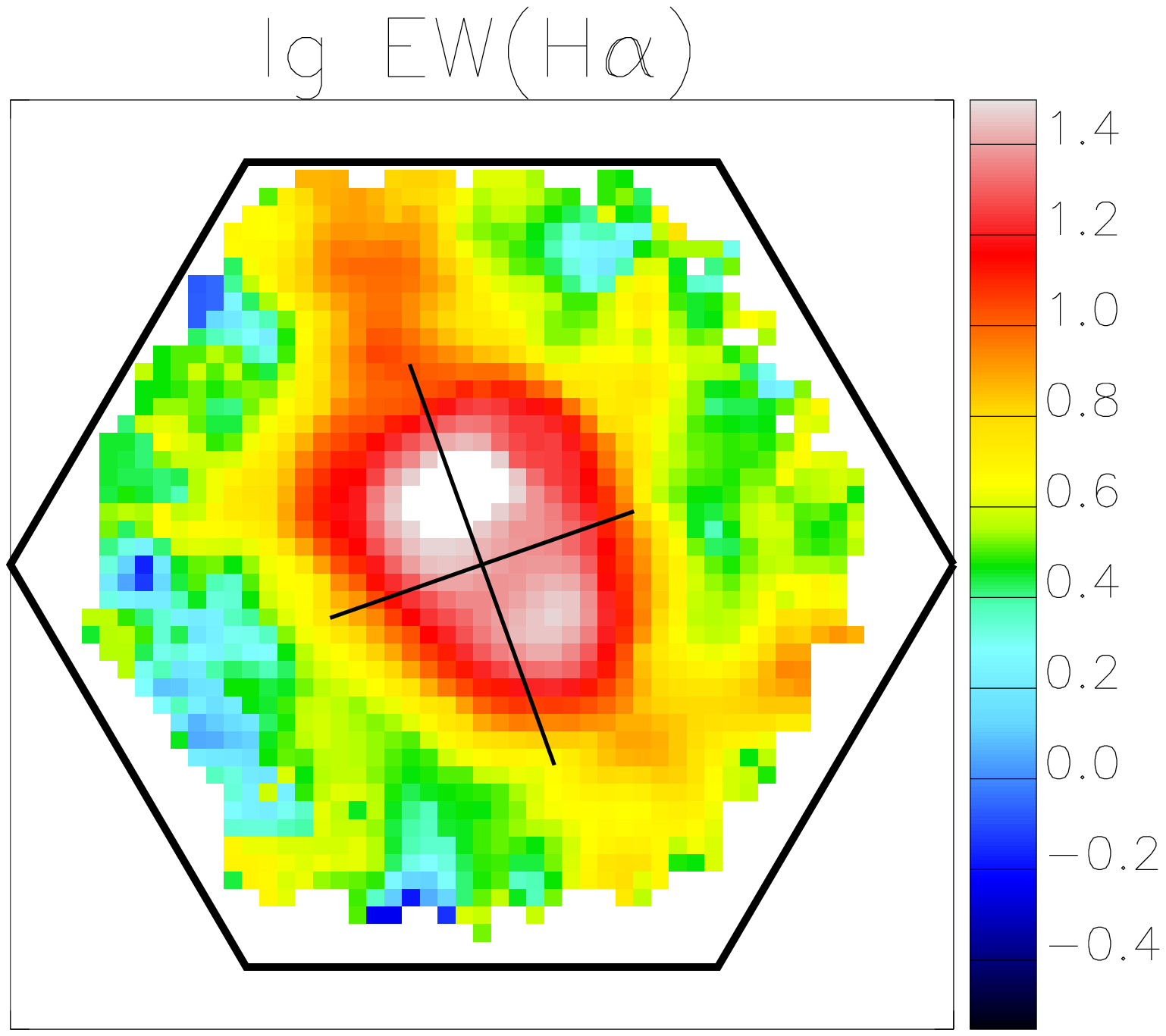,clip=true,height=0.140\textheight}
  \end{center}
  \begin{center}
    \epsfig{figure=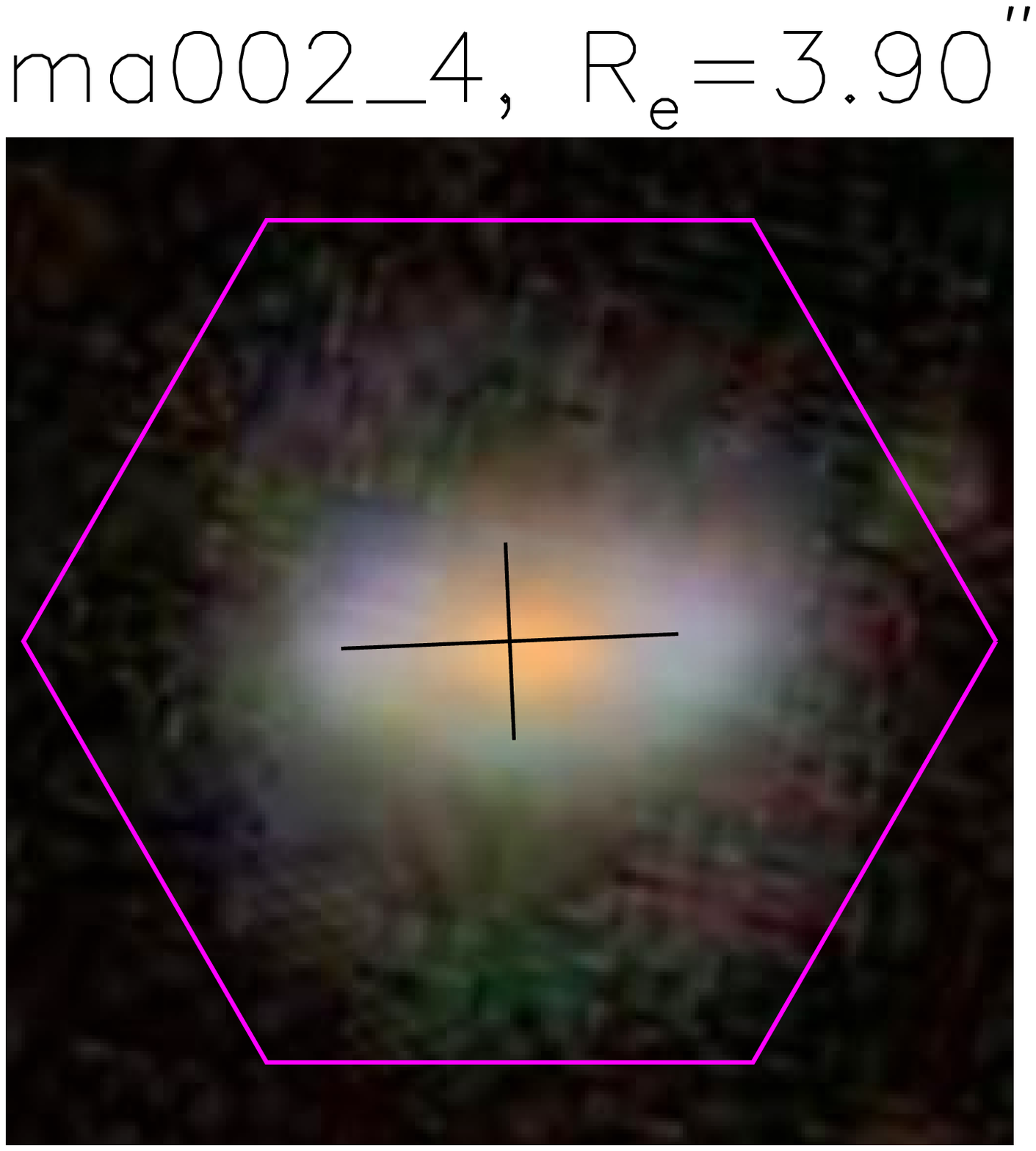,clip=true,height=0.140\textheight}
    \epsfig{figure=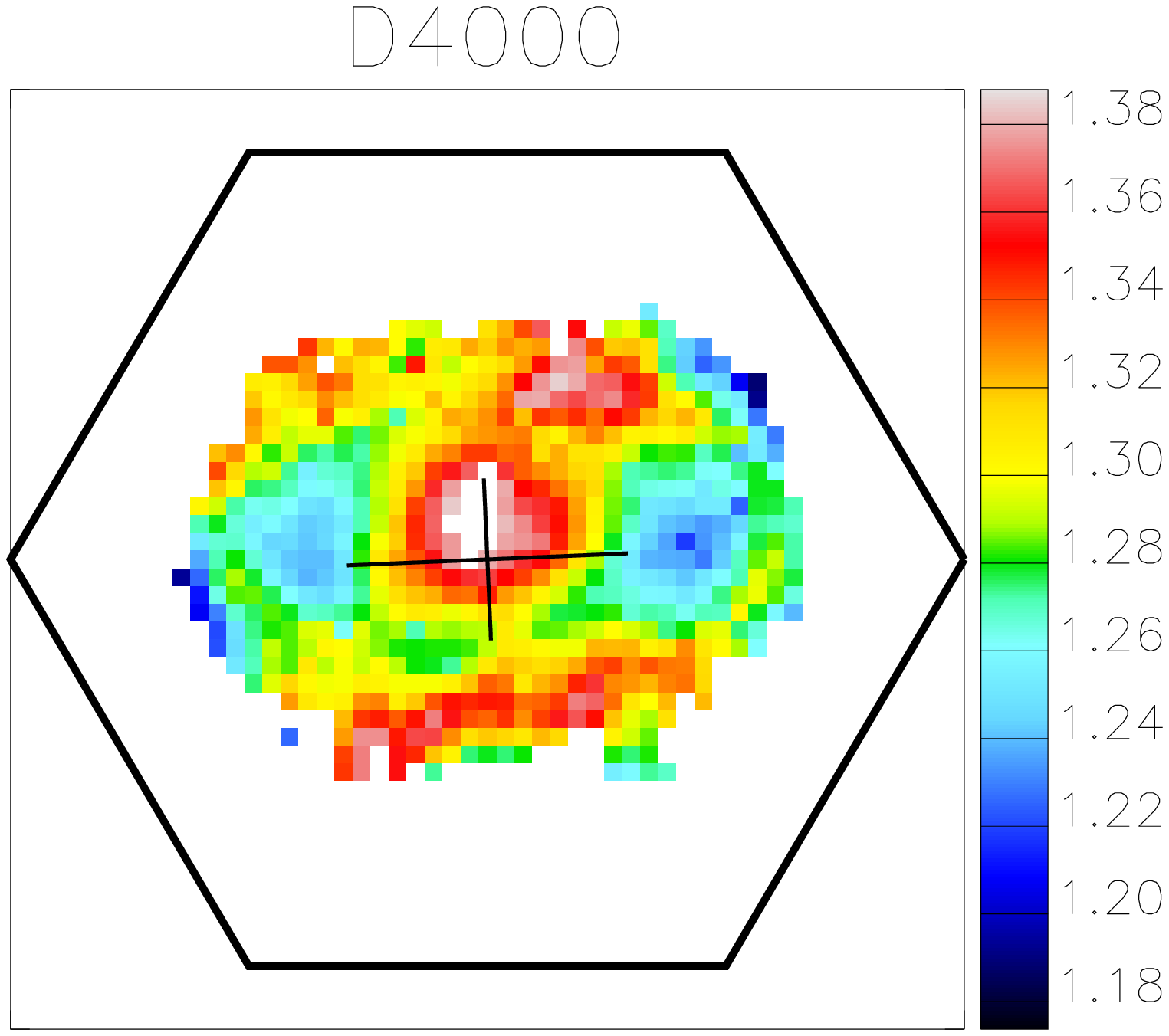,clip=true,height=0.140\textheight}
    \epsfig{figure=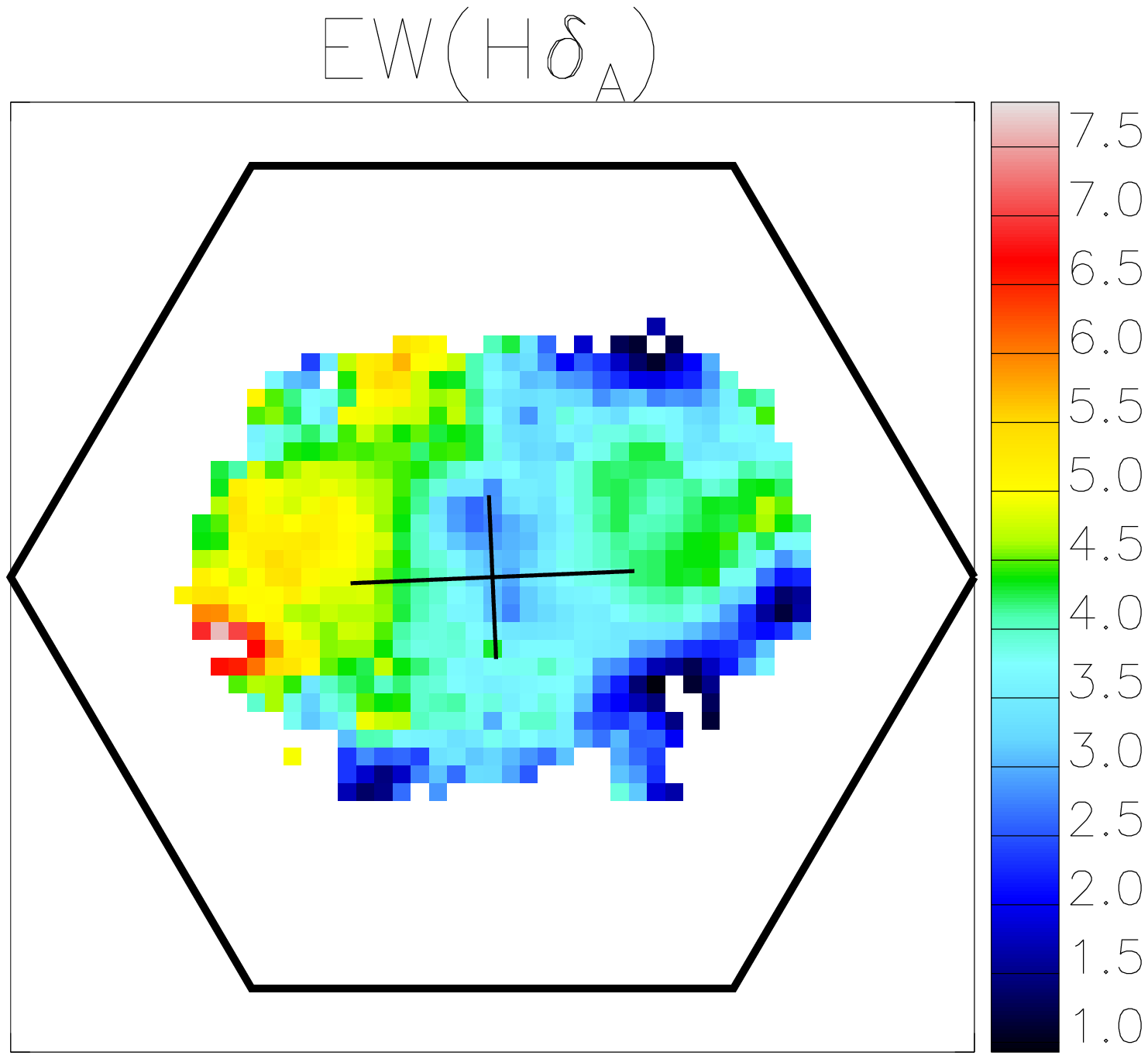,clip=true,height=0.140\textheight}
    \epsfig{figure=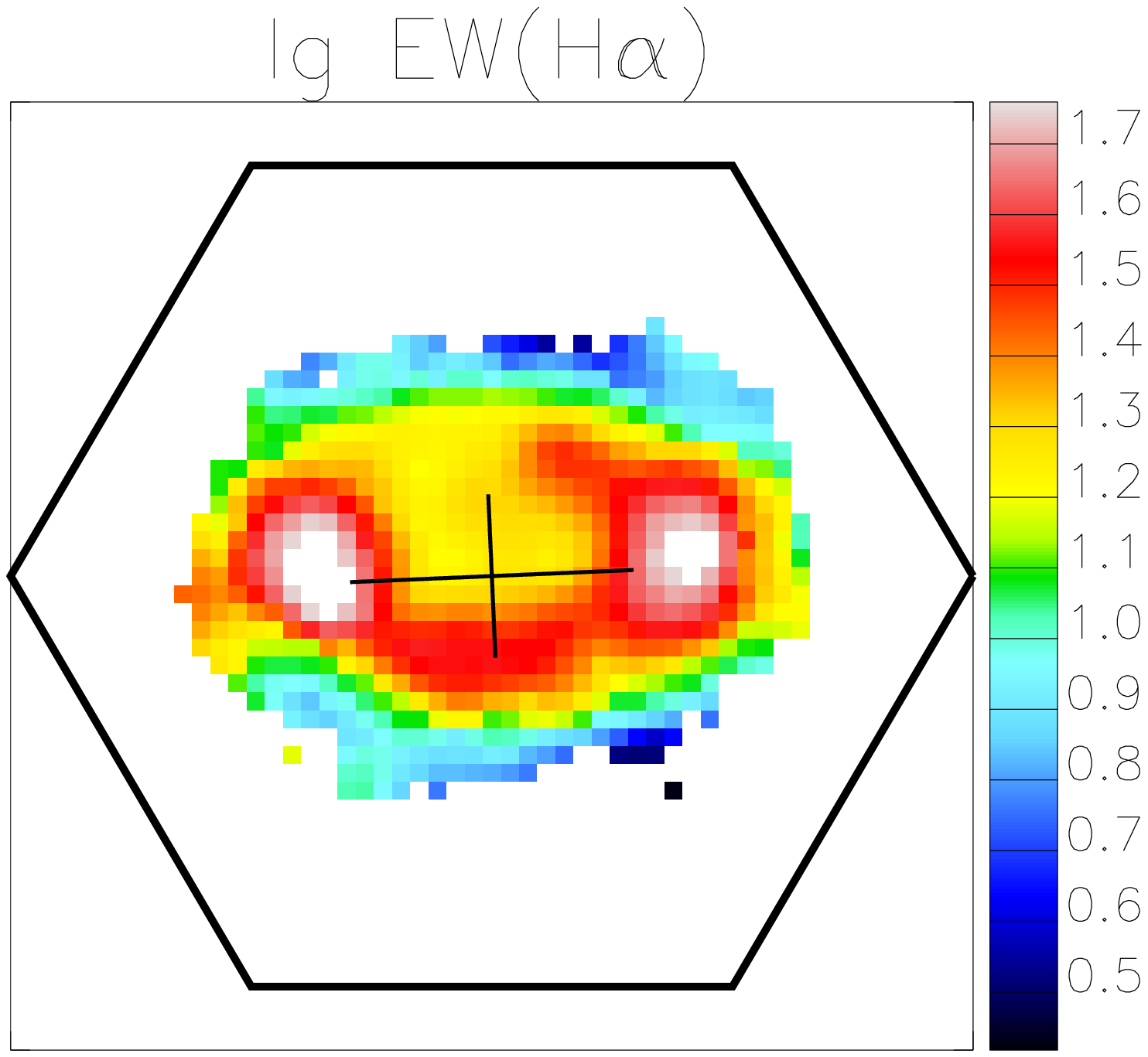,clip=true,height=0.140\textheight}
  \end{center}
  \caption{Panels from left to right show the SDSS image and maps of \dindex,  
    \ewhda\ and \lgewhae\ for the {\em centrally star-forming galaxies} in P-MaNGA. 
    From top to bottom panels the galaxies are  sorted by  increasing the
    \dindex\  measured   from  the  central  spaxel   of  the  P-MaNGA
    datacube. The longer black line in  each panel indicates the direction of
    the major axis, with the line length equal to twice the effective radius 
    of the galaxy (\Reff). The shorter line perpendicular to the major axis
    indicates the minor  axis, and  the line  length is set to $b/a$ times 
    \Reff, where $b/a$ is the minor-to-major  axis ratio determined from the 
    $r$-band SDSS image.
    The P-MaNGA galaxy name, as well as the value of \Reff\ is indicated 
    above the left-most panel, where the size of the SDSS images is set to 
    $6$\Reff$\times6$\Reff. The hexagon in each panel indicates the field of
    view of the IFU bundle. The color scales in the \dindex, \ewhda\ and \lgewhae\
    maps are tailored to each galaxy for the sake of contrast.}
  \label{fig:maps1}
\end{figure*}

\begin{figure*}
  \begin{center}
    \epsfig{figure=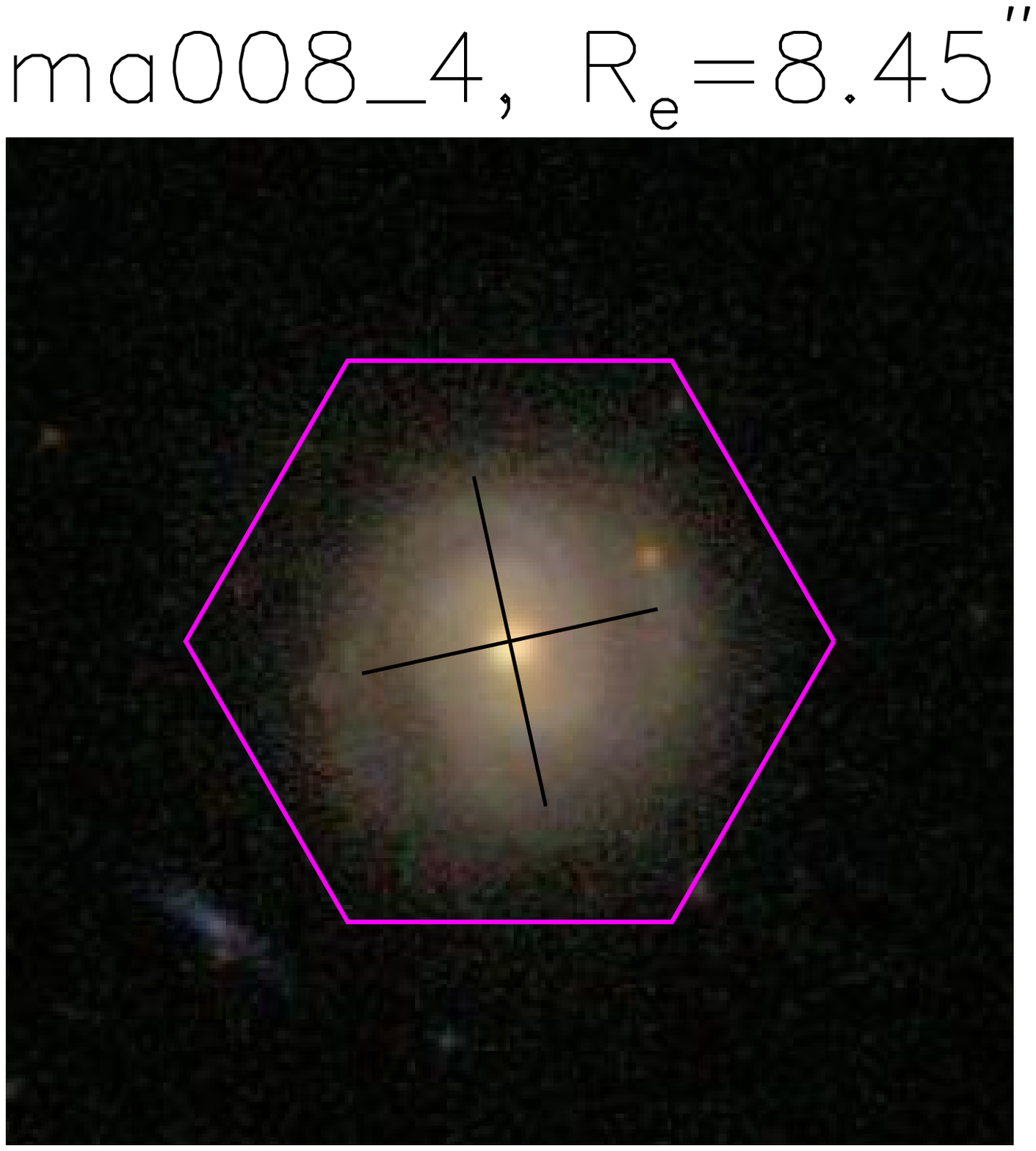,clip=true,height=0.140\textheight}
    \epsfig{figure=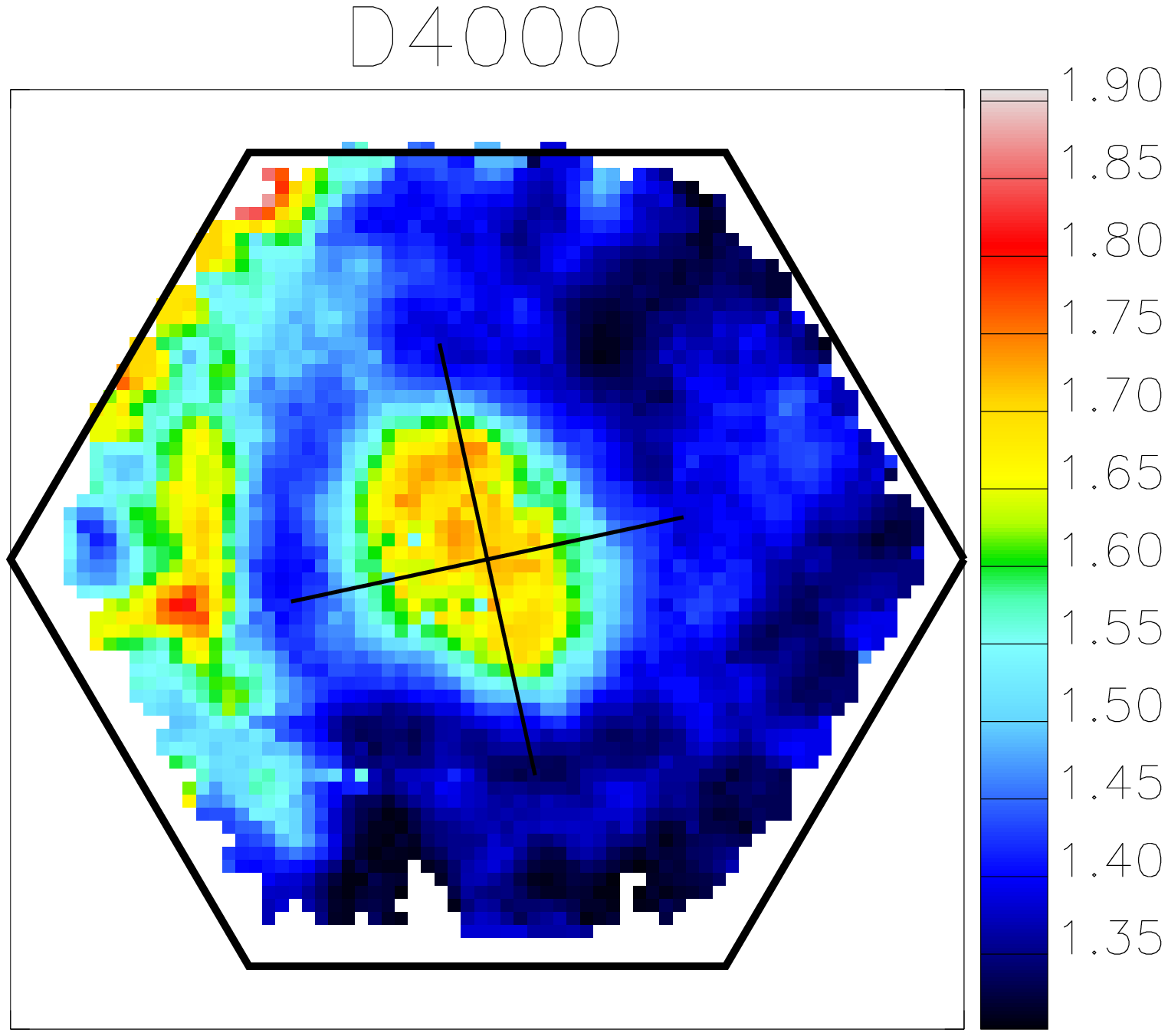,clip=true,height=0.140\textheight}
    \epsfig{figure=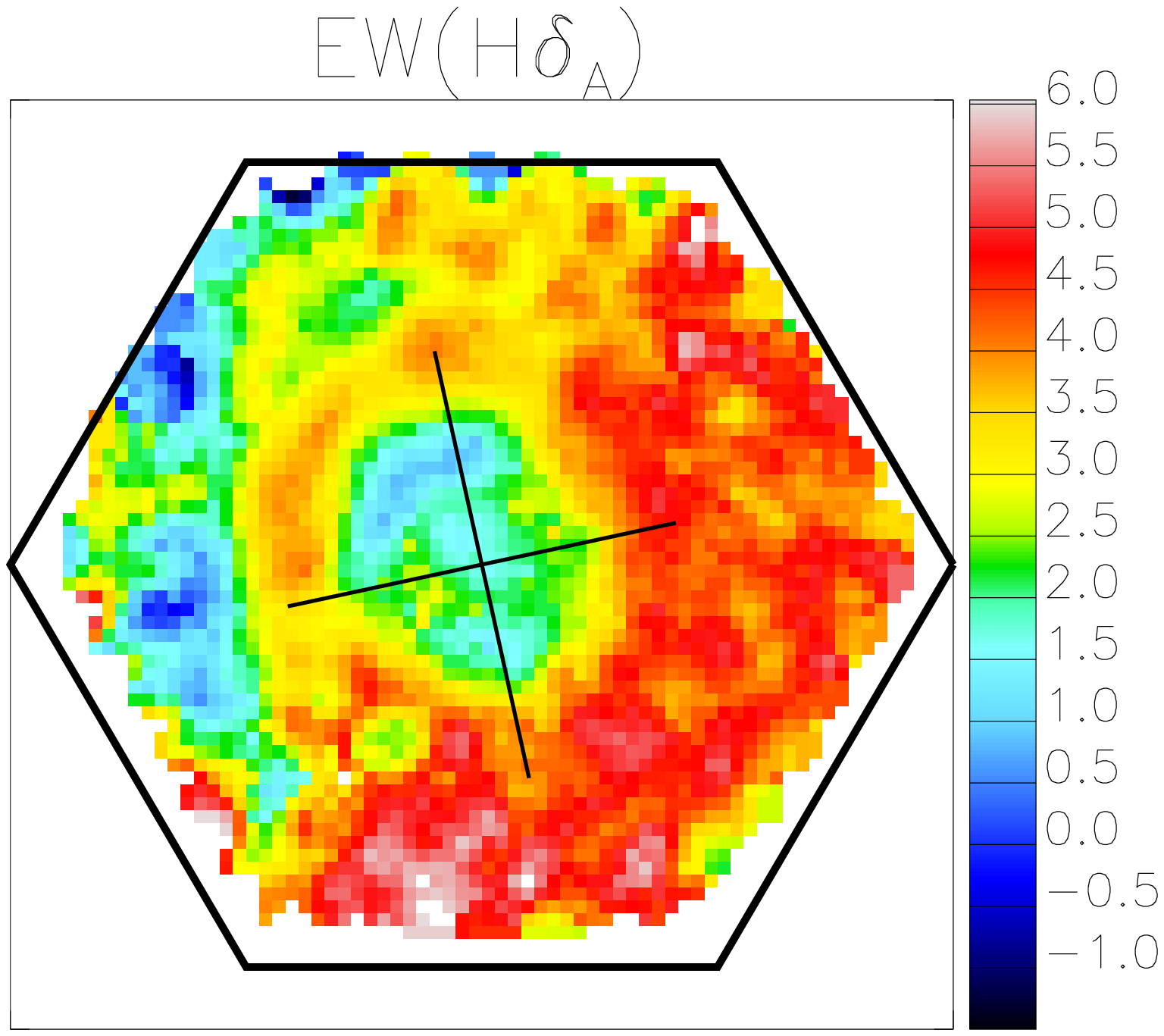,clip=true,height=0.140\textheight}
    \epsfig{figure=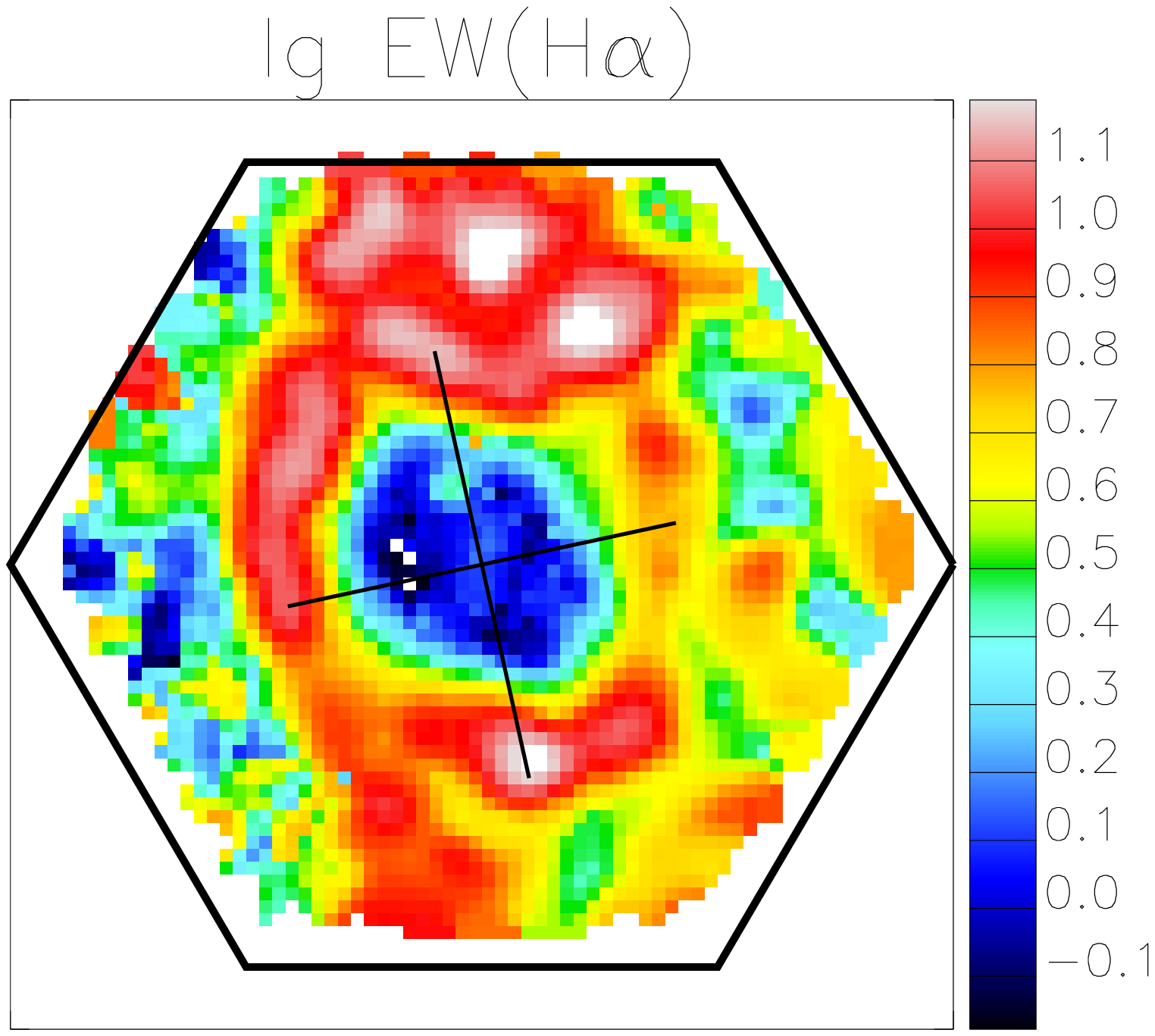,clip=true,height=0.140\textheight}
  \end{center}
  \begin{center}
    \epsfig{figure=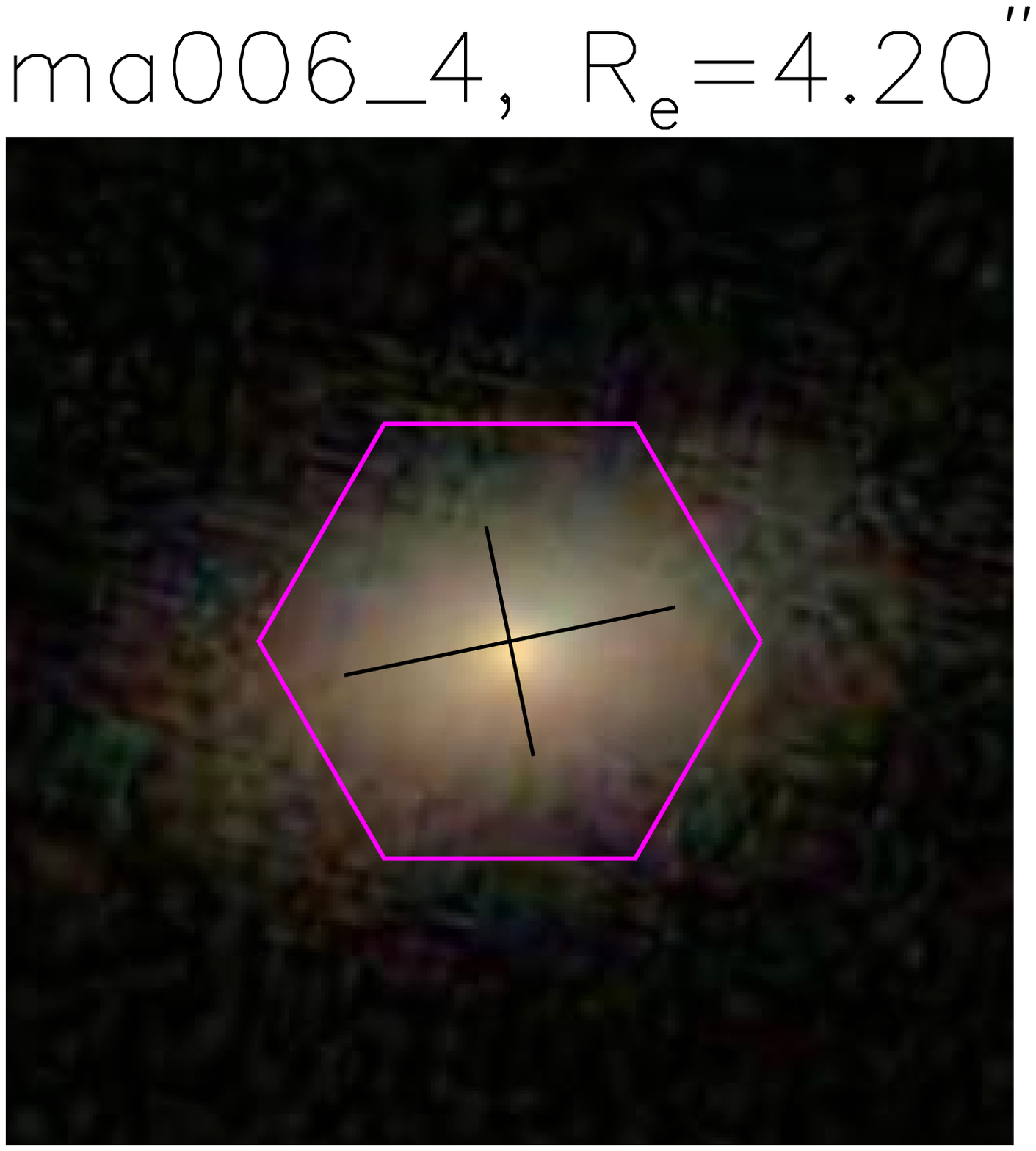,clip=true,height=0.140\textheight}
    \epsfig{figure=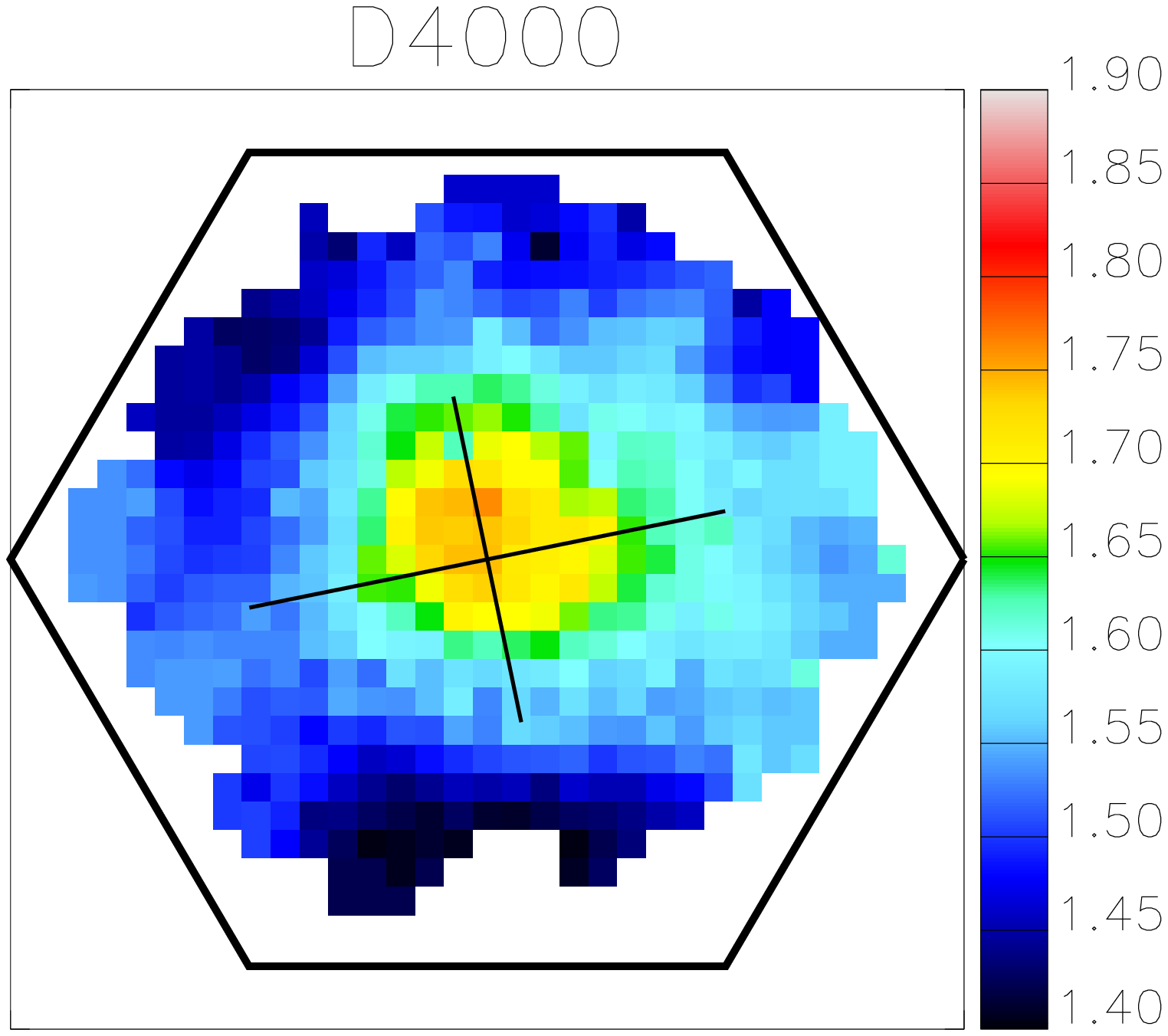,clip=true,height=0.140\textheight}
    \epsfig{figure=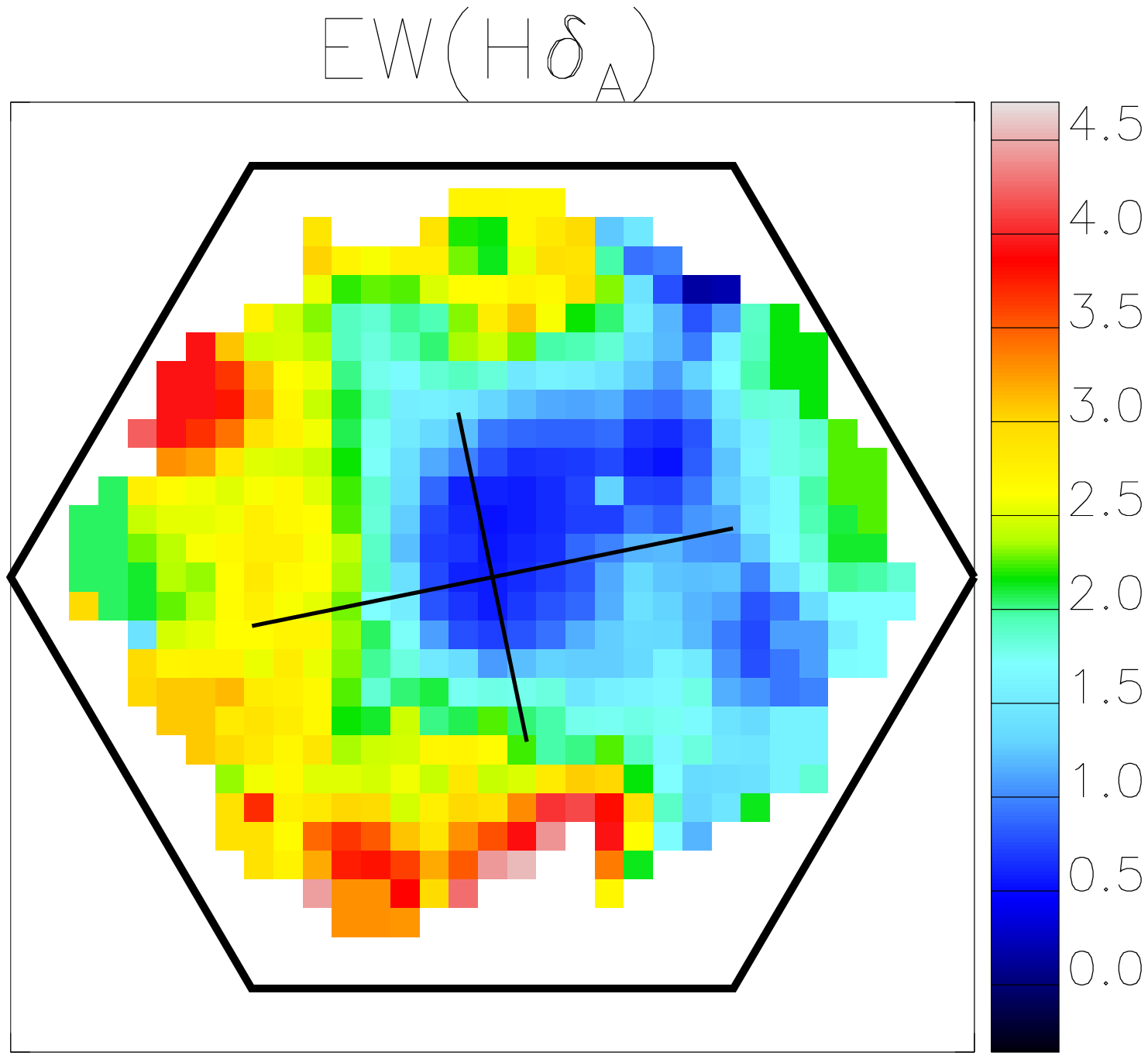,clip=true,height=0.140\textheight}
    \epsfig{figure=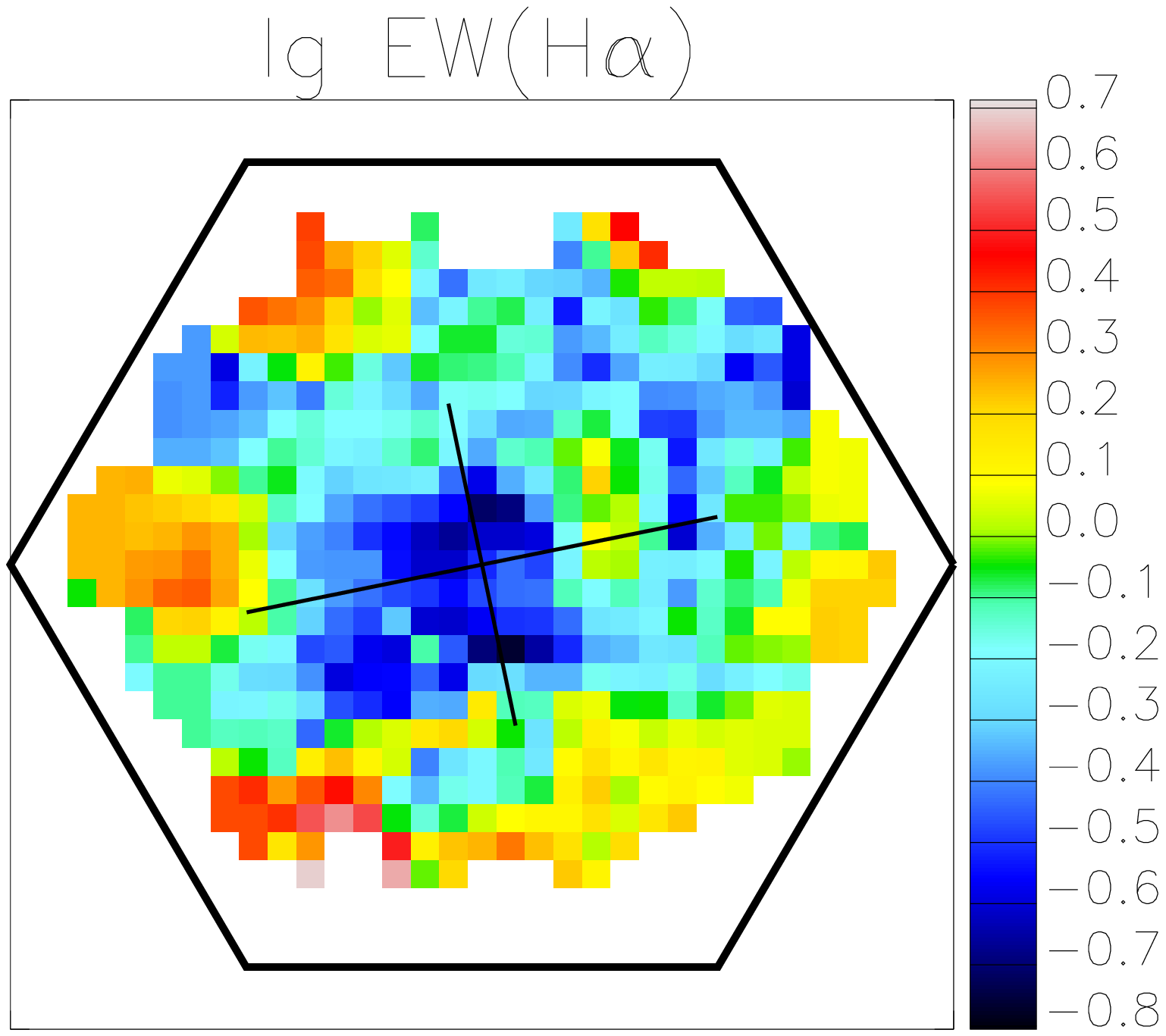,clip=true,height=0.140\textheight}
  \end{center}
  \begin{center}
    \epsfig{figure=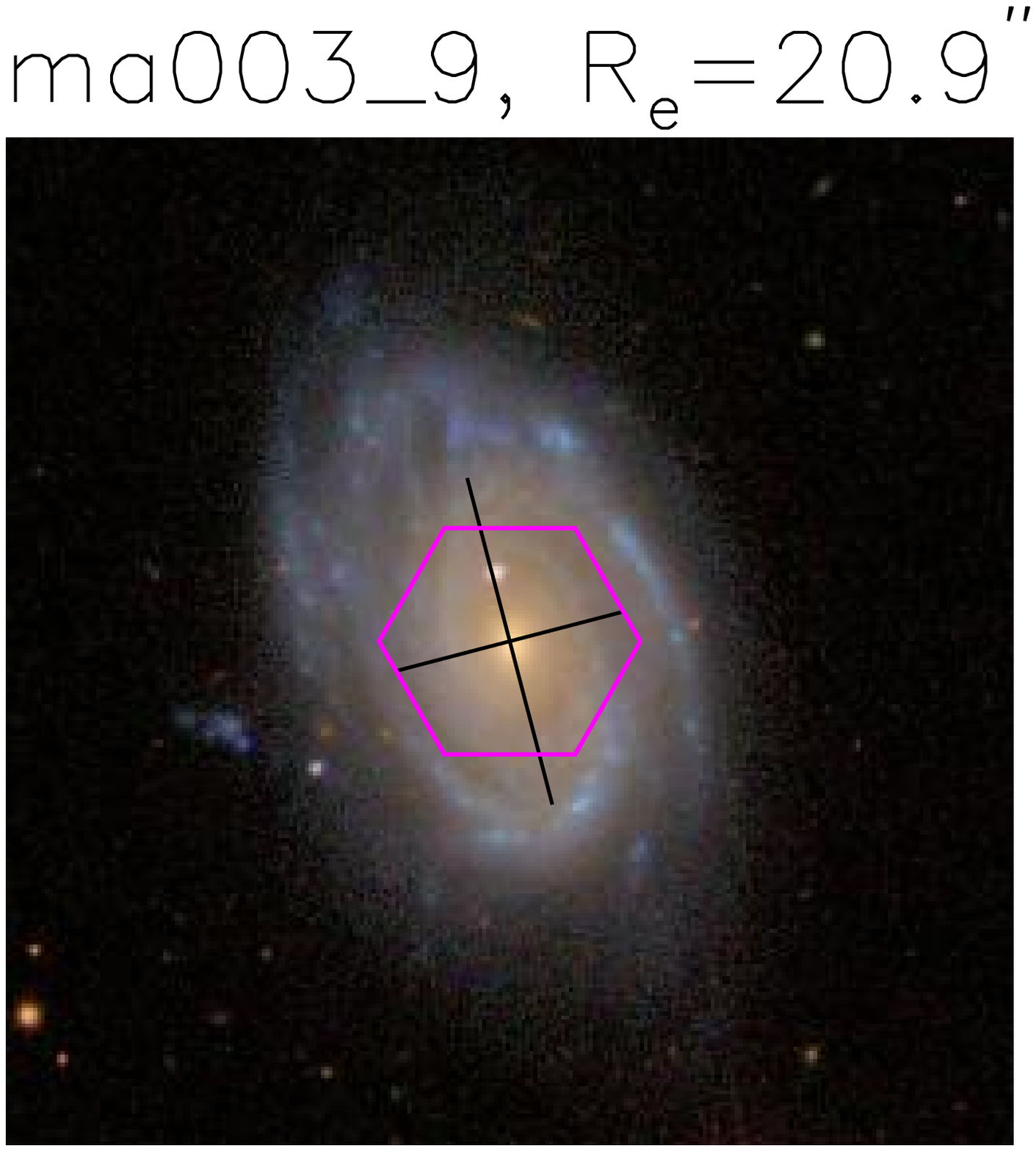,clip=true,height=0.140\textheight}
    \epsfig{figure=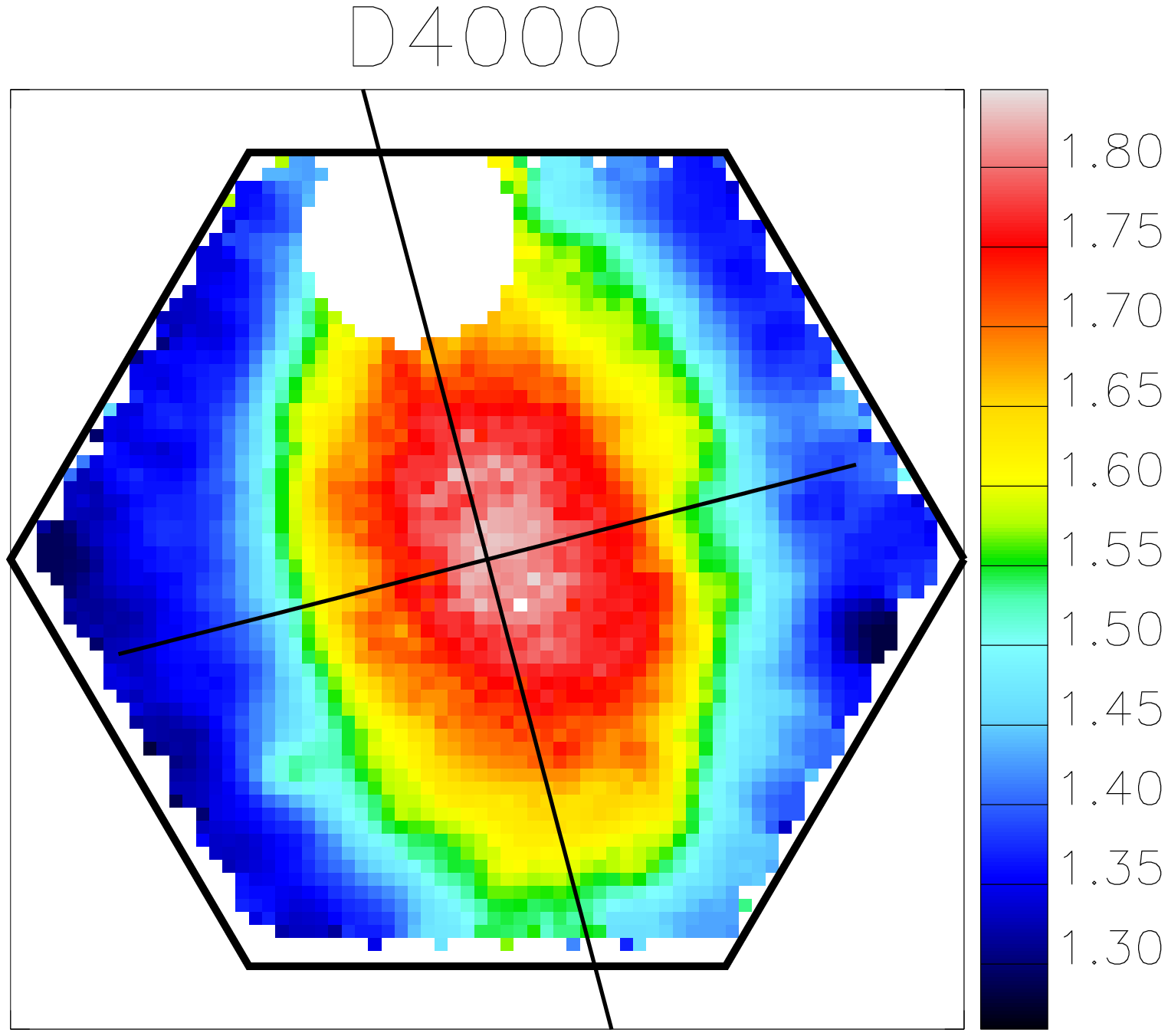,clip=true,height=0.140\textheight}
    \epsfig{figure=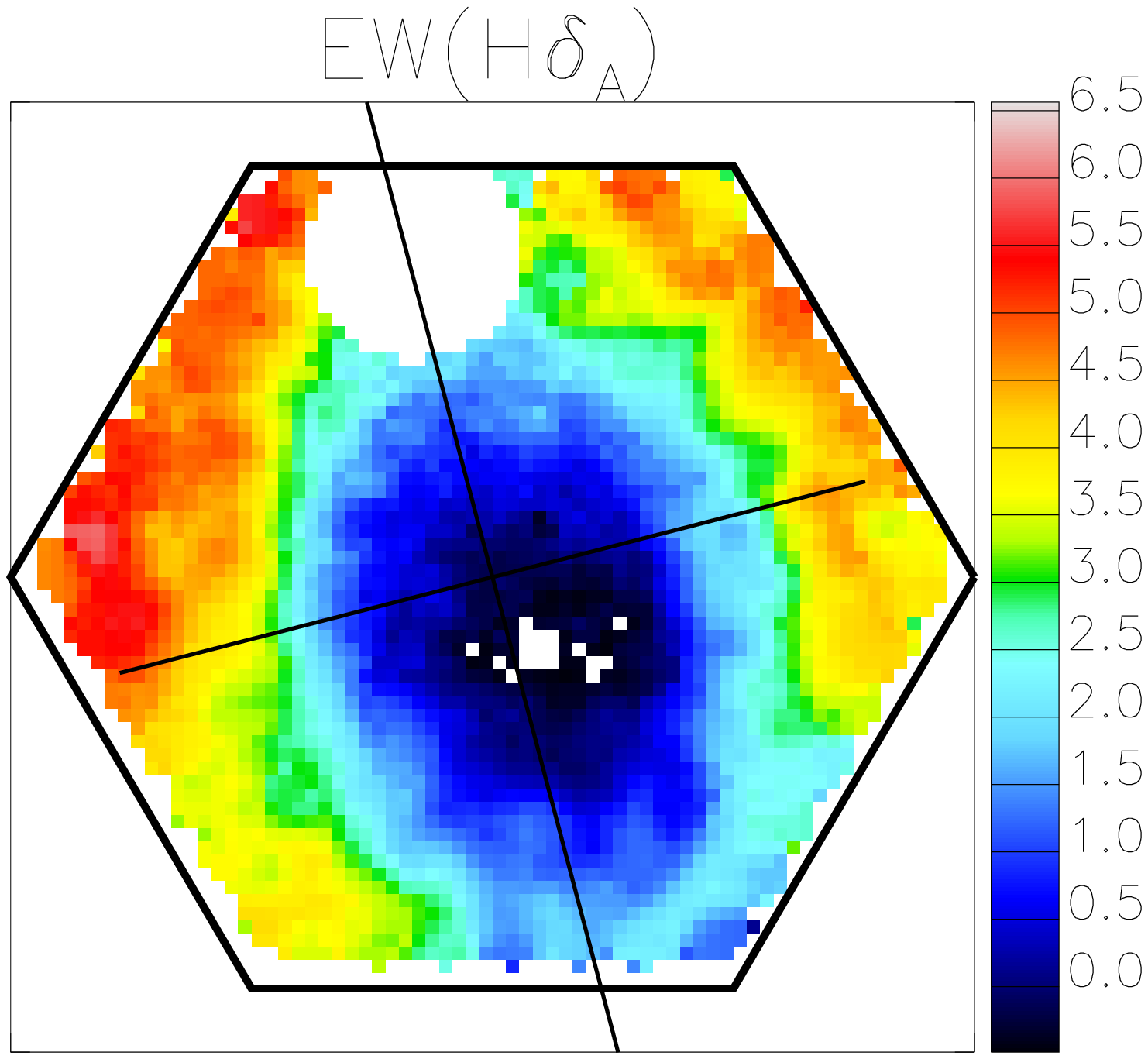,clip=true,height=0.140\textheight}
    \epsfig{figure=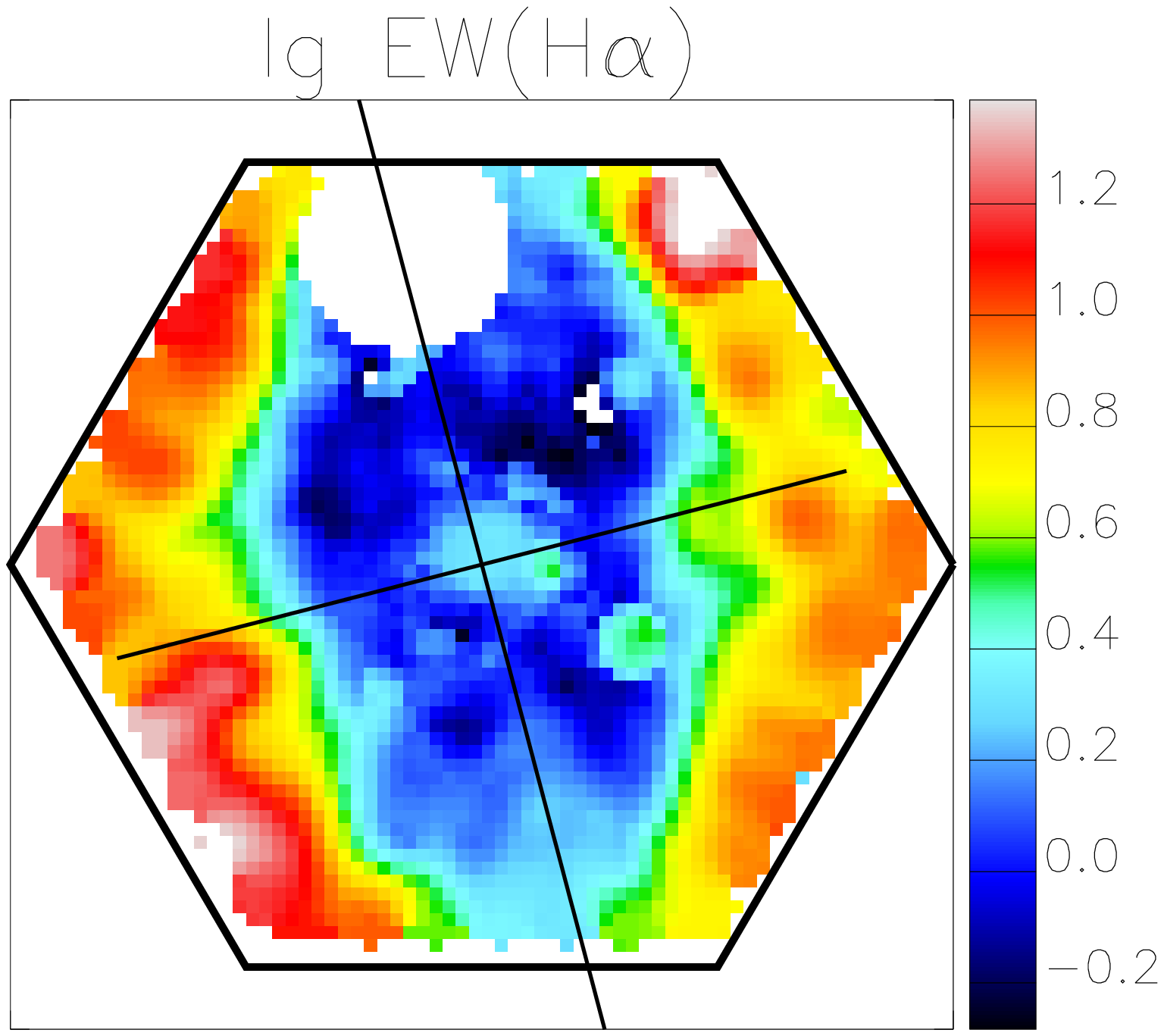,clip=true,height=0.140\textheight}
  \end{center}
  \begin{center}
    \epsfig{figure=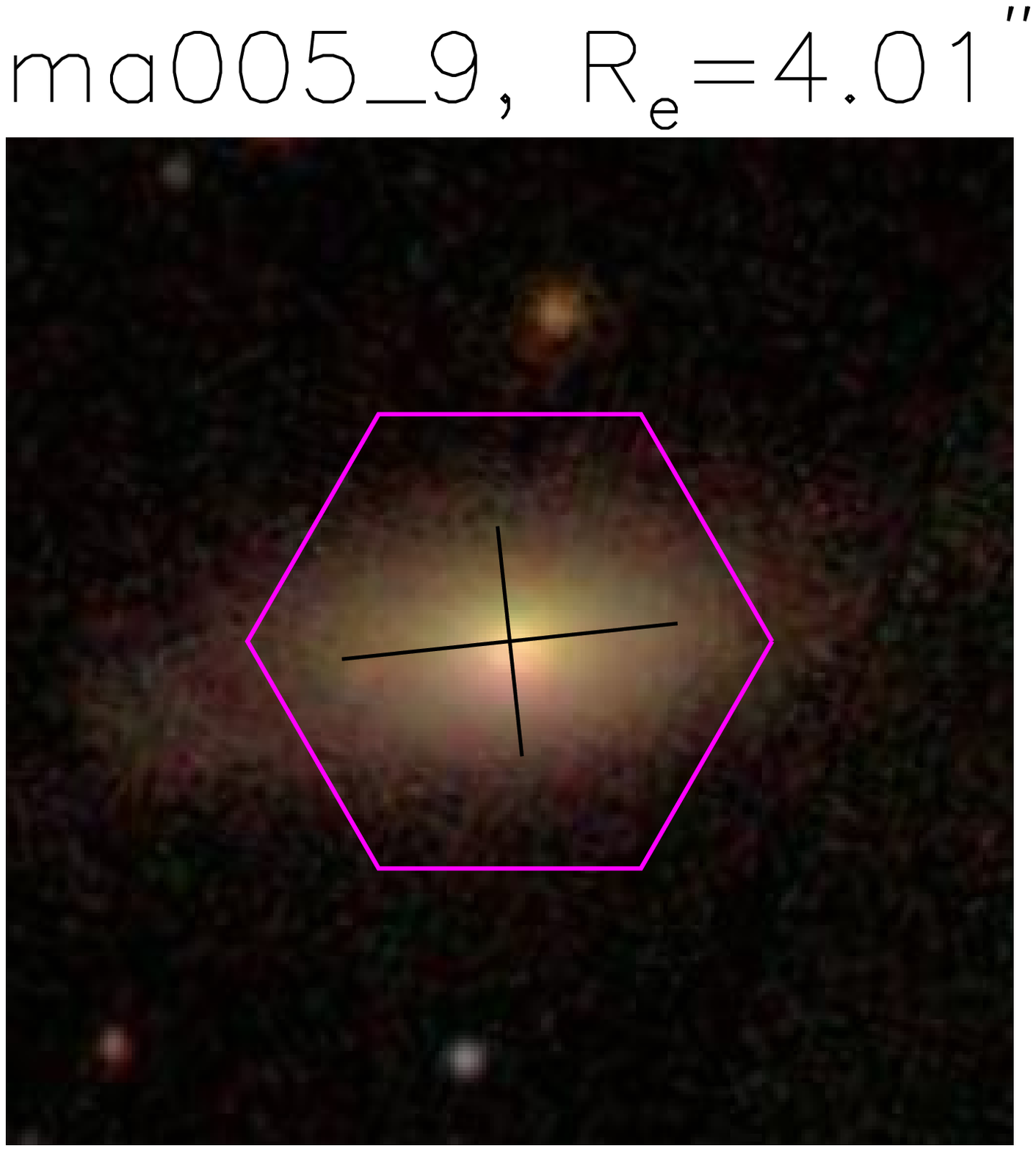,clip=true,height=0.140\textheight}
    \epsfig{figure=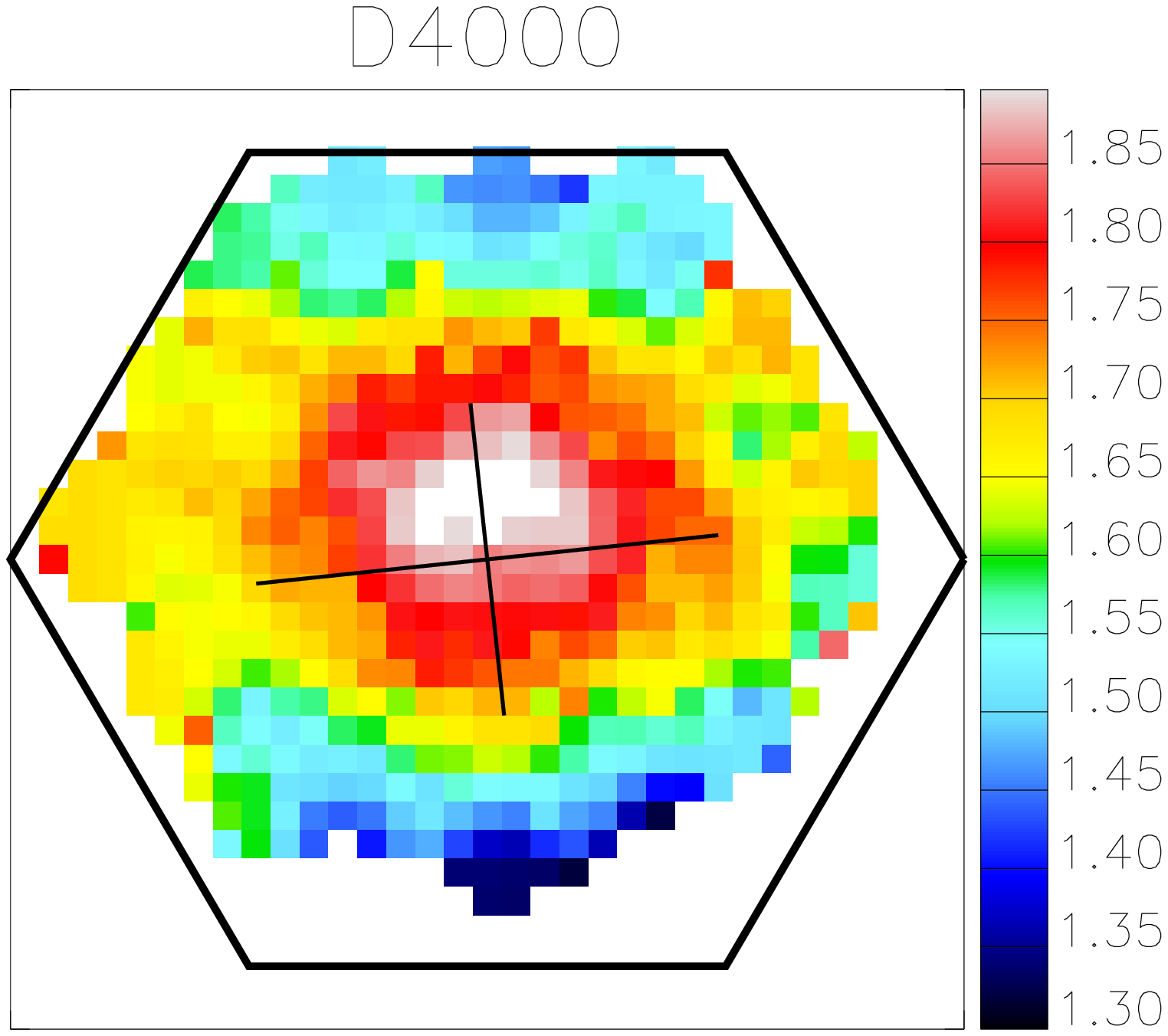,clip=true,height=0.140\textheight}
    \epsfig{figure=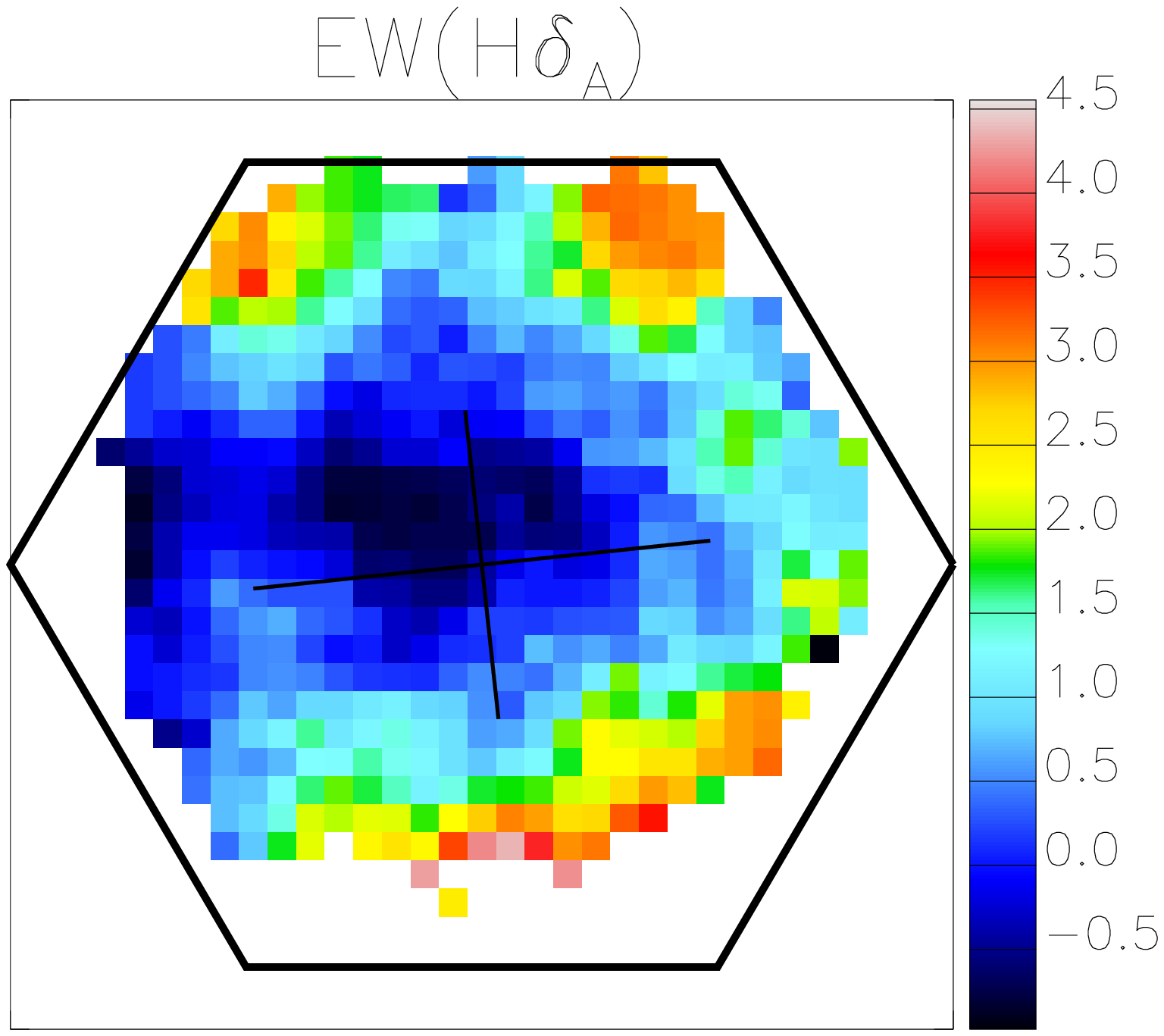,clip=true,height=0.140\textheight}
    \epsfig{figure=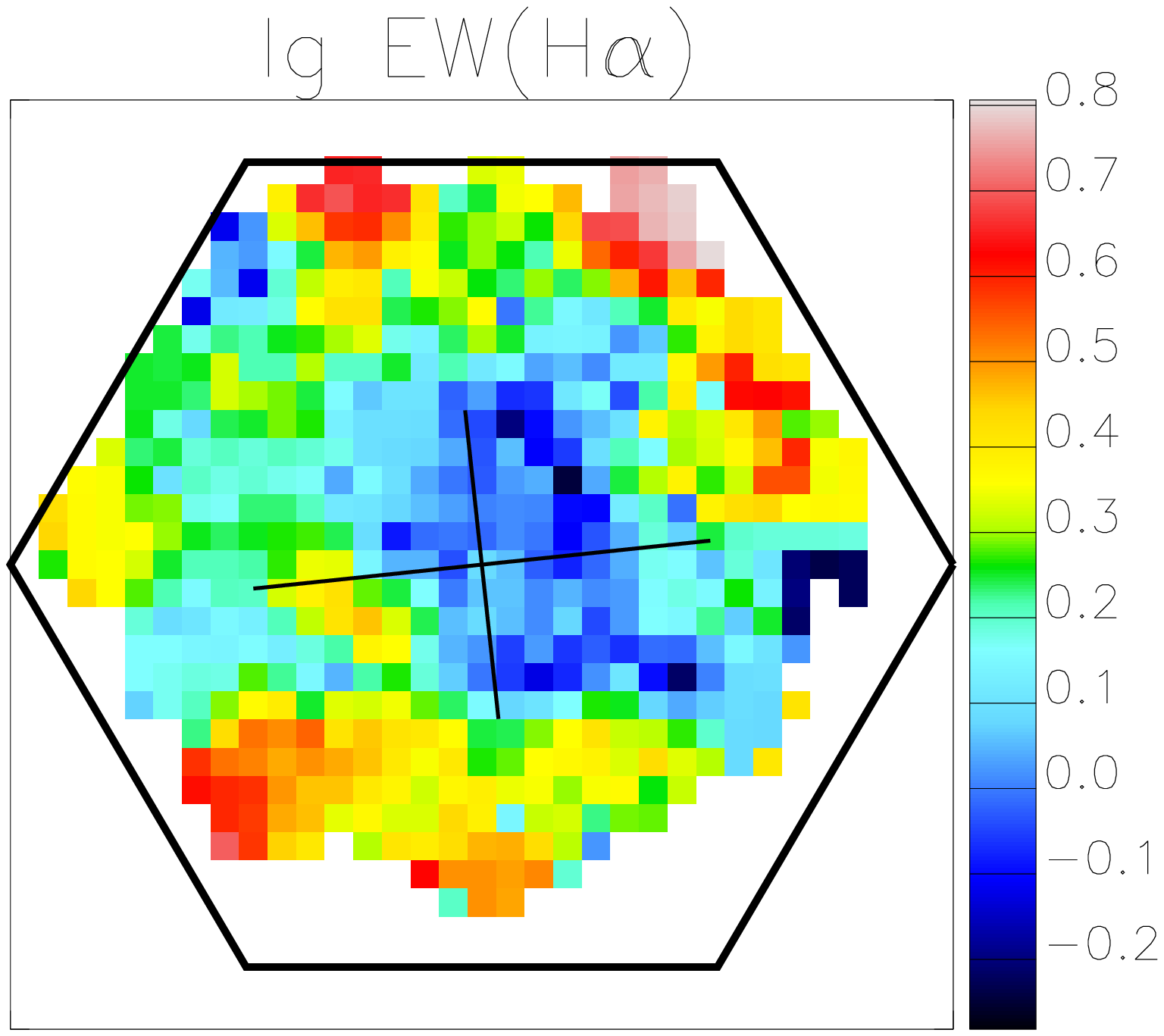,clip=true,height=0.140\textheight}
  \end{center}
  \begin{center}
    \epsfig{figure=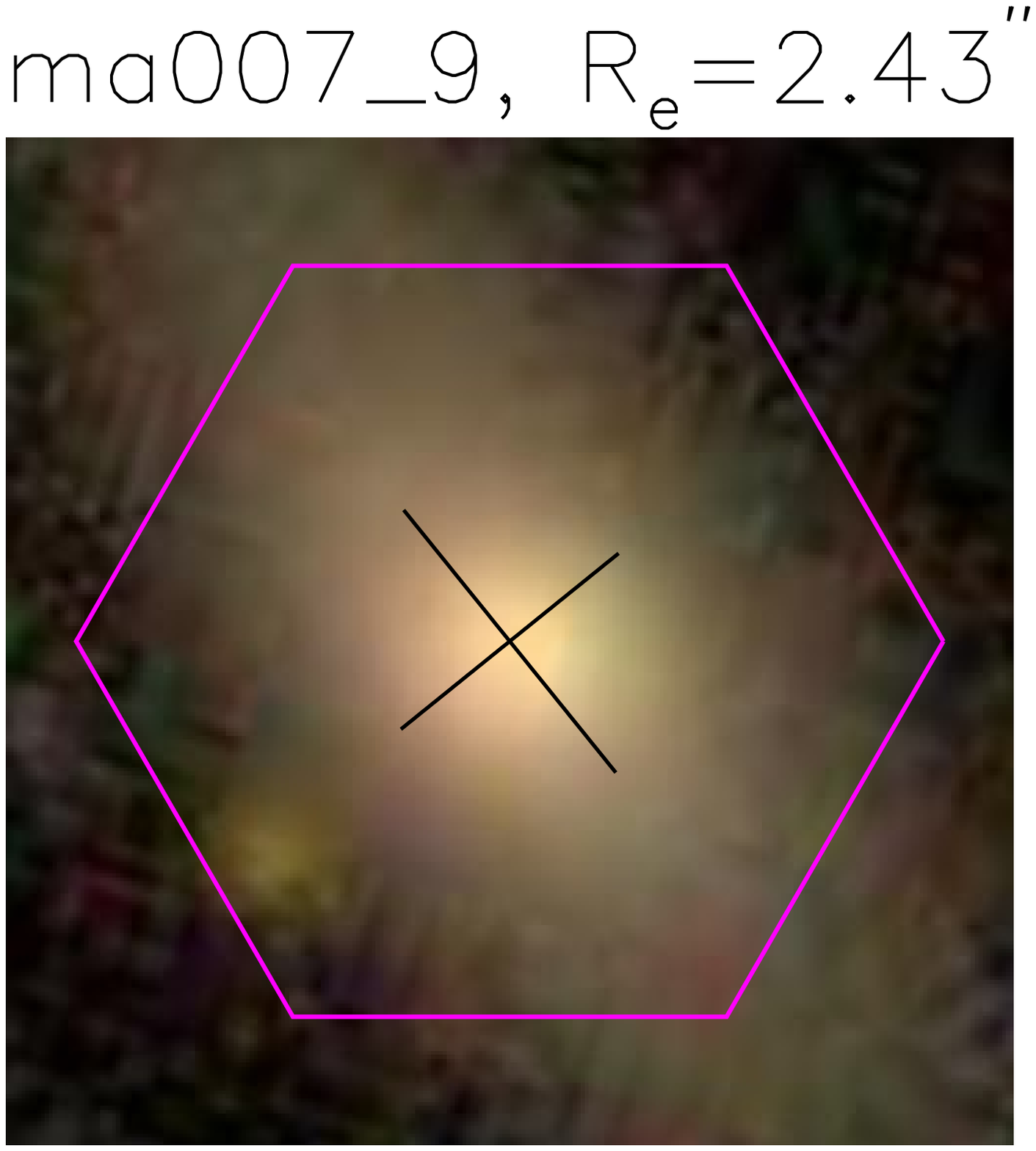,clip=true,height=0.140\textheight}
    \epsfig{figure=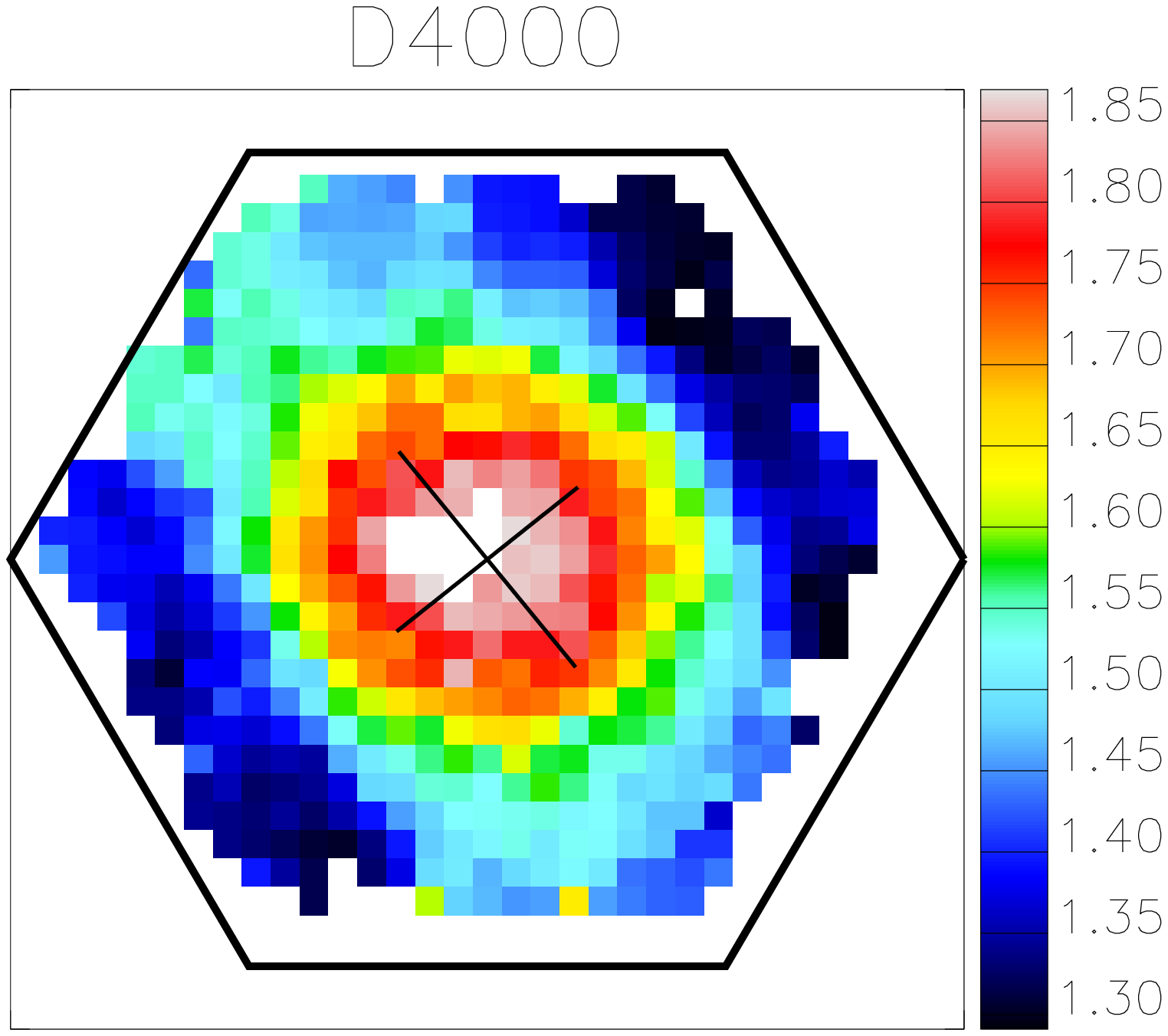,clip=true,height=0.140\textheight}
    \epsfig{figure=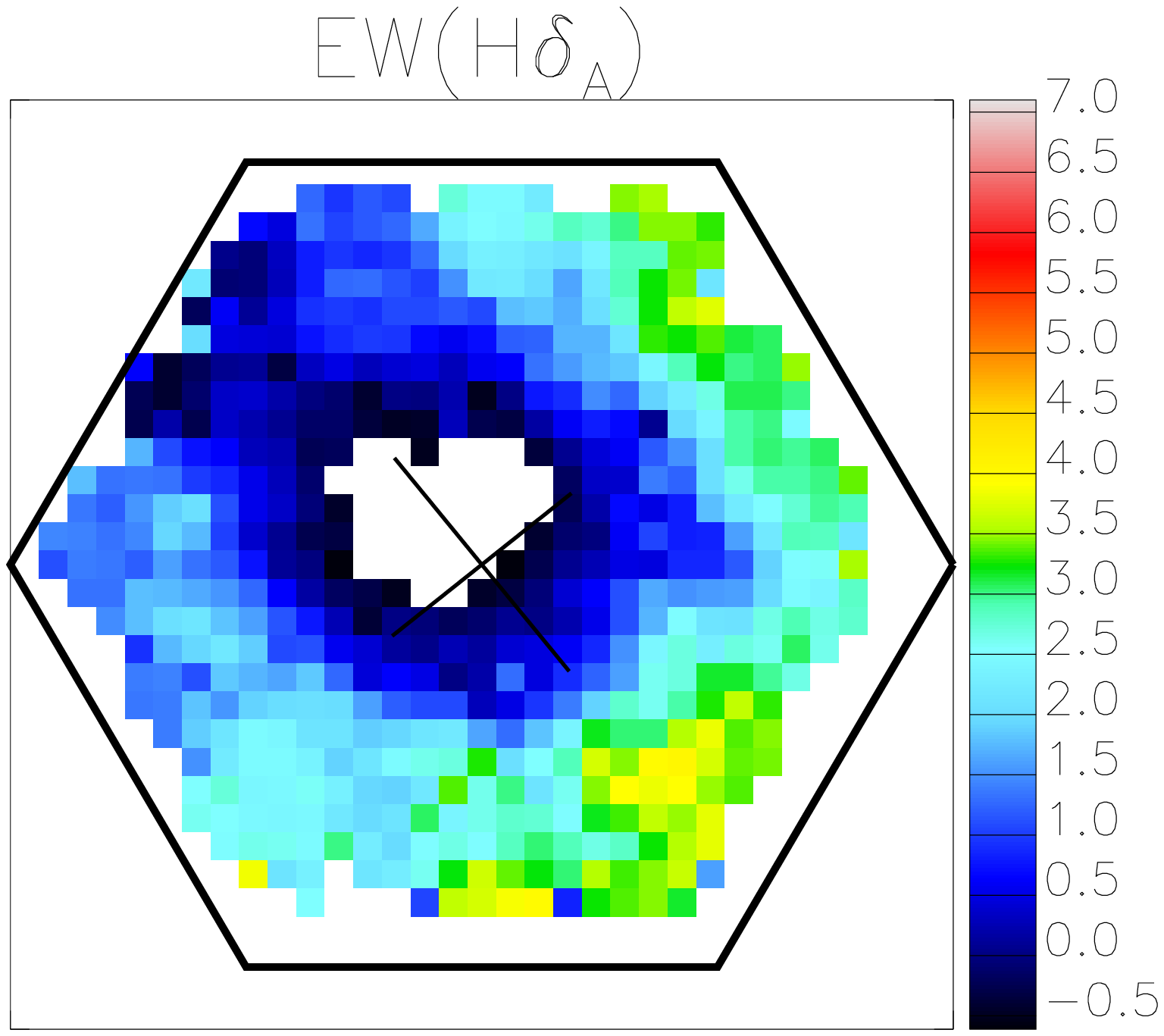,clip=true,height=0.140\textheight}
    \epsfig{figure=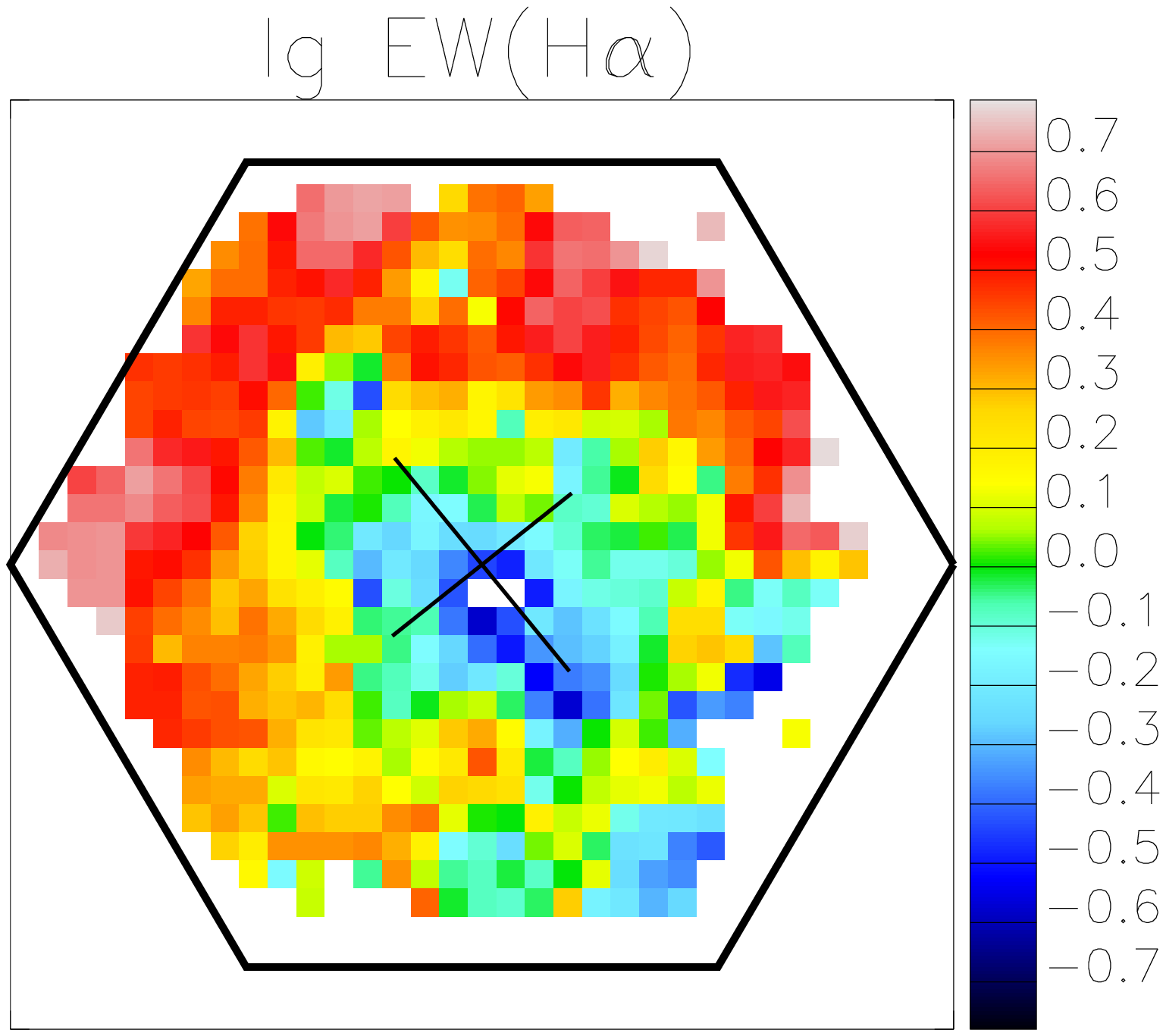,clip=true,height=0.140\textheight}
  \end{center}
  \begin{center}
    \epsfig{figure=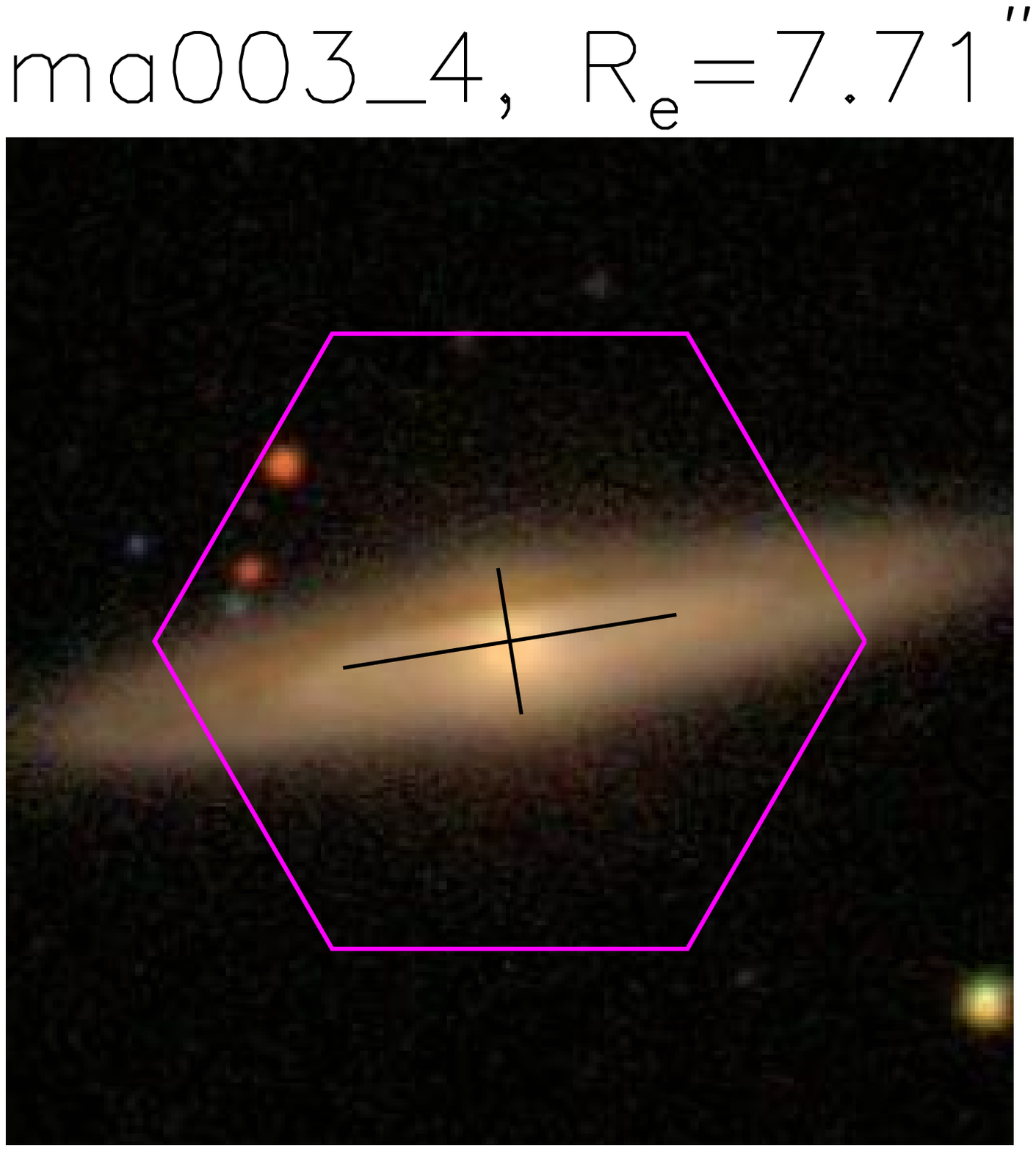,clip=true,height=0.140\textheight}
    \epsfig{figure=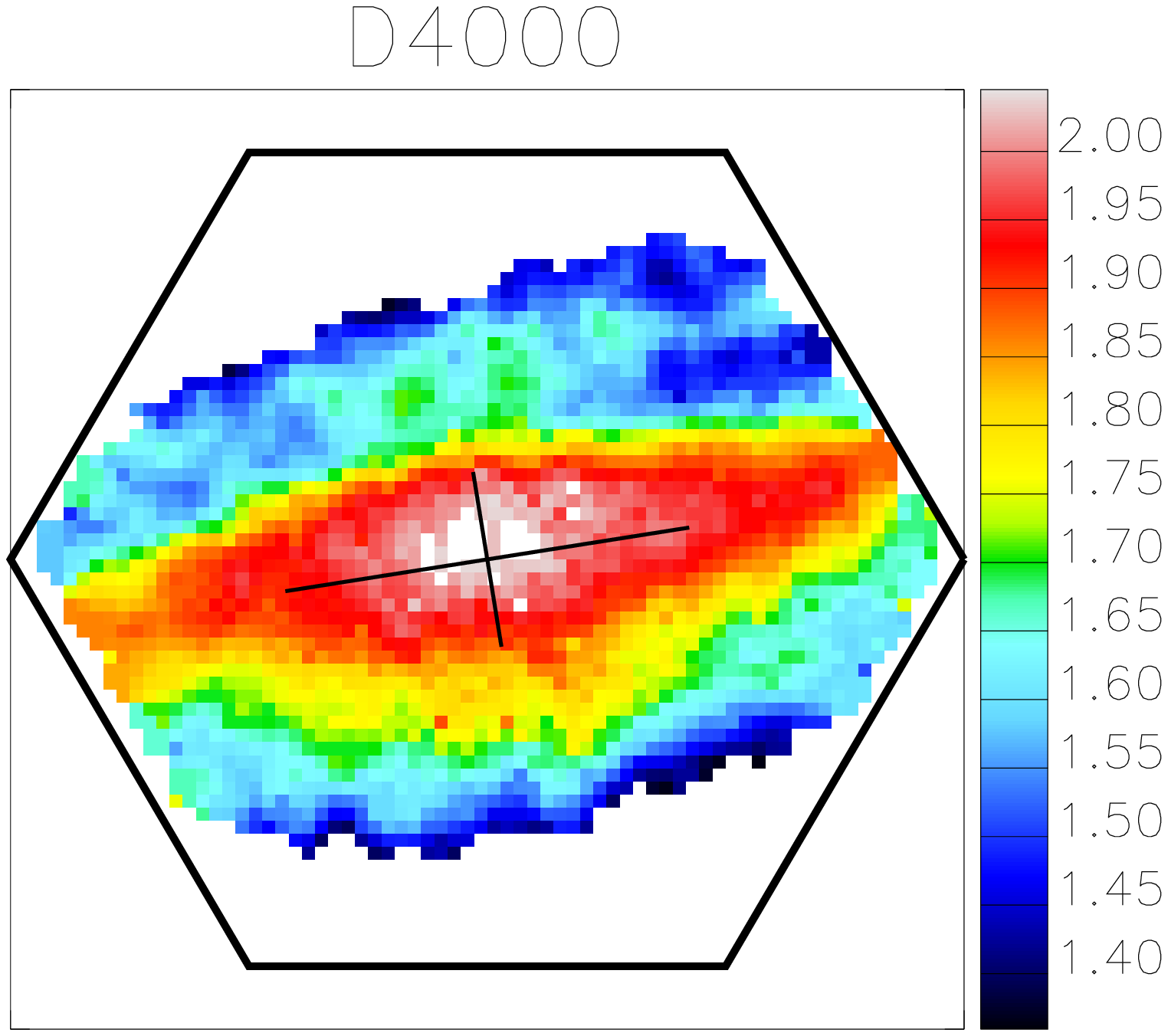,clip=true,height=0.140\textheight}
    \epsfig{figure=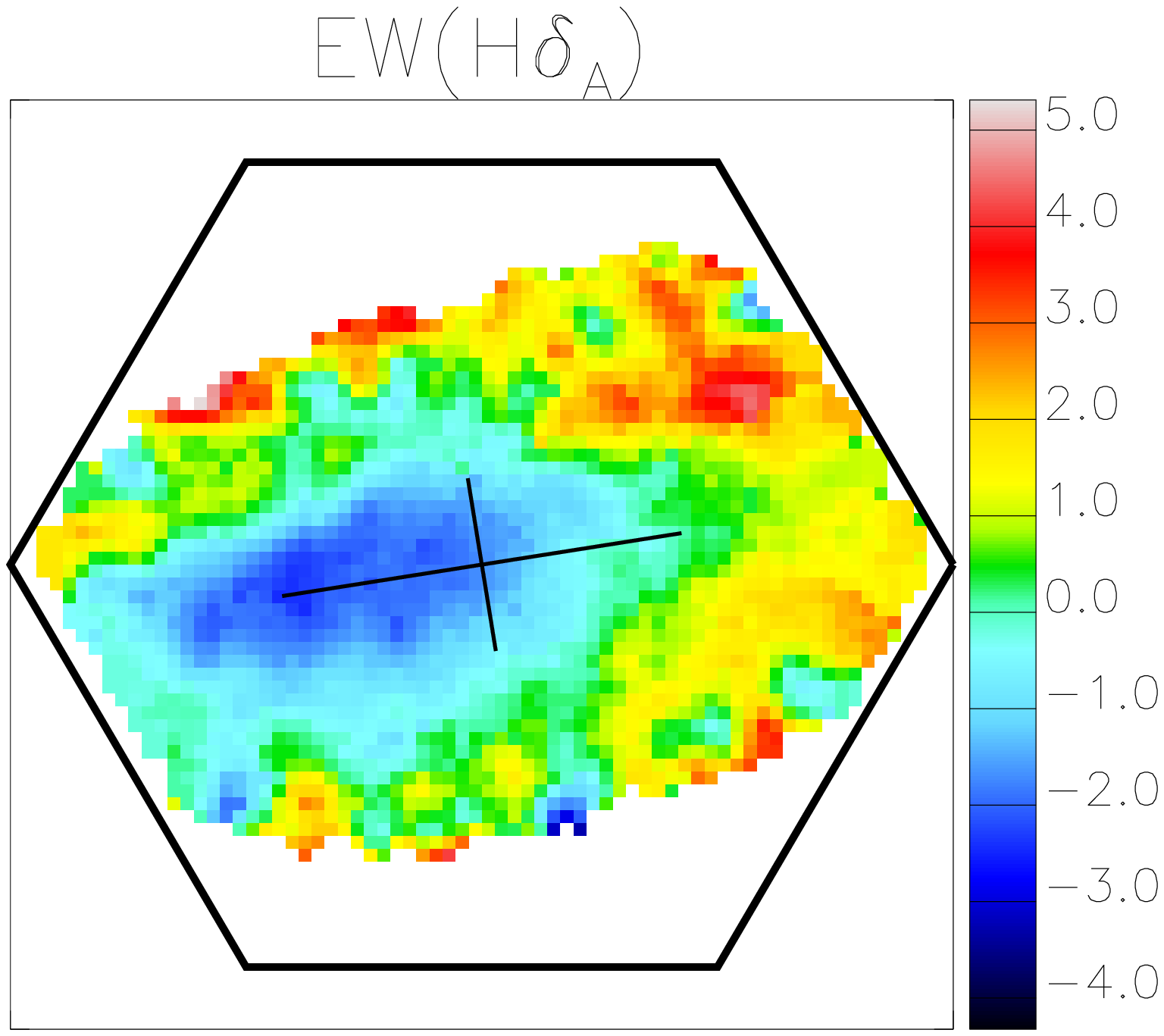,clip=true,height=0.140\textheight}
    \epsfig{figure=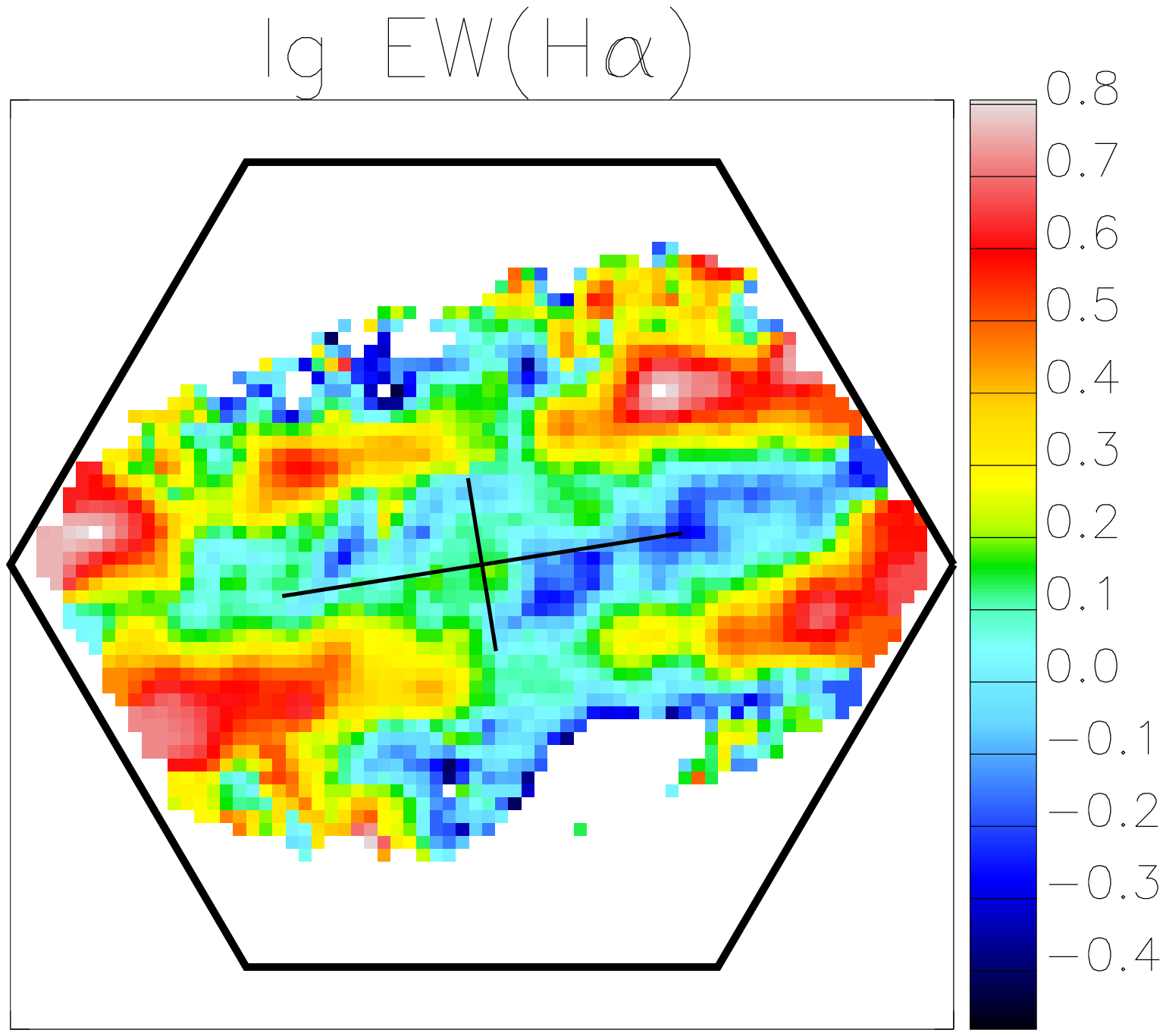,clip=true,height=0.140\textheight}
  \end{center}
  \caption{SDSS image  (left-most panels) and  maps of \dindex,  \ewhda\ and
    \lgewhae\   for  the   {\em  centrally   quiescent   galaxies}  in
    P-MaNGA. The galaxies from top to bottom are ordered by increasing
    their central  \dindex. Symbols and lines  are the same  as in the
    previous figure.}
  \label{fig:maps2}
\end{figure*}

\section{Results}

\subsection{Global versus central parameters}

Before investigating the spatially-resolved properties of these
galaxies' star-formation histories, we take a first look at any
spatial variations by comparing the central values of the diagnostic
parameters to their integrated values across the field of view.  To this end,
we generated a global spectrum for each galaxy by stacking the entire
datacube, with each spaxel at given wavelength being weighted by
$S/N^2$ following \citet{Cappellari-Copin-03}, where $S/N$ is the
signal-to-noise ratio at 5500\AA\ in the continuum.  Spaxels with
$S/N<3$ were excluded from the stacking.  We then measure the
diagnostic parameters \dindex, \ewhda\ and \lgewhae\ in the same
manner as above.  Figure~\ref{fig:cen_vs_glb} compares the global
measurements to the same parameters obtained from the central spaxel
of the galaxies.  For CSF galaxies, the global parameters differ
little from the central ones.  In contrast, however, most of the CQ
galaxies show significant difference between central and global
parameters, with smaller \dindex\ and larger \ewhda\ and \lgewhae\ in
the latter.  This result implies that the outer regions of CQ galaxies
contain younger stellar populations and have therefore experienced
more recent star formation than their inner parts.  The difference in
\ewhae\ is relatively weak compared to the other two parameters,
suggesting that the central-to-global variation may be mainly driven 
by the radial variation in recent star formation history, but not the
current star formation rate. It may also be the case that \ha\ 
is not a reliable tracer of star formation in galaxies dominated 
by older stellar populations (see \S~\ref{sec:summary}).

A comparison of the global parameters with the central parameters on
the diagnostic diagrams of recent star formation history is shown in
Figure~\ref{fig:d4k_hda_hae}, where we plot the galaxies on the
\dindex-\ewhda\ and \ewhda-\lgewhae\ planes, using the central and
global parameters for the upper and lower panels, respectively.
Comparing the upper and lower panels, we identify two noticeable
effects.  First, the gap between the CSF and CQ galaxies as seen in
the upper panels is largely filled in the lower panels, and this
in-fill is made up exclusively of CQ galaxies due to the broad scatter
in their global values of \dindex\ and \ewhda.  Second, when globally
measured, both the CSF galaxies and the CQ galaxies appear to deviate
from the continuous star-formation models, moving by varying degrees
toward the regime of recent bursts.

\subsection{2D Maps and radial profiles}
\label{sec:maps_profiles}

In Figures~\ref{fig:maps1} and~\ref{fig:maps2} we present the 2D maps
of the three diagnostic parameters, \dindex, \ewhda\ and \lgewhae, for
all the 12 galaxies in the sample, with the CSF galaxies in one figure
and the CQ galaxies in the other.  In each case, the galaxies have
been ordered such that the central values of \dindex, as measured from
the central spaxel of the datacubes, increases from left to right.
The SDSS $gri$ image is also shown for each galaxy.  The effective
radius, length and direction of major and minor axes are indicated in
the images and maps. When generating the maps, we only employ spaxels
where the S/N in the continuum at 5500\AA\ is greater than three.  

\begin{figure*}
  \begin{center}
    \epsfig{figure=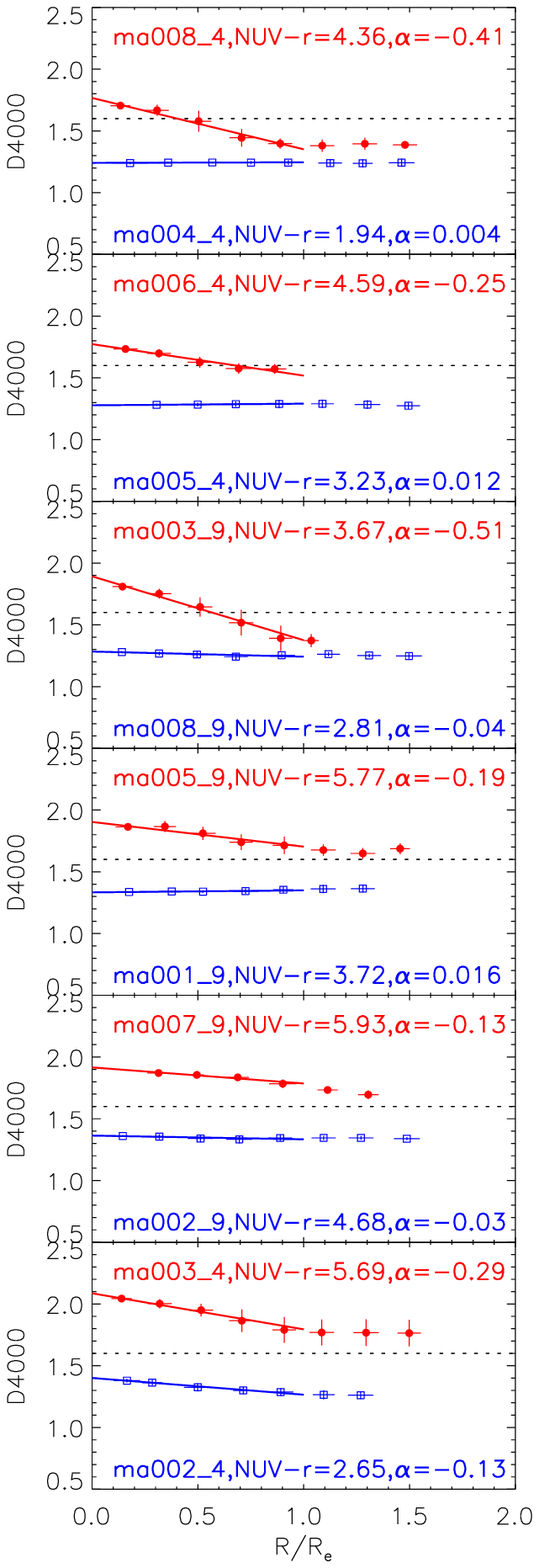,clip=true,width=0.33\textwidth}
    \epsfig{figure=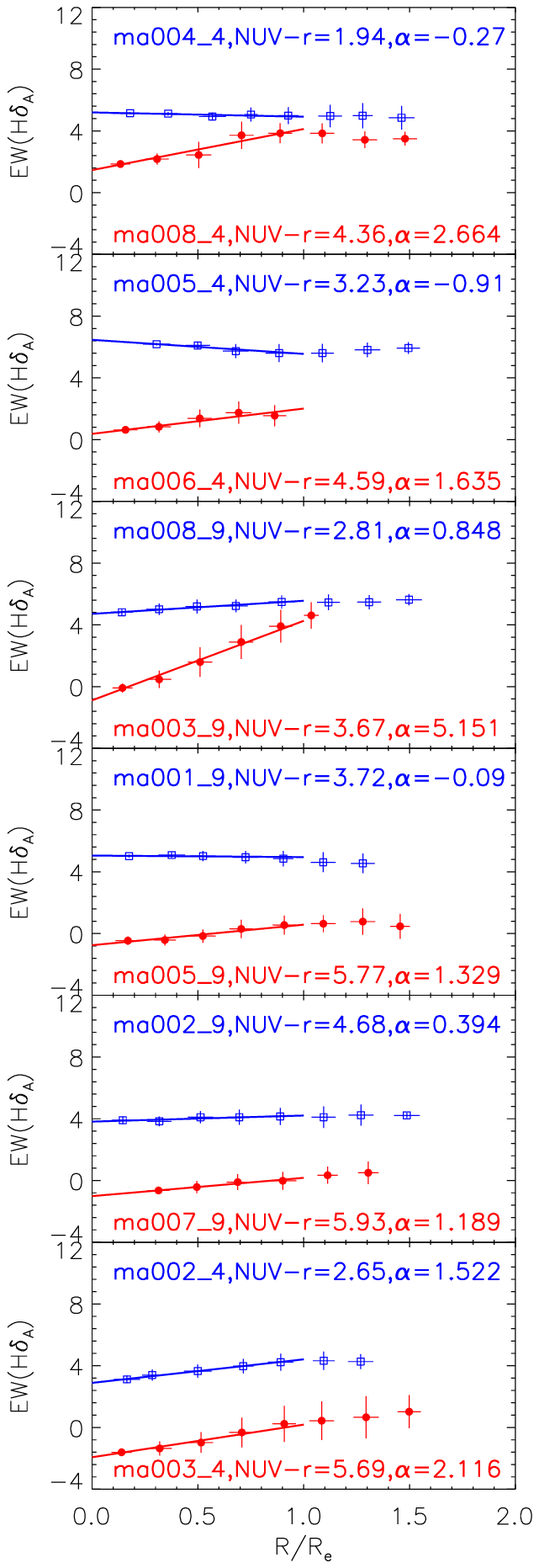,clip=true,width=0.33\textwidth}
    \epsfig{figure=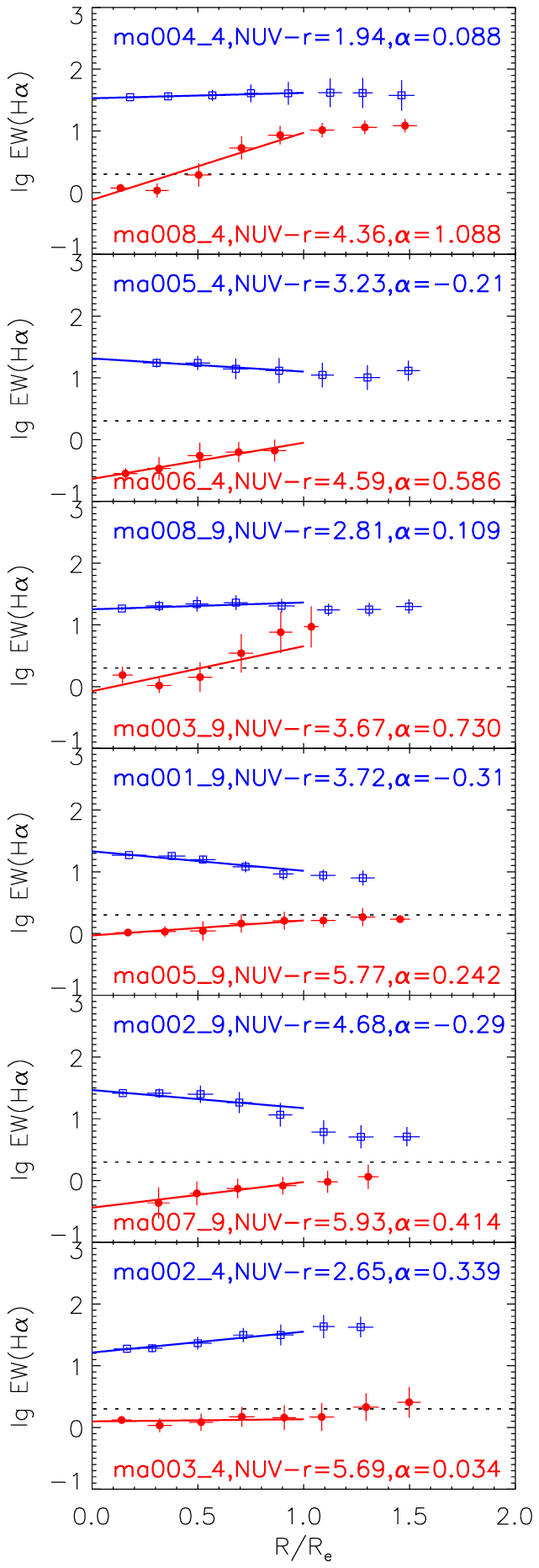,clip=true,width=0.33\textwidth}
  \end{center}
  \caption{Radial profiles of three diagnostic parameters measured for
    the P-MaNGA  galaxies (from left  to right): \dindex,  \ewhda\ and
    \lgewhae. Each  panel shows results for two  galaxies, a centrally
    star-forming  galaxy in blue  and a  centrally quiescent  galaxy in
    red, with their name (in  form of Unit\_Field) and global \nuvr\ 
    indicated. From top to  bottom panels  the galaxies  are ordered  by  
    increasing their central \dindex. A linear fit to each radial profile
    within the effective radius (\Reff) is plotted as the solid line,
    with the slope index $\alpha$ indicated.}
  \label{fig:profiles}
\end{figure*}

\begin{figure*}
  \begin{center}
    \epsfig{figure=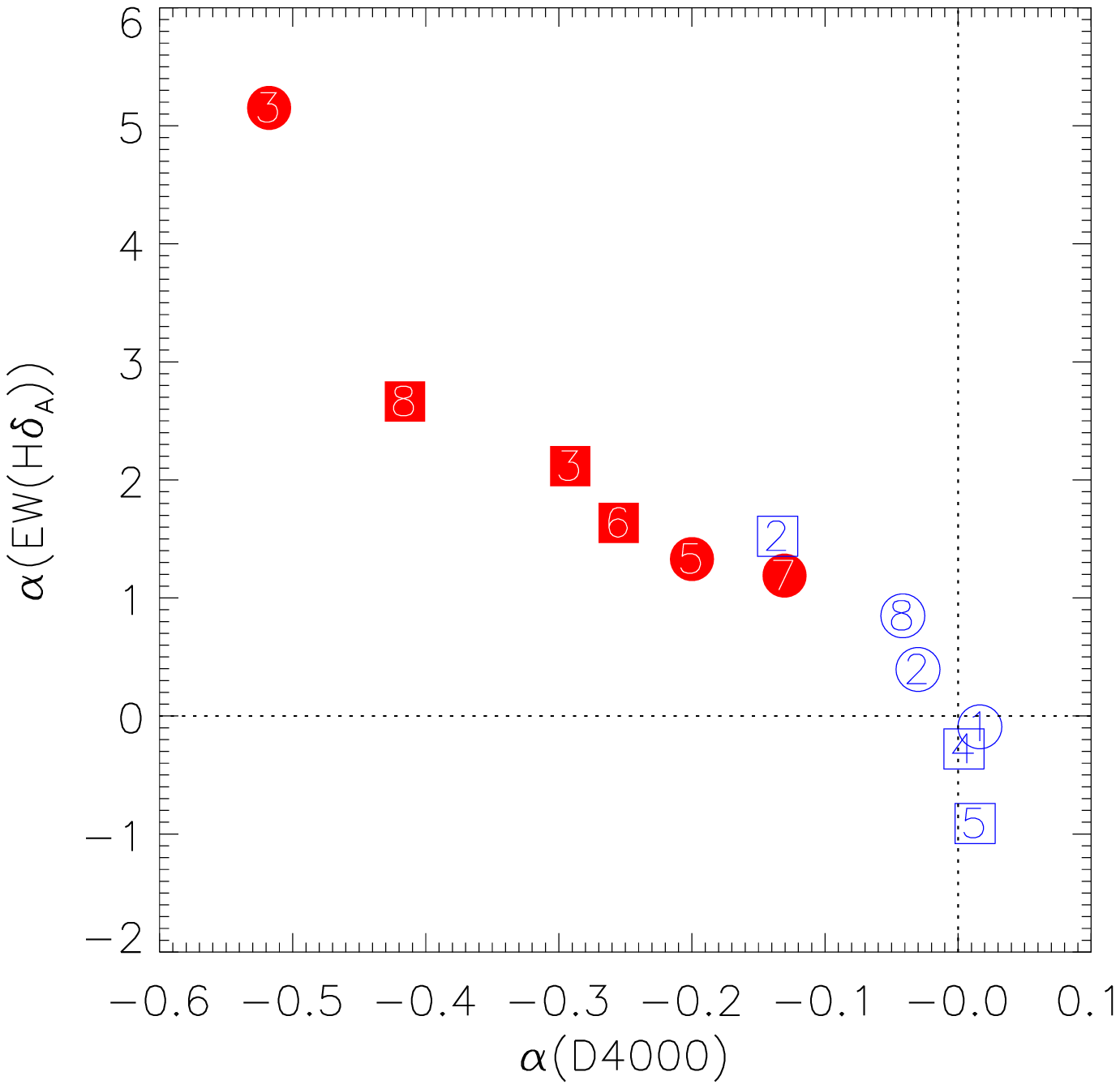,clip=true,width=0.33\textwidth}
    \epsfig{figure=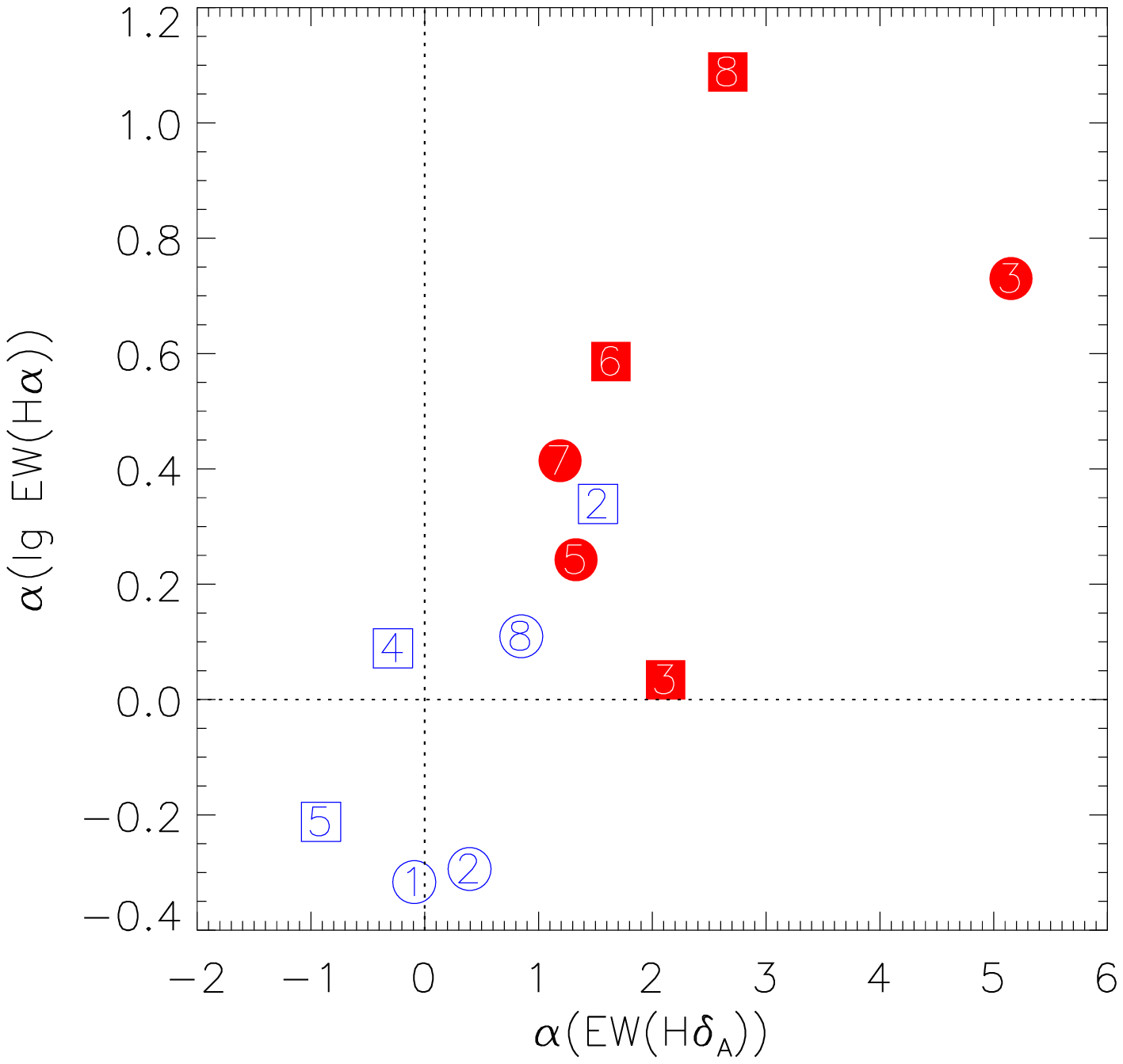,clip=true,width=0.33\textwidth}
    \epsfig{figure=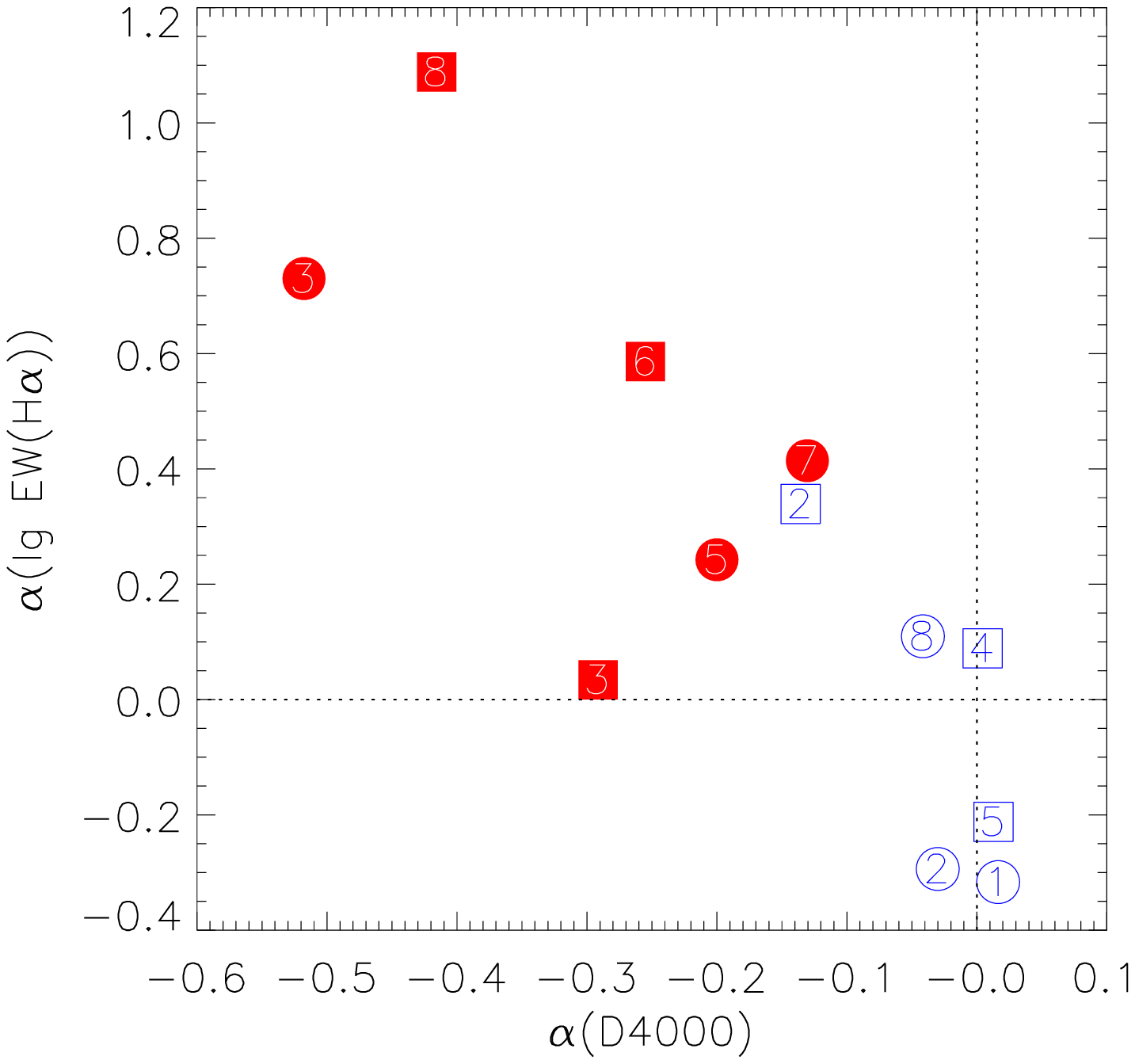,clip=true,width=0.33\textwidth}
  \end{center}
  \caption{Comparisons of  the slope index, determined  from the radial
    profile  at  radii  within  the  effective radius  for  the  three
    diagnostic parameters: \dindex, \ewhda\ and \lgewhae.}
  \label{fig:compare_slopes}
\end{figure*}

\begin{figure*}
  \begin{center}
    \epsfig{figure=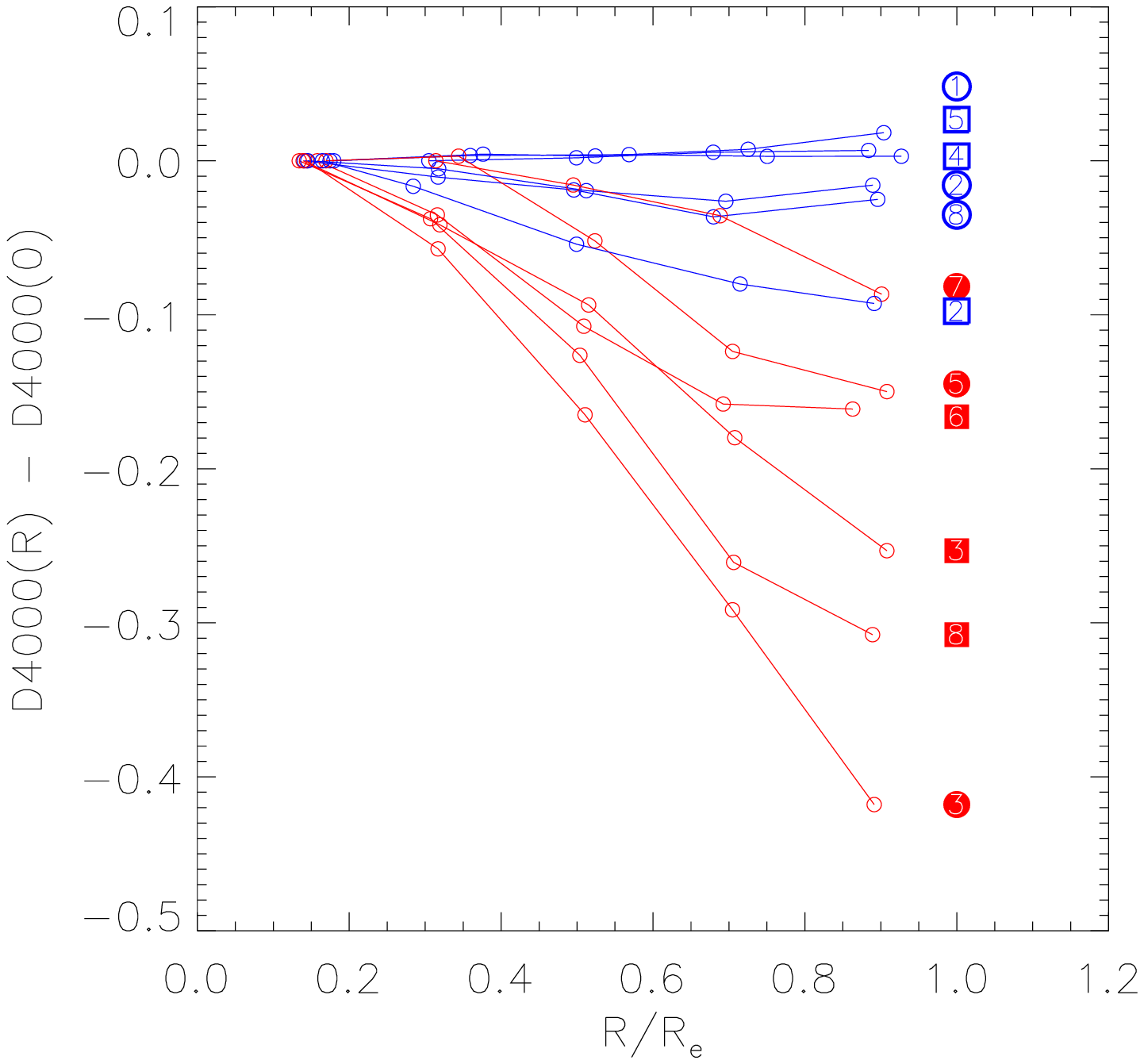,clip=true,width=0.33\textwidth}
    \epsfig{figure=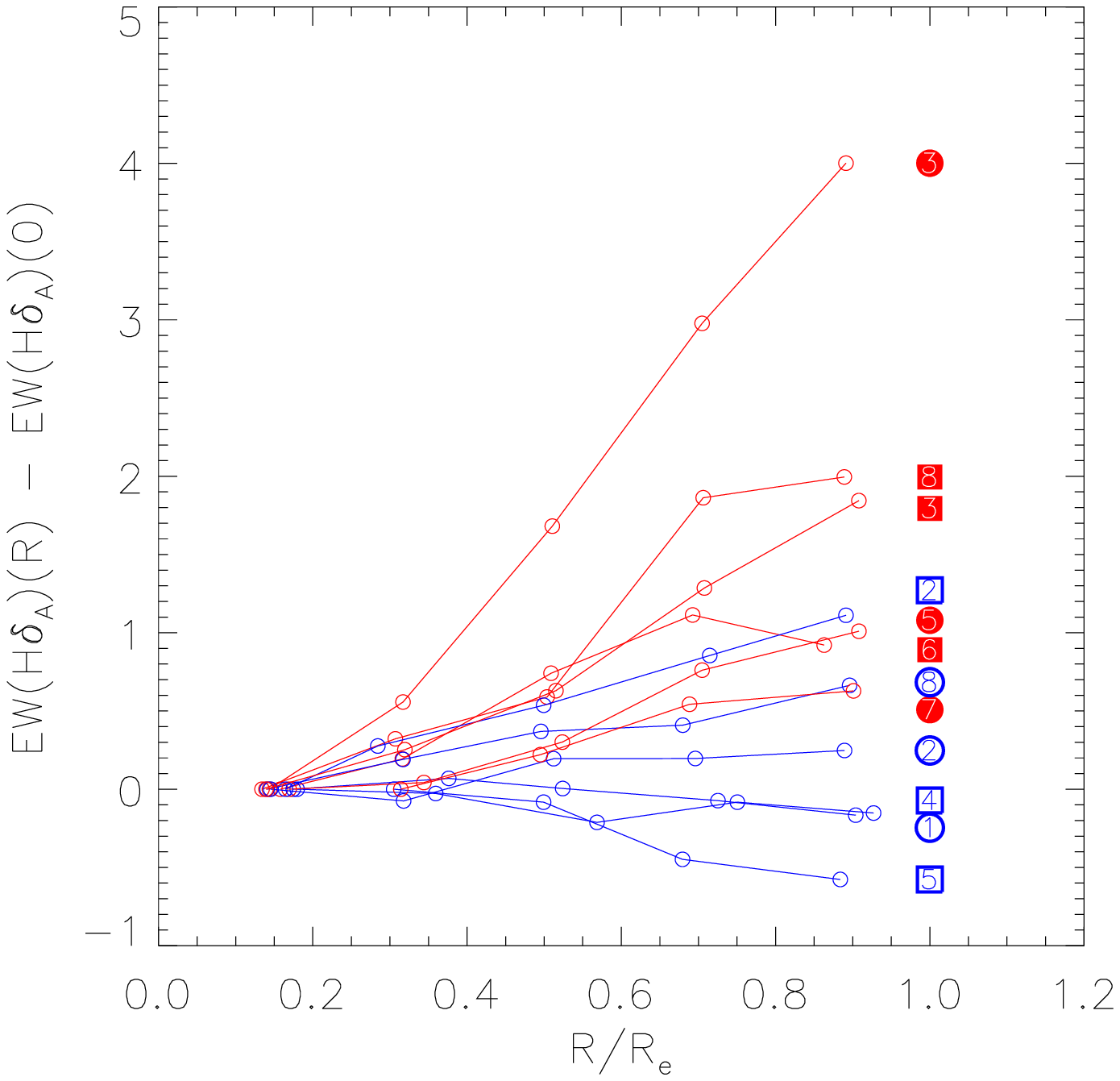,clip=true,width=0.33\textwidth}
    \epsfig{figure=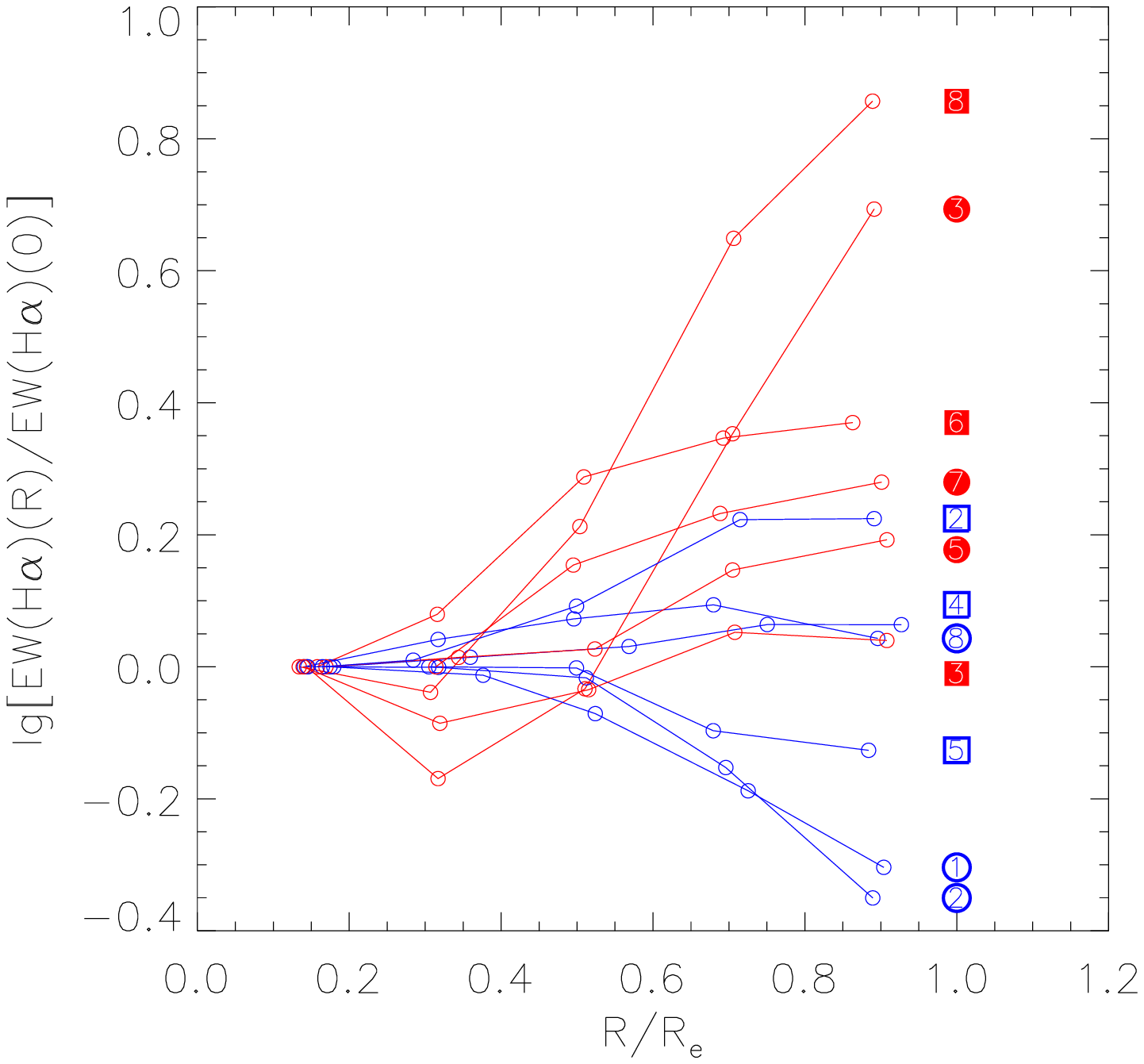,clip=true,width=0.33\textwidth}
  \end{center}
  \caption{Radial profiles of three diagnostic parameters measured for
    the P-MaNGA  galaxies (from left  to right): \dindex,  \ewhda\ and
    \lgewhae.  For a given parameter  and a given galaxy, each profile
    is normalized by the inner-most  radial bin and the radii are scaled
    by the  effective radius.  Results for  centrally star-forming and
    centrally  quiescent   galaxies  are  plotted  in   blue  and  red
    symbols/lines. The big symbols indicate  the field and IFU unit of
    the P-MaNGA observation.}
  \label{fig:profiles_in_a_panel}
\end{figure*}

\begin{figure*}
  \begin{center}
    \epsfig{figure=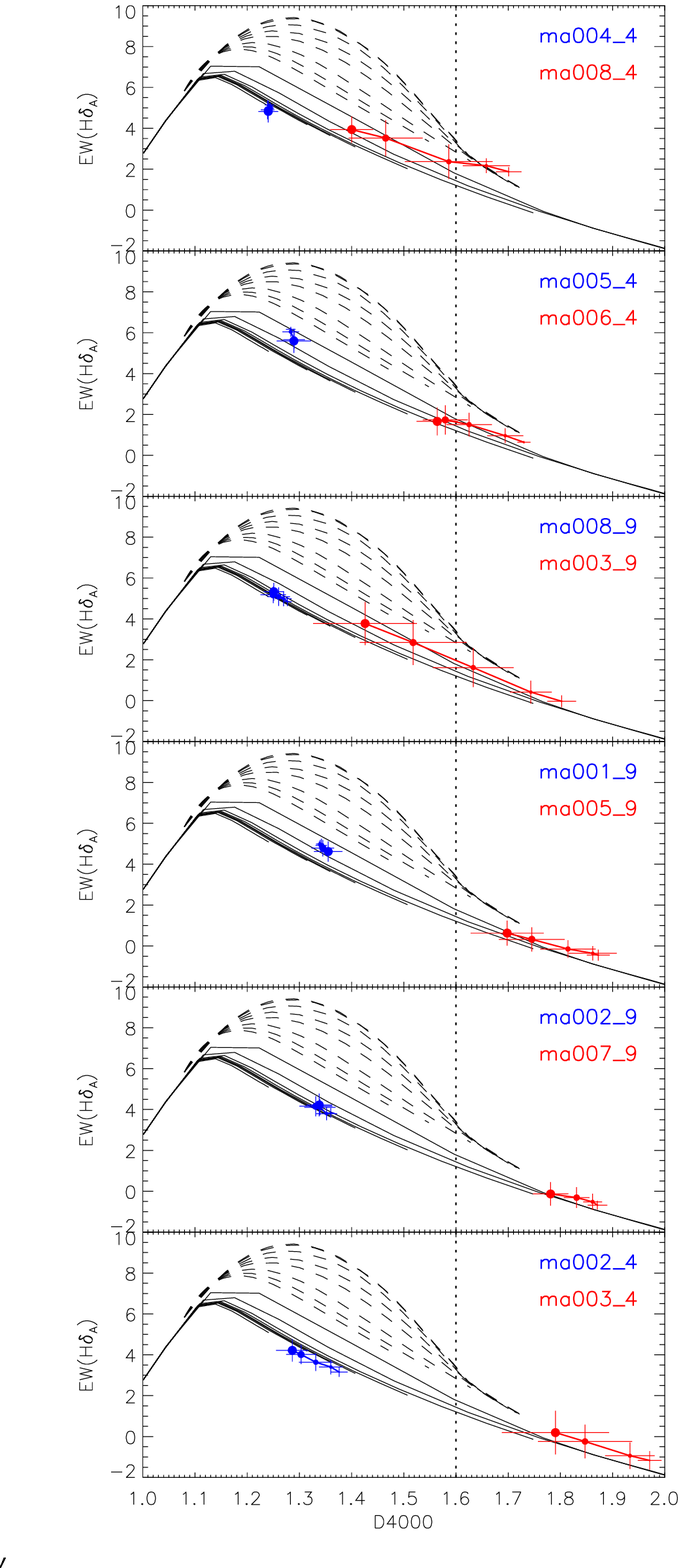,clip=true,width=0.48\textwidth}
    \epsfig{figure=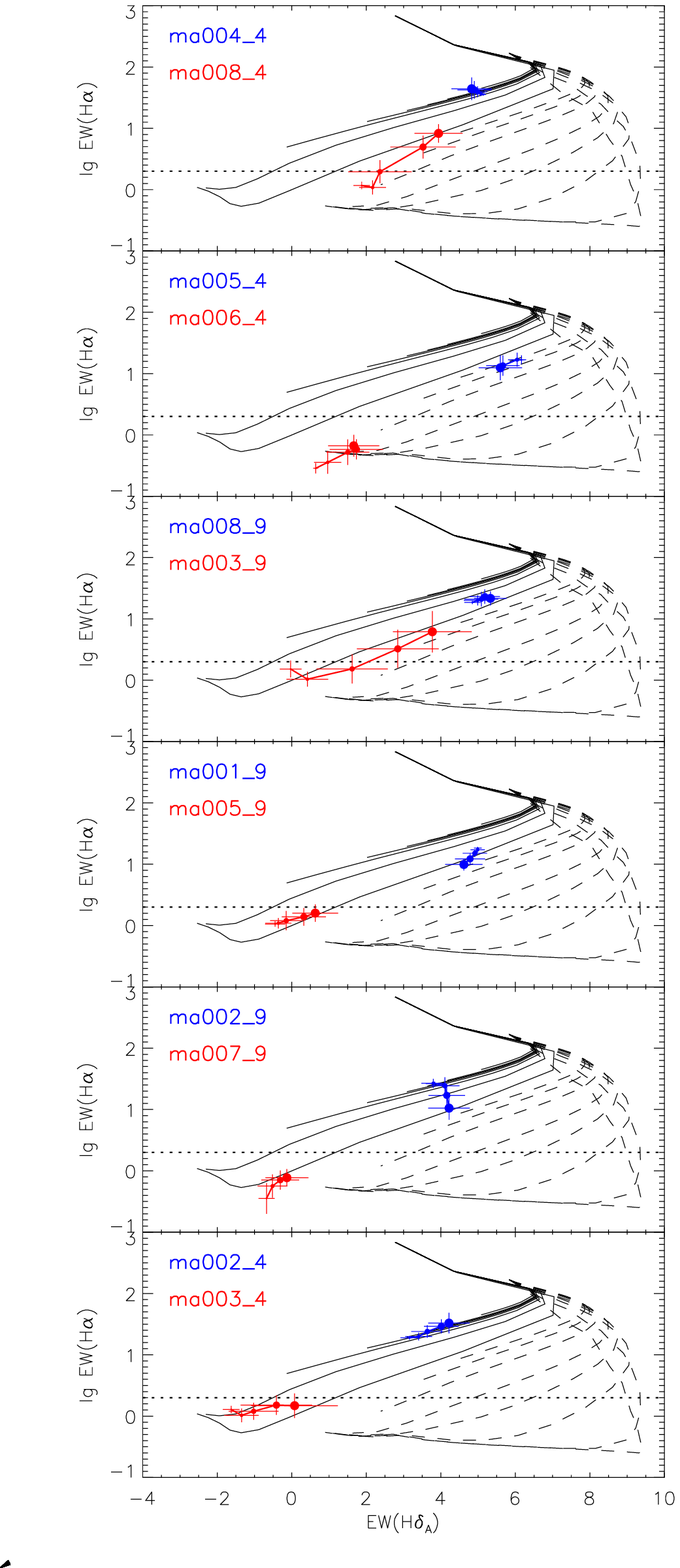,clip=true,width=0.48\textwidth}
  \end{center}
  \caption{Radial variation of the P-MaNGA galaxies in the plane
    of  \dindex\  vs.   \ewhda\   (left  panels),  and  the  plane  of
    \ewhda\ vs.   \lgewhae\ (right panels).  Each  panel shows results
    for two  galaxies: a centrally star-forming  galaxy (blue symbols)
    and a  centrally quiescent galaxy (red symbols),  which are defined
    to have \dindex$\le1.6$ and \dindex$>1.6$ in their center.  Larger
    data points  correspond to larger  radius.  Overplotted in  solid and
    dashed lines  are the same set  of stellar synthesis  models as in
    previous figures. Galaxy  name and plate ID are  indicated in each
    panel.}
  \label{fig:diagnostics}
\end{figure*}

Consistent with what was found from the global measurements in the
previous subsection, the overall impression from the 2D maps is that
all the galaxies show variation within the probed radius in all three
parameters, but to varying degrees and differing radial extents, with
the variations more pronounced in the CQ galaxies.  As noted above in
\S\ref{sec:sdss_data}, galaxy {\tt ma003\_9} 
shows a blue outer disk in the optical image,
which causes it to be classified as a blue-cloud galaxy in the
color--mass diagram, although its red core suggests that this is
intrinsically a CQ galaxy.  In agreement with this interpretation, the
galaxy shows strong contrast between the central and outer parts in
the maps of all the three diagnostic parameters.  Similar behavior is
seen in galaxy {\tt ma008\_4}, the most
massive galaxy in the P-MaNGA sample with
$M_\ast=2\times10^{11}M_\odot$, which is also a face-on spiral with a
relatively blue outer disk (although less prominent than in {\tt
  ma003\_9}) and strong radial variation in the diagnostic parameters.
The galaxy is classified as a green-valley galaxy on the color--mass
diagram (see Figure~\ref{fig:sample_properties}) due to its
intermediate color, \nuvr$=4.4$, which is unusually blue for its large
stellar mass.

The radial variations of the galaxies are shown more clearly in
Figure~\ref{fig:profiles}, where we plot the radial profiles of the
three parameters for all 12 galaxies. To save space we display two
galaxies in every panel, with one CSF galaxy (blue symbols) and one CQ
galaxy (red symbols).  This paired format also highlights the highly
different behavior of the two broad classes.  In constructing these
profiles, we have corrected for the effects of inclination on the
radius for each spaxel in the datacubes, using the minor-to-major axis
ratio from NSA, as determined from the SDSS $r$-band photometry.  When
estimating the radial profile, we bin all the spaxels with the
continuum $S/N>10$ at 5500\AA\ into a set of radial intervals with a
constant width of $\Delta\log_{10}(R/$\Reff$)=0.2$, according to the
deprojected radius of the spaxel ($r$) and the effective radius of the
galaxy (\Reff).  The value of the profile at given radius is then
estimated by the median of the spaxels falling in the radial interval,
and the error is given by the $1\sigma$ scatter between the
spaxels. 

Figure~\ref{fig:profiles} clearly shows that CQ galaxies generally
present significant radial gradients in all the parameters, in the
sense that the outer part of the galaxies show weaker \dindex\ (thus a
larger fraction of stellar populations with ages of $< 1\,{\rm Gyr}$),
stronger \hd\ absorption (thus more recent star formation in the past
a few Myr), and stronger \ha\ emission (thus likely stronger ongoing 
star formation). In contrast, CSF galaxies possess quite flat profiles 
in all cases, revealing little by way of gradient in star-formation
properties. At least out to \re, the radial profiles are close to
linear for all the galaxies. For each galaxy we apply a linear fit to
the radial profile within \re\ and show the best fit as a solid line
in the figure.  The slope of the best-fit line is indicated in each
panel. Figure~\ref{fig:compare_slopes} compares the slope index for
the three parameters, $\alpha($\dindex$)$, $\alpha($\ewhda$)$ and
$\alpha($\lgewhae$)$.  These plots confirm that the radial gradients
in the CSF galaxies are small, but also shows that the different
parameters are well correlated across galaxies of both types.  This
correlation is particularly strong for $\alpha($\dindex$)$ and
$\alpha($\ewhda$)$, which are beautifully (albeit non-linearly)
correlated.  However, given the small sample size, these apparent
correlations should probably not be over-interpreted.

Figure~\ref{fig:profiles_in_a_panel} compares the slopes of the radial
profiles in a different way, where we plot all the profiles in the
same panel for a given parameter, normalizing each profile by the
inner-most radial bin and scaling the radii by the effective radius.
The differing behavior of the two classes of galaxies is striking.
The CSF galaxies show only weak or no radial variation, while the CQ
galaxies display steeper profiles, spanning a wide range in slope.  It
is, however, interesting to note that although the average properties
of the two types differ dramatically, there is a degree of overlap
between the classes in all panels, suggesting that a continuum of the
same underlying physical processes has driven star formation in all
these galaxies.

As indicated from the 2D maps, the galaxies with a blue outer disk
({\tt ma003\_9} and {\tt ma008\_4}) indeed show the strongest radial
variation in all parameters. These two galaxies manifest their
peculiarity in the profile of \lgewhae, which is flat (for {\tt
  ma008\_4}) or even negative (for {\tt ma003\_9}) at $R<0.4$\Reff, before
rapidly increasing at larger radii.  When compared to these two
galaxies, the other CQ galaxies present much shallower profiles
(though still steeper than that of CSF galaxies), with \lgewhae\
increasing smoothly at $R<0.6$\Reff\ and becoming flat at larger radii.
The large values of \lgewhae\ at large radii indicates that these two
galaxies are strongly forming stars in their outer disks.  This
measurement is consistent with blue \nuvr\ color in these outer disks,
which is known to be a sensitive indicator of the cold gas mass
fraction in galaxies \citep[e.g.][]{Catinella-10,Li-12a}. 

The negative \lgewhae\ profile at $R<0.4$\Reff\ seen in {\tt ma003\_9} is
also found in {\tt ma003\_4}, and may be caused by the presence of 
a bulge or AGN in the galactic center.  Indeed, according to the BPT diagram
(Figure~\ref{fig:sample_properties}), the two galaxies are the only
AGN candidates in the sample.  It is likely that an AGN contributes to
the H$\alpha$ emission in the central region, thus enhancing the
central \lgewhae\ and leading to a dip in the profile as seen at
$R\sim0.3$\Reff. Interestingly, apart from this feature, the radial
profiles of the two probably AGN hosts are similar to those of the
other CQ galaxies, suggesting that current, low levels of AGN activity
may not have a strong effect on recent star formation activity, at
least for these two examples.

It is interesting to note that the three galaxies discussed
above, {\tt ma003\_9}, {\tt ma008\_4} and {\tt ma003\_4}, share
a number of common properties: they are all spiral galaxies with a 
quiescent center, at the massive end of our sample with 
\mstar$\sim10^{11}$M$_\odot$, and showing the largest gradients in \dindex\
indicative of young stellar populations formed within the past a few Gyr. 
They all share some properties with the population of ``passive red spirals'', 
identified recently as an interesting set of possible transition objects 
with disk-like morphologies but red colors, at both low-z
\citep[e.g.][]{Wolf-Gray-Meisenheimer-05, Wolf-09, Bamford-09,
Masters-10a} and high-z \citep[e.g.][]{Bundy-10}. 
Statistical analyses on a larger sample from MaNGA 
(or other IFU surveys) is needed in order to better understand the 
physical link between our galaxies and the passive/quiescent spiral population.

    The 2D maps in Figures~\ref{fig:maps1}
    and~\ref{fig:maps2}, as well as the radial profiles in
    Figures~\ref{fig:profiles} and~\ref{fig:profiles_in_a_panel}, are
    in broad agreement with maps/profiles of stellar population age
    estimated from both CALIFA \citep{GonzalezDelgado-14a} and P-MaNGA
    \citep{Wilkinson-15}, in the sense that most galaxies in the nearby
    universe show flat to negative gradients in stellar age, although
    in those  studies the detailed behaviors of the age gradients
    depend on  galaxy mass or morphology. In this work we only
    consider the dependence of our measurements on the \dindex of  the
    galactic center, which is an indicator of the prominence of
    stellar populations younger than 1-2 Gyr, rather than the mean
    stellar age as investigated in those studies. Despite these
    differences, the IFU-based results obtained so far are all consistent 
    with the ``inside-out'' growth of galactic disks
    \citep{White-Frenk-91,Mo-Mao-White-98}, a picture that was
    supported previously by both numerical simulations
    \citep[e.g.][]{Brook-06} and abundant studies of broadband color
    gradients pioneered more than 50 years ago by \citet{Tifft-63}.

\subsection{Diagnostic diagrams of recent star formation history}

We now return to the comparison of diagnostic diagrams of recent star
formation history, based on the radial profiles obtained in the
previous subsection.  Figure~\ref{fig:diagnostics} presents the
\dindex\ vs.  \ewhda\ (left panels) and \ewhda\ vs.  \lgewhae\ (right
panels) planes, with each panel once again contrasting two galaxies,
one CSF system (blue symbols) and one CQ system (red symbols).  Panels
are ordered such that \dindex\ increases from top to bottom.  For each
galaxy, the radial profiles within the effective radius \Reff\ as shown
in the previous figure are plotted here, with larger data points
corresponding to larger radii.  The set of BC03 models in
Figure~\ref{fig:d4k_hda_hae} are also repeated here as the solid and
dashed lines, but for clarity these lines are not colored; they
provide a reference grid to register the relative position of the
different radial bins on the diagnostic diagrams.

\begin{figure*}
  \begin{center}
    \epsfig{figure=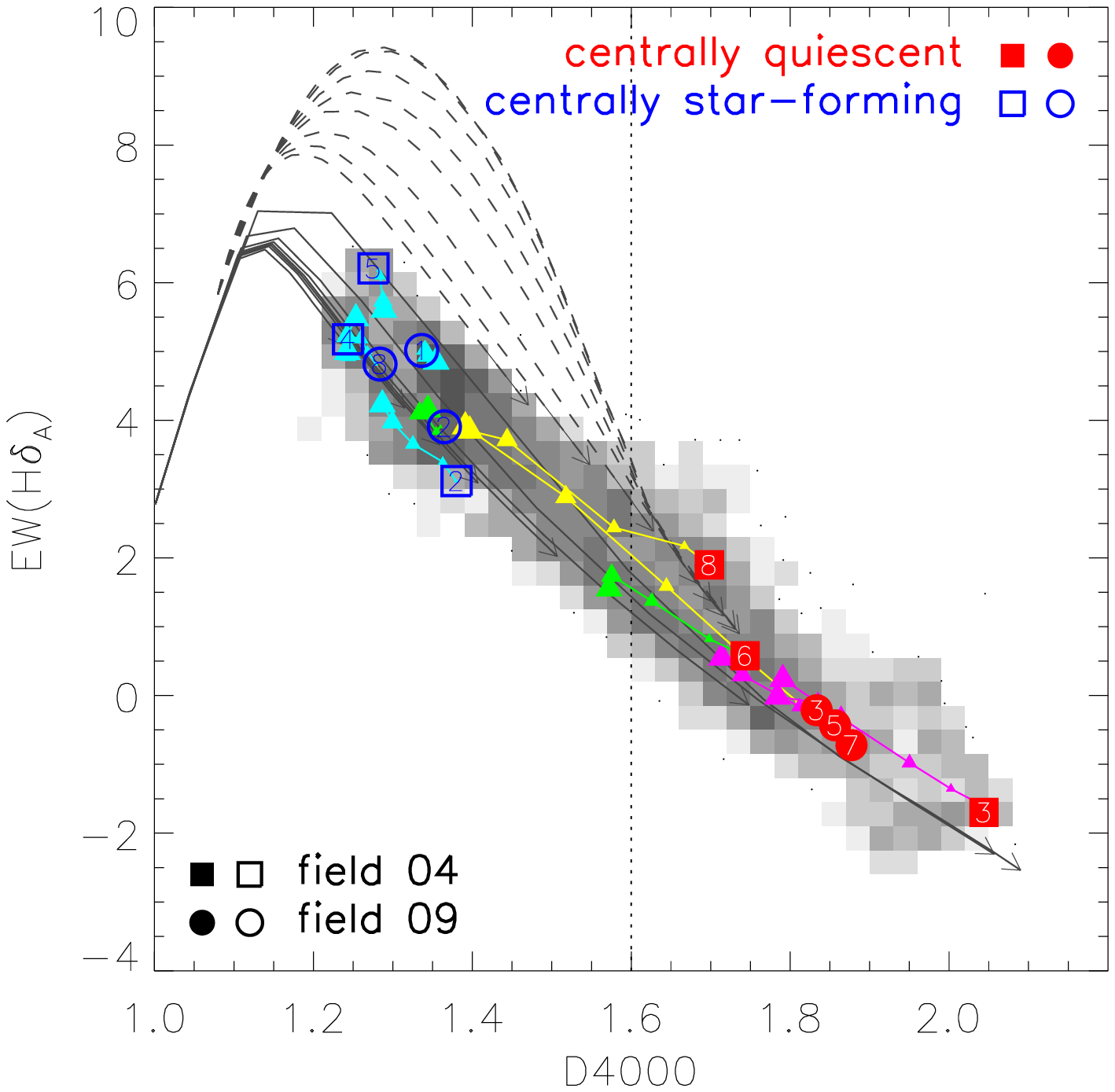,clip=true,width=0.4\textwidth}
    \epsfig{figure=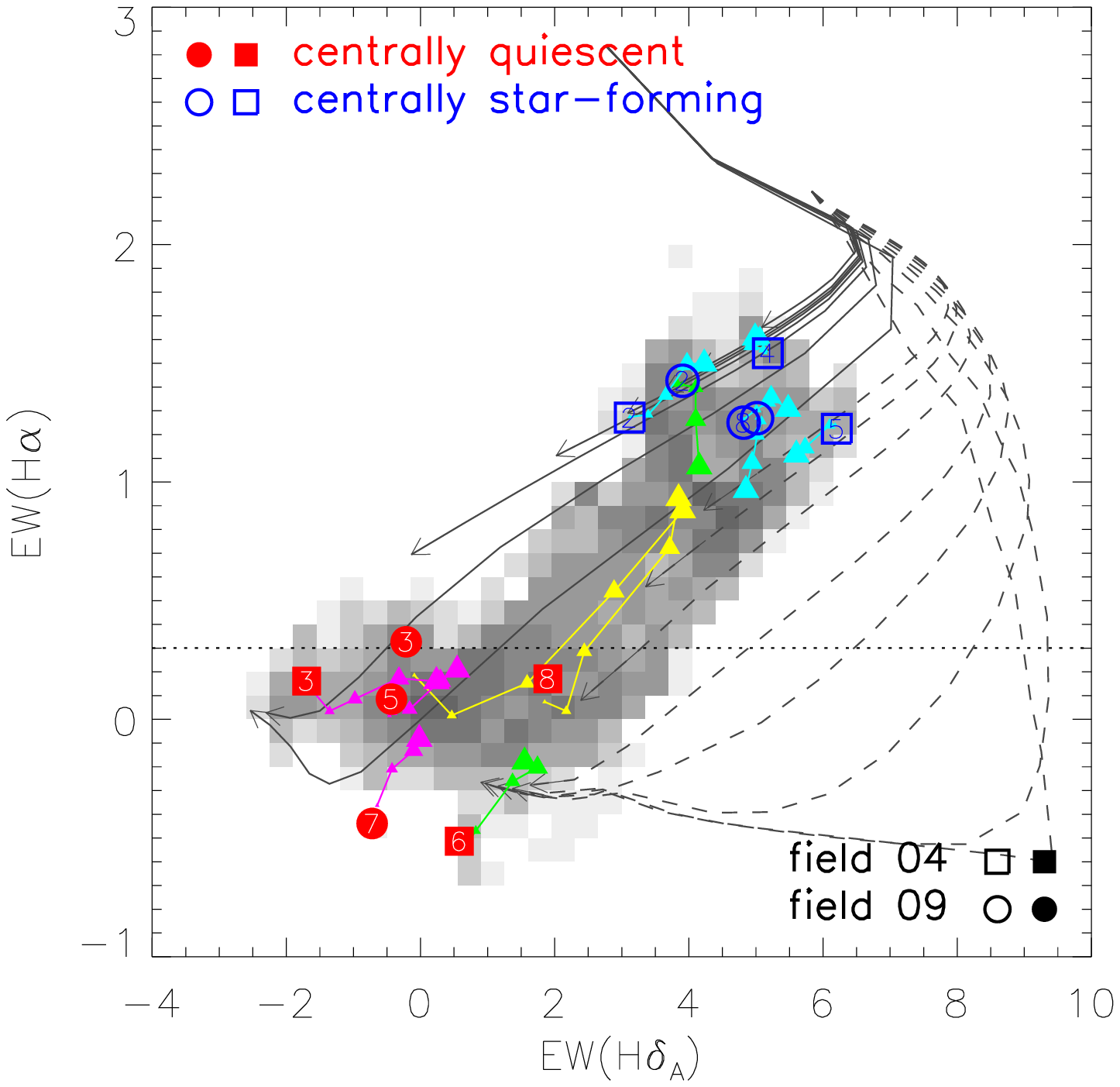,clip=true,width=0.4\textwidth}
  \end{center}
  \caption{Same  as   Figure~\ref{fig:d4k_hda_hae},  except  that  the
    radial  profiles  of  the P-MaNGA galaxies  are  additionally
    plotted,  with  magenta   (light  cyan)  triangles  for  centrally
    quiescent (star-forming)  galaxies. The profile for  a given galaxy
    is  connected with  a  solid line  and  the symbols  are sized  by
    radius, with larger triangles for larger radii. Plotted in yellow
    triangles are the two galaxies ({\tt ma003\_9} and {\tt ma008\_4}),
    which show the strongest radial gradients and unusual shape in the
    profile of \lgewhae, while green triangles highlight the green-valley
    galaxies with $4<$\nuvr$<5$. The individual spaxels of the P-MaNGA
    galaxies are plotted in the background as grey scale.}
  \label{fig:diagnostics_1panel}
\end{figure*}

Both the weak radial variations for CSF galaxies and the strong
variations for CQ galaxies are readily apparent in this figure.
Comparing the location of the radial bins on the two diagrams, as well
as their relative location with respect to the model curves, presents
a comprehensive picture of how and where star formation cessation is
occurring within the galaxies. The diagrams demonstrate that all the
CQ galaxies behave in the same manner: \dindex\ decreases, while
\ewhda\ and \lgewhae\ increase, as one moves outward in the galaxies.
This consistency is even clearer in
Figure~\ref{fig:diagnostics_1panel} where we overplot the radial
profiles of all 12 galaxies in a single set of diagnostic diagrams.
The profiles are plotted with cyan triangles for CSF galaxies, and
magenta triangles for CQ galaxies.  We highlight the two peculiar
galaxies discussed above, ({\tt ma003\_9} and {\tt ma008\_4}), which
show the strongest radial gradients and unusual shape in the profile
of \lgewhae, by identifying them with yellow triangles.  We also pick
out the green-valley galaxies ($4<$\nuvr$<5$) as green triangles.  For
comparison, the central spaxels are plotted with blue open symbols and
red solid symbols for the two classes of galaxies.  The CQ galaxies
indeed follow a tight track on the \dindex\ versus \ewhda\ plane.

These spatially-resolved data, reaching out to \Reff\ for all 12
galaxies, also show how the clear distinction between the two classes
of galaxies becomes blurred when one also considers their outer parts.
Figure~\ref{fig:diagnostics_1panel} shows that the gaps in the
parameter space between the regions occupied by the centers of the
galaxies are populated by these same galaxies' outer parts.  This
continuous sequence forms a tight relation in the
\dindex-\ewhda\ plane, closely following the continuous star formation
models.  In detail, though, the intermediate region on the diagrams is
mainly occupied by the two galaxies ({\tt ma003\_9} and {\tt ma008\_4})
that have strong radial variations and unusual profiles in \lgewhae.
In order to see this dominance more clearly, we have repeated the same 
panels of Figure~\ref{fig:diagnostics_1panel}, but excluding these two systems, 
and we find the gap between the galaxy types once again starts to emerge.
It is clear, however, that the CQ galaxies do still extend substantially
into the intermediate region, and that the effect appears to be stronger
for green-valley galaxies than for the other objects in the CQ galaxy class.

\subsection{Notes on individual galaxies}

\begin{description}

\item[ma008\_4] This system,
  CGCG~184-033, is the central galaxy of a group in the catalog of
  \citet{Yang-07}.  The system comprises 8 member galaxies with
  spectroscopically-measured redshift from SDSS, including the central
  galaxy itself.  This is the most massive galaxy in the sample with a
  stellar mass of $M_\ast=2\times10^{11}M_\odot$.  The SDSS
  spectroscopy and the central spaxel of the P-MaNGA datacube reveal
  that the central region of this galaxy shows relatively strong
  absorption in H$\delta$ for its \ewhae, implying that a
  starburst event has happened within the galaxy in the past 1 --
  $2\,{\rm Gyr}$. Considering the fact that the galaxy has a very
  close companion, a strongly star-forming spiral at the same redshift
  (z=0.042), the post-starburst feature may be the product of a tidal
  interaction with the companion, which caused the central part of the
  galaxy to have ceased star formation recently. The P-MaNGA data also
  reveal strong radial variation in all the diagnostic parameters in
  this galaxy: \dindex, \ewhda\ and \ewhae. In particular the profile
  of \lgewhae\ shows a very strange shape when compared to the other
  CQ galaxies.
  
\item[ma003\_9] This galaxy is a
  spiral with a red core but very blue outer disk in the optical image
  (see Figure~\ref{fig:maps2}).  It is classified as a blue-cloud
  galaxy because of its blue \nuvr\ color, but it is classified as a centrally
  quiescent galaxy due to its large central \dindex. The red core with old
  stellar populations and a blue global color seems to suggest that
  the galaxy might be a `Bluedisk' galaxy, a class of system with an
  unusually high fraction of H{\sc i} gas mass, as recently studied in
  detail by \citet{Wang-13b}. However, the H{\sc i} gas mass is
  $\log_{10}(M_{\rm HI}/M_\odot)=9.8$ according to the H{\sc i} global
  profile from \citet{Huchtmeier-Richter-89}, homogenized in 
  \citet{Paturel-03a}. This is comparable to the value expected from
  the \nuvr\ and stellar surface mass density for a normal spiral
  \citep{Catinella-10, Li-12c}. The BPT diagram shown in 
  Figure~\ref{fig:sample_properties} reveals strong nuclear activity 
  in the galactic center. A $\sim12$~mJy source coincident with this 
  galaxy is detected by the NVSS survey \citep{Condon-98}, but 
  no source is detected by the FIRST survey \citep{Becker-White-Helfand-95},
  suggesting that the radio emission from this galaxy is variable. 
  This is consistent with an AGN origin, albeit at a low luminosity 
  ($L_{\rm 1.4GHz} \sim 10^{21.6}$~W/Hz). See the related paper by 
  \citet{Belfiore-15} for a spatially-resolved BPT diagram of this galaxy,
  also based on the P-MaNGA data.

\item [ma003\_4] This system is one of the two
  AGNs in the sample (the other one is {\tt ma003\_9} above). The AGN
  component is likely responsible for the strange profiles of the
  \lgewhae\ diagnostic in these systems, as discussed above.

\item [ma006\_4] With a stellar mass of 
  $6.3\times10^{9}M_\odot$, this galaxy is the only low-mass CQ galaxy 
  in the P-MaNGA sample. It has unusually low \ewhae\ for its \ewhda
  (see Figure~\ref{fig:diagnostics}). This might be a metallicity effect,
  as the galaxy is much closer to the continuous models for a lower 
  metallicity of Z=0.008.

\end{description}

\section{Summary and discussion}
\label{sec:summary}

Using datacubes for 12 galaxies produced during the MaNGA prototype
(P-MaNGA) observations, we have obtained maps and radial profiles of
the \dindex, \ewhda\ and \ewhae\ spectral diagnostics.  These
parameters were derived for each spaxel by decomposing the spectrum
into an emission-line component and a continuum plus absorption-line
component, and measuring the relevant indices from these separate
components.  We study the spatially-resolved recent star formation
histories of these galaxies by displaying their radial profiles on the
\dindex\ versus \ewhda\ and \ewhda\ versus \lgewhae\ planes.  We have
classified the galaxies as either `centrally quiescent' (CQ) or
`centrally star-forming' (CSF) according to the value of \dindex\ in
the central spaxel, and we compare the 2D maps and radial profiles for
the two classes.  CQ galaxies present significant radial gradients in
all the three parameters, in the sense that \dindex\ decreases, while
both \ewhda\ and \ewhae\ increase as one goes from the galactic center
toward the outer region.  In contrast, CSF galaxies show very weak or
no radial variations.  The outer parts of the galaxies show greater
scatter on the diagnostic diagrams than their central regions.  In
particular, the separation between CQ and CSF galaxies in these
diagnostic planes is largely filled in by the outer parts of galaxies
whose global colors place them in the green valley.

The three spectral diagnostics are known to be sensitive indicators of
stellar populations of different ages.  Spatially resolving a galaxy
on the \dindex--\ewhda--\ewhae\ diagrams probes the radial variations
in recent star formation histories, thus revealing the way in which
the galaxy grows and its star formation ceases. When spatially
resolved, the P-MaNGA galaxies are found to form a tight sequence on
the \dindex\ versus \ewhda\ plane, although the distribution of their
central regions is strongly bimodal
(Figure~\ref{fig:diagnostics_1panel}).  This sequence closely follows
the continuous star formation locus predicted by current stellar
population models, and covers a very similar area to the distribution
of the large sample of galactic centers from SDSS
(Figure~\ref{fig:sdss_d4k_hda_hae}).  This relation suggests that, at
least for the systems being studied here, galaxy growth has been a
smooth process, and is likely regulated by a common set of physical
drivers.  This result is consistent with the `inside-out' picture of
galaxy growth, where the stellar mass assembly starts in the galactic
center and gradually extends to the outer regions
\citep{White-Frenk-91, Mo-Mao-White-98, Brook-06}.
In this picture, the shutdown of star formation also first occurs in 
the central region and slowly propagates out to ever larger radii. 
However, this result should not be overemphasized given the small size
of our sample. The tight sequence on the \dindex--\ewhda\ diagram might
be just a result of small number statistic. It could also be the case
that there are a variety of physical processes behind this sequence, 
which can be discriminated only when a larger sample becomes available.

Radially resolved stellar populations have been 
studied previously to some extent, based on multi-wavelength 
broadband photometry at both low-z
\citep[e.g.][]{deJong-96, Bell-deJong-00, Taylor-05,
Munoz-Mateos-07,Zibetti-Charlot-Rix-09,Roche-Bernardi-Hyde-10,
Suh-10,Tortora-10a,
Gonzalez-Perez-Castander-Kauffmann-11,Tortora-11,Kauffmann-15}
and high-z 
\citep[e.g.][]{Abraham-99,Azzollini-Beckman-Trujillo-09,
Szomoru-12,Wuyts-12,Szomoru-13,Wuyts-13}, 
as well as long-slit 
spectroscopy for nearby galaxies \citep[e.g.][]{Moran-12, Huang-13b}.
Most spirals are found to have stellar population gradients, 
with the inner regions being older and more metal-rich than 
their outer regions, supporting the inside-out scenario of disk 
galaxy formation. The majority of the early-type galaxies show 
flat color profiles, with only a small fraction being ``blue-core'' 
systems with negative color gradients (i.e. bluer outer parts). 
The star formation history of galactic disks is found to be primarily 
driven by its structural properties, as usually quantified by the 
surface stellar mass density \citep[\mustar,][]{Bell-deJong-00} 
and stellar light concentration \citep[][]{Gonzalez-Perez-Castander-Kauffmann-11},
while the total stellar mass is a secondary parameter, correlated 
more closely with metallicity than with age. For massive galaxies,
recent short-lived episodes or bursts of star formation occurring
in their outer regions are required to interpret the radially 
resolved spectral indices as measured from long-slit spectroscopy
\citep{Huang-13b}, a result that is consistent with the existence 
of many early-type galaxies in the nearby universe which display 
extended star-forming discs, as originally discovered by ultraviolet 
imaging \citep[e.g.][]{Yi-05, Kauffmann-07, Fang-12}.
The overall radial dependence of the diagnostic parameters presented
in our work is apparently consistent with these previous findings,
although the small sample size doesn't allow the correlation of the
radial profiles with galaxy morphology, stellar mass and structural
properties to be examined in a statistical manner. 

More recent studies using data from CALIFA have provided the most 
comprehensive results so far regarding the radial variations of the 
stellar population and star formation history of nearby galaxies
\citep{Perez-13,GonzalezDelgado-14a,Sanchez-14,Sanchez-Blazquez-14}.
\citet{Perez-13} analyzed the first 105 galaxies of the CALIFA survey,
quantifying their spatially resolved history of stellar mass assembly
and demonstrating how massive galaxies grow their stellar mass 
inside-out. Oxygen abundances estimated from H{\sc ii} regions in 
306 CALIFA galaxies present a common negative gradient in the oxygen 
abundance between 0.3 and 2 times \re, independent of morphology, 
the presence of bars, absolute magnitude or mass, provided that the analysis
is limited to those galaxies without clear evidence of an interaction
\citep{Sanchez-14}. This result also supports the inside-out picture 
for the growth of galactic disks, and is consistent with the stellar 
age and metallicity gradients obtained by \citet{Sanchez-Blazquez-14}
based on CALIFA data for a smaller sample of 62 nearly face-on spirals.

Using a sample of 107 galaxies from CALIFA, \citet{GonzalezDelgado-14a}
performed a detailed study of the radial structure of the stellar mass
surface density (\mustar) and stellar population age, as well as their
dependence on the total stellar mass and morphology. In support of
the inside-out formation picture, negative gradients of the stellar
age are present in most of their galaxies. The behaviors of the 
\dindex\ profiles in the P-MaNGA galaxies (see Fig.~\ref{fig:profiles_in_a_panel})
are in good agreement with this result, although the \dindex\ is not
a direct measure of stellar age, but rather an indicator of the prominence
of young stellar populations. Confirming the previous broadband photometry-based
findings, the CALIFA data reveals that \mustar\ is more important than
\mstar\ in driving the star formation history in disks, while \mstar\
plays a more fundamental role in spheroidal systems. The CALIFA
results, together with the results from broadband photometry and 
long-slit spectroscopy, strongly suggest that galaxy growth and
death are driven by multiple parameters (at least the local surface density,
total stellar mass and morphology), and that further studies
of much larger samples are needed in order to isolate the role of each
parameter, and the physical process behind it. The increasing MaNGA 
sample is expected to provide much better results in the near future.

Another interesting result from our work is that, when spatially resolved, 
the distribution of galaxies on the diagnostic diagrams is no longer bimodal, 
and the separation between the CQ and CSF populations is filled in by 
those galaxies that are classified as `green-valley' according to their
optical/UV color. In a recent complementary study, \citet{Mendel-13}
selected from SDSS a large sample of `recently quenched' galaxies that
have a significant fraction of young stars within the central
$3^{\prime\prime}$ region, and found them to populate the region of
green-valley galaxies in \dindex\ and optical colors, so their central
parts seem to be at the same evolutionary stage as the outer parts of
the P-MaNGA galaxies.  The two studies combine to suggest that the
green valley does, indeed, pick out a transition phase between
star-forming and passive populations, but that different parts of
different galaxies enter the green valley at different times.

There are a few caveats to this analysis that we will address in future
work. We have used relative measurements of \dindex, \ewhda\ and
\ewhae\ as indicators of recent star formation history. We have not
presented quantitative measurements of mean stellar ages due to the
sensitivity of these indices to stellar metallicity, element abundance
and dust attenuation \citep{Worthey-Ottaviani-97, Bruzual-Charlot-03, 
Thomas-04, Thomas-Davies-06, Sanchez-12}. Throughout this paper we use
solar metallicity models for comparison.
We have compared the data to models with three different metallicities
(Z=0.4, 1, 2.5Z$_\odot$), finding that the dependence on metallicity is
very small for younger stellar populations, but become more important at
older ages. In \citet{Wilkinson-15} we explore the possibility
of simultaneously deriving the age and metallicity of P-MaNGA galaxies
using stellar population model fits to the full spectrum.
In the present paper, we use the \ha\ emission line to characterize the
instantaneous star formation rate. However, for galaxies with dominant
old stellar populations, OB stars may not be the only source of
ionizing photons \citep[e.g.][]{CidFernandes-10,Sanchez-14}.
Indeed, \ewhae\ lower than 6\AA\ was found to be more compatible with
light from post-AGB stars \citep{Papaderos-13,Sarzi-13}, and it would
be interesting to evaluate the contribution of such stars to the
emission-line properties of the spectra once a larger sample is available.

In a recent SDSS-based study \citet{Kauffmann-14} has nicely demonstrated 
that a combination of \dindex, \ewhda\ and the specific star formation
rate (SFR/M$_\ast$) provides stringent constraints on the recent
star formation histories of the {\em central region} of galaxies 
in the stellar mass range $10^8-10^{10}$M$_\odot$. In particular, this
methodology is powerful in separating out the galaxies with ongoing or
past starbursts from those with a continuous star formation history. 
By applying this new methodology to the spatially resolved spectroscopy
for a large sample from MaNGA, we may well expect to better quantify
the fractions of the mass produced by ongoing and recent bursts, as
well as their radial variation across the whole galaxy, thus leading to
more reliable constraints on the maps and radial profiles of the recent 
star formation history. 

Previous studies have proposed a variety of galaxy properties as
indicators of the termination of star formation, including both color
indices (e.g. \nuvr) and structural parameters like the central
surface mass density and stellar velocity dispersion
\citep[e.g.][]{Fang-13}, the effective radius \re\ and the ratio of
stellar mass relative to \re, \re$^2$, or \re$^{1.5}$
\citep[e.g.][]{Omand-Balogh-Poggianti-14}, and Sersic profile index
$n$ \citep[e.g.][]{Bell-08}. Since different indicators may trace
different physical processes, it is important to ask which galaxy
property is most closely linked to the end of star formation.  Again,
once larger samples of galaxies become available, it will be very
interesting to correlate the kind of spectral diagnostic gradient
study presented here with these other parameters, to see which is the
trigger, and even how the trigger varies between galaxies of different
types or in different environments.  

Despite these limitations, it is heartening to see how much  can be
learned about the cessation of star formation from even a sample of 12
galaxies, thanks to the resolved spectroscopic observations  that
P-MaNGA IFUs have provided.  The full MaNGA survey began on 2014 July
1, offering new opportunities to extend these kinds of studies, as
observations of its 10,000 galaxies continue.

\acknowledgments

CL acknowledges the support of  the 100  Talents Program of  Chinese
Academy    of    Sciences    (CAS). This work is supported by NSFC
(11173045,11233005,11325314,11320101002), by the Strategic Priority
Research Program ``The Emergence of Cosmological Structures'' of CAS
(Grant No. XDB09000000), and by World Premier International Research
Center Initiative (WPI Initiative), MEXT, Japan. We have made use of
data from HyperLeda for our work.  
Funding for SDSS-III and SDSS-IV has been provided by the Alfred
P.~Sloan Foundation and Participating Institutions.  Additional
funding for SDSS-III comes from the National Science Foundation and
the U.S.~Department of Energy Office of Science.  Further information
about both projects is available at {\tt www.sdss3.org}.

SDSS is managed by the Astrophysical Research Consortium for the
Participating Institutions in both collaborations.  In SDSS-III these
include the University of Arizona, the Brazilian Participation Group,
Brookhaven National Laboratory, Carnegie Mellon University, University
of Florida, the French Participation Group, the German Participation
Group, Harvard University, the Instituto de Astrofisica de Canarias,
the Michigan State/Notre Dame/JINA Participation Group, Johns Hopkins
University, Lawrence Berkeley National Laboratory, Max Planck
Institute for Astrophysics, Max Planck Institute for Extraterrestrial
Physics, New Mexico State University, New York University, Ohio State
University, Pennsylvania State University, University of Portsmouth,
Princeton University, the Spanish Participation Group, University of
Tokyo, University of Utah, Vanderbilt University, University of
Virginia, University of Washington, and Yale University.

The Participating Institutions in SDSS-IV are Carnegie Mellon
University, Colorado University, Boulder, Harvard-Smithsonian Center
for Astrophysics Participation Group, Johns Hopkins University, Kavli
Institute for the Physics and Mathematics of the Universe,
Max-Planck-Institut fuer Astrophysik (MPA Garching),
Max-Planck-Institut fuer Extraterrestrische Physik (MPE),
Max-Planck-Institut fuer Astronomie (MPIA Heidelberg), National
Astronomical Observatory of China, New Mexico State University, New
York University, The Ohio State University, Penn State University,
Shanghai Astronomical Observatory, United Kingdom Participation Group,
University of Portsmouth, University of Utah, University of Wisconsin,
and Yale University.


\label{lastpage}

\begin{thebibliography}{129}
\expandafter\ifx\csname natexlab\endcsname\relax\def\natexlab#1{#1}\fi

\bibitem[{{Abadi} {et~al.}(1999){Abadi}, {Moore}, \&
  {Bower}}]{Abadi-Moore-Bower-99}
{Abadi}, M.~G., {Moore}, B., \& {Bower}, R.~G. 1999, \mnras, 308, 947

\bibitem[{{Abraham} {et~al.}(1999){Abraham}, {Ellis}, {Fabian}, {Tanvir}, \&
  {Glazebrook}}]{Abraham-99}
{Abraham}, R.~G., {Ellis}, R.~S., {Fabian}, A.~C., {Tanvir}, N.~R., \&
  {Glazebrook}, K. 1999, \mnras, 303, 641

\bibitem[{{Azzollini} {et~al.}(2009){Azzollini}, {Beckman}, \&
  {Trujillo}}]{Azzollini-Beckman-Trujillo-09}
{Azzollini}, R., {Beckman}, J.~E., \& {Trujillo}, I. 2009, \aap, 501, 119

\bibitem[{{Bacon} {et~al.}(2001){Bacon}, {Copin}, {Monnet}, {Miller},
  {Allington-Smith}, {Bureau}, {Carollo}, {Davies}, \& {et al.}}]{Bacon-01}
{Bacon}, R., {Copin}, Y., {Monnet}, G., {et~al.} 2001, \mnras, 326, 23

\bibitem[{{Baldry} {et~al.}(2004){Baldry}, {Glazebrook}, {Brinkmann},
  {Ivezi{\'c}}, {Lupton}, {Nichol}, \& {Szalay}}]{Baldry-04}
{Baldry}, I.~K., {Glazebrook}, K., {Brinkmann}, J., {et~al.} 2004, \apj, 600,
  681

\bibitem[{{Baldwin} {et~al.}(1981){Baldwin}, {Phillips}, \&
  {Terlevich}}]{Baldwin-Phillips-Terlevich-81}
{Baldwin}, J.~A., {Phillips}, M.~M., \& {Terlevich}, R. 1981, \pasp, 93, 5

\bibitem[{{Balogh} {et~al.}(1999){Balogh}, {Morris}, {Yee}, {Carlberg}, \&
  {Ellingson}}]{Balogh-99}
{Balogh}, M.~L., {Morris}, S.~L., {Yee}, H.~K.~C., {Carlberg}, R.~G., \&
  {Ellingson}, E. 1999, \apj, 527, 54

\bibitem[{{Bamford} {et~al.}(2009){Bamford}, {Nichol}, {Baldry}, {Land},
  {Lintott}, {Schawinski}, {Slosar}, {Szalay}, \& {et al.}}]{Bamford-09}
{Bamford}, S.~P., {Nichol}, R.~C., {Baldry}, I.~K., {et~al.} 2009, \mnras, 393,
  1324

\bibitem[{{Becker} {et~al.}(1995){Becker}, {White}, \&
  {Helfand}}]{Becker-White-Helfand-95}
{Becker}, R.~H., {White}, R.~L., \& {Helfand}, D.~J. 1995, \apj, 450, 559

\bibitem[{{Belfiore} {et~al.}(2015){Belfiore}, {Maiolino}, {Bundy}, {Thomas},
  {Maraston}, {Wilkinson}, {S{\'a}nchez}, {Bershady}, {Blanc}, {Bothwell},
  {Cales}, {Coccato}, {Drory}, {Emsellem}, {Fu}, {Gelfand}, {Law}, {Masters},
  {Parejko}, {Tremonti}, {Wake}, {Weijmans}, {Yan}, {Xiao}, {Zhang}, {Zheng},
  {Bizyaev}, {Kinemuchi}, {Oravetz}, \& {Simmons}}]{Belfiore-15}
{Belfiore}, F., {Maiolino}, R., {Bundy}, K., {et~al.} 2015, \mnras\ submitted,
arXiv:1410.7781

\bibitem[{{Bell}(2008)}]{Bell-08}
{Bell}, E.~F. 2008, \apj, 682, 355

\bibitem[{{Bell} \& {de Jong}(2000)}]{Bell-deJong-00}
{Bell}, E.~F., \& {de Jong}, R.~S. 2000, \mnras, 312, 497

\bibitem[{{Bell} {et~al.}(2004){Bell}, {McIntosh}, {Barden}, {Wolf},
  {Caldwell}, {Rix}, {Beckwith}, {Borch}, \& {et al.}}]{Bell-04b}
{Bell}, E.~F., {McIntosh}, D.~H., {Barden}, M., {et~al.} 2004, \apjl, 600, L11

\bibitem[{{Bell} {et~al.}(2012){Bell}, {van der Wel}, {Papovich}, {Kocevski},
  {Lotz}, {McIntosh}, {Kartaltepe}, {Faber}, \& {et al.}}]{Bell-12}
{Bell}, E.~F., {van der Wel}, A., {Papovich}, C., {et~al.} 2012, \apj, 753, 167

\bibitem[{{Bershady} {et~al.}(2010){Bershady}, {Verheijen}, {Swaters},
  {Andersen}, {Westfall}, \& {Martinsson}}]{Bershady-10}
{Bershady}, M.~A., {Verheijen}, M.~A.~W., {Swaters}, R.~A., {et~al.} 2010,
  \apj, 716, 198

\bibitem[{{Birnboim} \& {Dekel}(2003)}]{Birnboim-Dekel-03}
{Birnboim}, Y., \& {Dekel}, A. 2003, \mnras, 345, 349

\bibitem[{{Blanton} {et~al.}(2005{\natexlab{a}}){Blanton}, {Eisenstein},
  {Hogg}, {Schlegel}, \& {Brinkmann}}]{Blanton-05b}
{Blanton}, M.~R., {Eisenstein}, D., {Hogg}, D.~W., {Schlegel}, D.~J., \&
  {Brinkmann}, J. 2005{\natexlab{a}}, \apj, 629, 143

\bibitem[{{Blanton} {et~al.}(2011){Blanton}, {Kazin}, {Muna}, {Weaver}, \&
  {Price-Whelan}}]{Blanton-11}
{Blanton}, M.~R., {Kazin}, E., {Muna}, D., {Weaver}, B.~A., \& {Price-Whelan},
  A. 2011, \aj, 142, 31

\bibitem[{{Blanton} {et~al.}(2005{\natexlab{b}}){Blanton}, {Lupton},
  {Schlegel}, {Strauss}, {Brinkmann}, {Fukugita}, \& {Loveday}}]{Blanton-05a}
{Blanton}, M.~R., {Lupton}, R.~H., {Schlegel}, D.~J., {et~al.}
  2005{\natexlab{b}}, \apj, 631, 208

\bibitem[{{Blanton} \& {Roweis}(2007)}]{Blanton-Roweis-07}
{Blanton}, M.~R., \& {Roweis}, S. 2007, \aj, 133, 734

\bibitem[{{Blanton} {et~al.}(2003){Blanton}, {Hogg}, {Bahcall}, {Baldry},
  {Brinkmann}, {Csabai}, {Eisenstein}, {Fukugita}, \& {et al.}}]{Blanton-03a}
{Blanton}, M.~R., {Hogg}, D.~W., {Bahcall}, N.~A., {et~al.} 2003, \apj, 594,
  186

\bibitem[{{Blumenthal} {et~al.}(1984){Blumenthal}, {Faber}, {Primack}, \&
  {Rees}}]{Blumenthal-84}
{Blumenthal}, G.~R., {Faber}, S.~M., {Primack}, J.~R., \& {Rees}, M.~J. 1984,
  \nat, 311, 517

\bibitem[{{Brammer} {et~al.}(2009){Brammer}, {Whitaker}, {van Dokkum},
  {Marchesini}, {Labb{\'e}}, {Franx}, {Kriek}, {Quadri}, \& {et
  al.}}]{Brammer-09}
{Brammer}, G.~B., {Whitaker}, K.~E., {van Dokkum}, P.~G., {et~al.} 2009, \apjl,
  706, L173

\bibitem[{{Brinchmann} {et~al.}(2004){Brinchmann}, {Charlot}, {White},
  {Tremonti}, {Kauffmann}, {Heckman}, \& {Brinkmann}}]{Brinchmann-04}
{Brinchmann}, J., {Charlot}, S., {White}, S.~D.~M., {et~al.} 2004, \mnras, 351,
  1151

\bibitem[{{Brook} {et~al.}(2006){Brook}, {Kawata}, {Martel}, {Gibson}, \&
  {Bailin}}]{Brook-06}
{Brook}, C.~B., {Kawata}, D., {Martel}, H., {Gibson}, B.~K., \& {Bailin}, J.
  2006, \apj, 639, 126

\bibitem[{{Bruzual} \& {Charlot}(2003)}]{Bruzual-Charlot-03}
{Bruzual}, G., \& {Charlot}, S. 2003, \mnras, 344, 1000

\bibitem[{{Bundy} {et~al.}(2006){Bundy}, {Ellis}, {Conselice}, {Taylor},
  {Cooper}, {Willmer}, {Weiner}, {Coil}, \& {et al.}}]{Bundy-06}
{Bundy}, K., {Ellis}, R.~S., {Conselice}, C.~J., {et~al.} 2006, \apj, 651, 120

\bibitem[{{Bundy} {et~al.}(2010){Bundy}, {Scarlata}, {Carollo}, {Ellis},
  {Drory}, {Hopkins}, {Salvato}, {Leauthaud}, \& {et al.}}]{Bundy-10}
{Bundy}, K., {Scarlata}, C., {Carollo}, C.~M., {et~al.} 2010, \apj, 719, 1969

\bibitem[{{Bundy} {et~al.}(2015){Bundy}, {Bershady}, {Law}, {Yan}, {Drory},
  {MacDonald}, {Wake}, {Cherinka}, {S{\'a}nchez-Gallego}, {Weijmans}, {Thomas},
  {Tremonti}, {Masters}, {Coccato}, {Diamond-Stanic}, {Arag{\'o}n-Salamanca},
  {Avila-Reese}, {Badenes}, {Falc{\'o}n-Barroso}, {Belfiore}, {Bizyaev},
  {Blanc}, {Bland-Hawthorn}, {Blanton}, {Brownstein}, {Byler}, {Cappellari},
  {Conroy}, {Dutton}, {Emsellem}, {Etherington}, {Frinchaboy}, {Fu}, {Gunn},
  {Harding}, {Johnston}, {Kauffmann}, {Kinemuchi}, {Klaene}, {Knapen},
  {Leauthaud}, {Li}, {Lin}, {Maiolino}, {Malanushenko}, {Malanushenko}, {Mao},
  {Maraston}, {McDermid}, {Merrifield}, {Nichol}, {Oravetz}, {Pan}, {Parejko},
  {Sanchez}, {Schlegel}, {Simmons}, {Steele}, {Steinmetz}, {Thanjavur},
  {Thompson}, {Tinker}, {van den Bosch}, {Westfall}, {Wilkinson}, {Wright},
  {Xiao}, \& {Zhang}}]{Bundy-15}
{Bundy}, K., {Bershady}, M.~A., {Law}, D.~R., {et~al.} 2015, \apj, 798, 7

\bibitem[{{Cappellari} \& {Copin}(2003)}]{Cappellari-Copin-03}
{Cappellari}, M., \& {Copin}, Y. 2003, \mnras, 342, 345

\bibitem[{{Cappellari} {et~al.}(2011){Cappellari}, {Emsellem}, {Krajnovi{\'c}},
  {McDermid}, {Serra}, {Alatalo}, {Blitz}, {Bois}, \& {et
  al.}}]{Cappellari-11b}
{Cappellari}, M., {Emsellem}, E., {Krajnovi{\'c}}, D., {et~al.} 2011, \mnras,
  416, 1680

\bibitem[{{Catinella} {et~al.}(2010){Catinella}, {Schiminovich}, {Kauffmann},
  {Fabello}, {Wang}, {Hummels}, {Lemonias}, {Moran}, \& {et
  al.}}]{Catinella-10}
{Catinella}, B., {Schiminovich}, D., {Kauffmann}, G., {et~al.} 2010, \mnras,
  403, 683

\bibitem[{{Cattaneo} {et~al.}(2006){Cattaneo}, {Dekel}, {Devriendt},
  {Guiderdoni}, \& {Blaizot}}]{Cattaneo-06}
{Cattaneo}, A., {Dekel}, A., {Devriendt}, J., {Guiderdoni}, B., \& {Blaizot},
  J. 2006, \mnras, 370, 1651

\bibitem[{{Chabrier}(2003)}]{Chabrier-03}
{Chabrier}, G. 2003, \pasp, 115, 763

\bibitem[{{Cheung} {et~al.}(2012){Cheung}, {Faber}, {Koo}, {Dutton}, {Simard},
  {McGrath}, {Huang}, {Bell}, \& {et al.}}]{Cheung-12}
{Cheung}, E., {Faber}, S.~M., {Koo}, D.~C., {et~al.} 2012, \apj, 760, 131

\bibitem[{{Chung} {et~al.}(2009){Chung}, {van Gorkom}, {Kenney}, {Crowl}, \&
  {Vollmer}}]{Chung-09}
{Chung}, A., {van Gorkom}, J.~H., {Kenney}, J.~D.~P., {Crowl}, H., \&
  {Vollmer}, B. 2009, \aj, 138, 1741

\bibitem[{{Cid Fernandes} {et~al.}(2004){Cid Fernandes}, {Gu}, {Melnick},
  {Terlevich}, {Terlevich}, {Kunth}, {Rodrigues Lacerda}, \&
  {Joguet}}]{CidFernandes-04b}
{Cid Fernandes}, R., {Gu}, Q., {Melnick}, J., {et~al.} 2004, \mnras, 355, 273

\bibitem[{{Cid Fernandes} {et~al.}(2010){Cid Fernandes}, {Stasi{\'n}ska},
  {Schlickmann}, {Mateus}, {Vale Asari}, {Schoenell}, \&
  {Sodr{\'e}}}]{CidFernandes-10}
{Cid Fernandes}, R., {Stasi{\'n}ska}, G., {Schlickmann}, M.~S., {et~al.} 2010,
  \mnras, 403, 1036

\bibitem[{{Cirasuolo} {et~al.}(2007){Cirasuolo}, {McLure}, {Dunlop}, {Almaini},
  {Foucaud}, {Smail}, {Sekiguchi}, {Simpson}, \& {et al.}}]{Cirasuolo-07}
{Cirasuolo}, M., {McLure}, R.~J., {Dunlop}, J.~S., {et~al.} 2007, \mnras, 380,
  585

\bibitem[{{Colless} {et~al.}(2001){Colless}, {Dalton}, {Maddox}, {Sutherland},
  {Norberg}, {Cole}, {Bland-Hawthorn}, {Bridges}, \& {et al.}}]{Colless-01}
{Colless}, M., {Dalton}, G., {Maddox}, S., {et~al.} 2001, \mnras, 328, 1039

\bibitem[{{Condon} {et~al.}(1998){Condon}, {Cotton}, {Greisen}, {Yin},
  {Perley}, {Taylor}, \& {Broderick}}]{Condon-98}
{Condon}, J.~J., {Cotton}, W.~D., {Greisen}, E.~W., {et~al.} 1998, \aj, 115,
  1693

\bibitem[{{Cooper} {et~al.}(2008){Cooper}, {Newman}, {Weiner}, {Yan},
  {Willmer}, {Bundy}, {Coil}, {Conselice}, \& {et al.}}]{Cooper-08}
{Cooper}, M.~C., {Newman}, J.~A., {Weiner}, B.~J., {et~al.} 2008, \mnras, 383,
  1058

\bibitem[{{Cowie} \& {Barger}(2008)}]{Cowie-Barger-08}
{Cowie}, L.~L., \& {Barger}, A.~J. 2008, \apj, 686, 72

\bibitem[{{Croom} {et~al.}(2012)}]{Croom2012}
{Croom}, S.~M., {et~al.} 2012, \mnras, 421, 872

\bibitem[{{de Jong}(1996)}]{deJong-96}
{de Jong}, R.~S. 1996, \aap, 313, 377

\bibitem[{{Dekel} \& {Birnboim}(2006)}]{Dekel-Birnboim-06}
{Dekel}, A., \& {Birnboim}, Y. 2006, \mnras, 368, 2

\bibitem[{{Di Matteo} {et~al.}(2005){Di Matteo}, {Springel}, \&
  {Hernquist}}]{DiMatteo-Springel-Hernquist-05}
{Di Matteo}, T., {Springel}, V., \& {Hernquist}, L. 2005, \nat, 433, 604

\bibitem[Drory et al.(2015)]{Drory-15} Drory, N., MacDonald, N., 
Bershady, M.~A., et al.\ 2015, \aj, 149, 77 

\bibitem[{{Eisenstein} {et~al.}(2011){Eisenstein}, {Weinberg}, {Agol},
  {Aihara}, {Allende Prieto}, {Anderson}, {Arns}, {Aubourg}, \& {et
  al.}}]{Eisenstein-11}
{Eisenstein}, D.~J., {Weinberg}, D.~H., {Agol}, E., {et~al.} 2011, \aj, 142, 72

\bibitem[{{Fabello} {et~al.}(2011){Fabello}, {Catinella}, {Giovanelli},
  {Kauffmann}, {Haynes}, {Heckman}, \& {Schiminovich}}]{Fabello-11}
{Fabello}, S., {Catinella}, B., {Giovanelli}, R., {et~al.} 2011, \mnras, 411,
  993

\bibitem[{{Fabello} {et~al.}(2012){Fabello}, {Kauffmann}, {Catinella}, {Li},
  {Giovanelli}, \& {Haynes}}]{Fabello-12}
{Fabello}, S., {Kauffmann}, G., {Catinella}, B., {et~al.} 2012, \mnras, 427,
  2841

\bibitem[{{Faber} {et~al.}(2007){Faber}, {Willmer}, {Wolf}, {Koo}, {Weiner},
  {Newman}, {Im}, {Coil}, \& {et al.}}]{Faber-07}
{Faber}, S.~M., {Willmer}, C.~N.~A., {Wolf}, C., {et~al.} 2007, \apj, 665, 265

\bibitem[{{Fang} {et~al.}(2013){Fang}, {Faber}, {Koo}, \& {Dekel}}]{Fang-13}
{Fang}, J.~J., {Faber}, S.~M., {Koo}, D.~C., \& {Dekel}, A. 2013, \apj, 776, 63

\bibitem[{{Fang} {et~al.}(2012){Fang}, {Faber}, {Salim}, {Graves}, \&
  {Rich}}]{Fang-12}
{Fang}, J.~J., {Faber}, S.~M., {Salim}, S., {Graves}, G.~J., \& {Rich}, R.~M.
  2012, \apj, 761, 23

\bibitem[{{Franx} {et~al.}(2008){Franx}, {van Dokkum}, {Schreiber}, {Wuyts},
  {Labb{\'e}}, \& {Toft}}]{Franx-08}
{Franx}, M., {van Dokkum}, P.~G., {Schreiber}, N.~M.~F., {et~al.} 2008, \apj,
  688, 770

\bibitem[{{Giovanelli} {et~al.}(2005){Giovanelli}, {Haynes}, {Kent},
  {Perillat}, {Saintonge}, {Brosch}, {Catinella}, {Hoffman}, \& {et
  al.}}]{Giovanelli-05}
{Giovanelli}, R., {Haynes}, M.~P., {Kent}, B.~R., {et~al.} 2005, \aj, 130, 2598

\bibitem[{{Gon{\c c}alves} {et~al.}(2012){Gon{\c c}alves}, {Martin},
  {Men{\'e}ndez-Delmestre}, {Wyder}, \& {Koekemoer}}]{Goncalves-12}
{Gon{\c c}alves}, T.~S., {Martin}, D.~C., {Men{\'e}ndez-Delmestre}, K.,
  {Wyder}, T.~K., \& {Koekemoer}, A. 2012, \apj, 759, 67

\bibitem[{{Gonz{\'a}lez Delgado} {et~al.}(2014){Gonz{\'a}lez Delgado},
  {P{\'e}rez}, {Cid Fernandes}, {Garc{\'{\i}}a-Benito}, {de Amorim},
  {S{\'a}nchez}, {Husemann}, {Cortijo-Ferrero}, \& {et
  al.}}]{GonzalezDelgado-14a}
{Gonz{\'a}lez Delgado}, R.~M., {P{\'e}rez}, E., {Cid Fernandes}, R., {et~al.}
  2014, \aap, 562, A47

\bibitem[{{Gonzalez-Perez} {et~al.}(2011){Gonzalez-Perez}, {Castander}, \&
  {Kauffmann}}]{Gonzalez-Perez-Castander-Kauffmann-11}
{Gonzalez-Perez}, V., {Castander}, F.~J., \& {Kauffmann}, G. 2011, \mnras, 411,
  1151

\bibitem[{{Gunn} \& {Gott}(1972)}]{Gunn-Gott-72}
{Gunn}, J.~E., \& {Gott}, III, J.~R. 1972, \apj, 176, 1

\bibitem[{{Hopkins} {et~al.}(2006){Hopkins}, {Hernquist}, {Cox}, {Di Matteo},
  {Robertson}, \& {Springel}}]{Hopkins-06}
{Hopkins}, P.~F., {Hernquist}, L., {Cox}, T.~J., {et~al.} 2006, \apjs, 163, 1

\bibitem[{{Huang} {et~al.}(2013{\natexlab{a}}){Huang}, {Faber}, {Willmer},
  {Rigopoulou}, {Koo}, {Newman}, {Shu}, {Ashby}, \& {et al.}}]{Huang-13a}
{Huang}, J.-S., {Faber}, S.~M., {Willmer}, C.~N.~A., {et~al.}
  2013{\natexlab{a}}, \apj, 766, 21

\bibitem[{{Huang} {et~al.}(2013{\natexlab{b}}){Huang}, {Kauffmann}, {Chen},
  {Moran}, {Heckman}, {Dav{\'e}}, \& {Johansson}}]{Huang-13b}
{Huang}, M.-L., {Kauffmann}, G., {Chen}, Y.-M., {et~al.} 2013{\natexlab{b}},
  \mnras, 431, 2622

\bibitem[{{Huchtmeier} \& {Richter}(1989)}]{Huchtmeier-Richter-89}
{Huchtmeier}, W.~K., \& {Richter}, O.-G. 1989, \aap, 210, 1

\bibitem[{{Hummer} \& {Storey}(1987)}]{Hummer-Storey-87}
{Hummer}, D.~G., \& {Storey}, P.~J. 1987, \mnras, 224, 801

\bibitem[{{Hunter} \& {Elmegreen}(2004)}]{Hunter-Elmegreen-04}
{Hunter}, D.~A., \& {Elmegreen}, B.~G. 2004, \aj, 128, 2170

\bibitem[{{Kauffmann}(2014)}]{Kauffmann-14}
{Kauffmann}, G. 2014, ArXiv e-prints

\bibitem[{{Kauffmann}(2015)}]{Kauffmann-15}
---. 2015, ArXiv e-prints

\bibitem[{{Kauffmann} {et~al.}(2006){Kauffmann}, {Heckman}, {De Lucia},
  {Brinchmann}, {Charlot}, {Tremonti}, {White}, \& {Brinkmann}}]{Kauffmann-06}
{Kauffmann}, G., {Heckman}, T.~M., {De Lucia}, G., {et~al.} 2006, \mnras, 367,
  1394

\bibitem[{{Kauffmann} {et~al.}(2003{\natexlab{a}}){Kauffmann}, {Heckman},
  {White}, {Charlot}, {Tremonti}, {Peng}, {Seibert}, {Brinkmann}, \& {et
  al.}}]{Kauffmann-03b}
{Kauffmann}, G., {Heckman}, T.~M., {White}, S.~D.~M., {et~al.}
  2003{\natexlab{a}}, \mnras, 341, 54

\bibitem[{{Kauffmann} {et~al.}(2003{\natexlab{b}}){Kauffmann}, {Heckman},
  {Tremonti}, {Brinchmann}, {Charlot}, {White}, {Ridgway}, {Brinkmann}, \& {et
  al.}}]{Kauffmann-03c}
{Kauffmann}, G., {Heckman}, T.~M., {Tremonti}, C., {et~al.} 2003{\natexlab{b}},
  \mnras, 346, 1055

\bibitem[{{Kauffmann} {et~al.}(2007){Kauffmann}, {Heckman}, {Budav{\'a}ri},
  {Charlot}, {Hoopes}, {Martin}, {Seibert}, {Barlow}, \& {et
  al.}}]{Kauffmann-07}
{Kauffmann}, G., {Heckman}, T.~M., {Budav{\'a}ri}, T., {et~al.} 2007, \apjs,
  173, 357

\bibitem[{{Kere{\v s}} {et~al.}(2005){Kere{\v s}}, {Katz}, {Weinberg}, \&
  {Dav{\'e}}}]{Keres-05}
{Kere{\v s}}, D., {Katz}, N., {Weinberg}, D.~H., \& {Dav{\'e}}, R. 2005,
  \mnras, 363, 2

\bibitem[{{Kewley} {et~al.}(2006){Kewley}, {Groves}, {Kauffmann}, \&
  {Heckman}}]{Kewley-06}
{Kewley}, L.~J., {Groves}, B., {Kauffmann}, G., \& {Heckman}, T. 2006, \mnras,
  372, 961

\bibitem[{{Law} {et~al.}(2015)}]{Law-15}
{Law}, D.~R., {Yan}, R., {Bershady}, M.~A., {et~al.} 2015, \aj\ submitted

\bibitem[{{Li} {et~al.}(2012{\natexlab{a}}){Li}, {Jing}, {Mao}, {Han}, {Peng},
  {Yang}, {Mo}, \& {van den Bosch}}]{Li-12a}
{Li}, C., {Jing}, Y.~P., {Mao}, S., {et~al.} 2012{\natexlab{a}}, \apj, 758, 50

\bibitem[{{Li} {et~al.}(2012{\natexlab{b}}){Li}, {Kauffmann}, {Fu}, {Wang},
  {Catinella}, {Fabello}, {Schiminovich}, \& {Zhang}}]{Li-12c}
{Li}, C., {Kauffmann}, G., {Fu}, J., {et~al.} 2012{\natexlab{b}}, \mnras, 424,
  1471

\bibitem[{{Li} {et~al.}(2005){Li}, {Wang}, {Zhou}, {Dong}, \& {Cheng}}]{Li-05}
{Li}, C., {Wang}, T.-G., {Zhou}, H.-Y., {Dong}, X.-B., \& {Cheng}, F.-Z. 2005,
  \aj, 129, 669

\bibitem[{{Lin} {et~al.}(2014){Lin}, {Jian}, {Foucaud}, {Norberg}, {Bower},
  {Cole}, {Arnalte-Mur}, {Chen}, {Coupon}, {Hsieh}, {Heinis}, {Phleps}, {Chen},
  {Lee}, {Burgett}, {Chambers}, {Denneau}, {Draper}, {Flewelling}, {Hodapp},
  {Huber}, {Kaiser}, {Kudritzki}, {Magnier}, {Metcalfe}, {Price}, {Tonry},
  {Wainscoat}, \& {Waters}}]{Lin-14}
{Lin}, L., {Jian}, H.-Y., {Foucaud}, S., {et~al.} 2014, \apj, 782, 33

\bibitem[{{Martig} {et~al.}(2009){Martig}, {Bournaud}, {Teyssier}, \&
  {Dekel}}]{Martig-09}
{Martig}, M., {Bournaud}, F., {Teyssier}, R., \& {Dekel}, A. 2009, \apj, 707,
  250

\bibitem[{{Martin} {et~al.}(2005){Martin}, {Fanson}, {Schiminovich},
  {Morrissey}, {Friedman}, {Barlow}, {Conrow}, {Grange}, \& {et
  al.}}]{Martin-05}
{Martin}, D.~C., {Fanson}, J., {Schiminovich}, D., {et~al.} 2005, \apjl, 619,
  L1

\bibitem[{{Martin} {et~al.}(2007){Martin}, {Wyder}, {Schiminovich}, {Barlow},
  {Forster}, {Friedman}, {Morrissey}, {Neff}, \& {et al.}}]{Martin-07}
{Martin}, D.~C., {Wyder}, T.~K., {Schiminovich}, D., {et~al.} 2007, \apjs, 173,
  342

\bibitem[{{Masters} {et~al.}(2010){Masters}, {Mosleh}, {Romer}, {Nichol},
  {Bamford}, {Schawinski}, {Lintott}, {Andreescu}, \& {et al.}}]{Masters-10a}
{Masters}, K.~L., {Mosleh}, M., {Romer}, A.~K., {et~al.} 2010, \mnras, 405, 783

\bibitem[{{Mendel} {et~al.}(2013){Mendel}, {Simard}, {Ellison}, \&
  {Patton}}]{Mendel-13}
{Mendel}, J.~T., {Simard}, L., {Ellison}, S.~L., \& {Patton}, D.~R. 2013,
  \mnras, 429, 2212

\bibitem[{{Mendez} {et~al.}(2011){Mendez}, {Coil}, {Lotz}, {Salim},
  {Moustakas}, \& {Simard}}]{Mendez-11}
{Mendez}, A.~J., {Coil}, A.~L., {Lotz}, J., {et~al.} 2011, \apj, 736, 110

\bibitem[{{Merluzzi} {et~al.}(2013){Merluzzi}, {Busarello}, {Dopita}, {Haines},
  {Steinhauser}, {Mercurio}, {Rifatto}, {Smith}, \& {et al.}}]{Merluzzi-13}
{Merluzzi}, P., {Busarello}, G., {Dopita}, M.~A., {et~al.} 2013, \mnras, 429,
  1747

\bibitem[{{Mo} {et~al.}(1998){Mo}, {Mao}, \& {White}}]{Mo-Mao-White-98}
{Mo}, H.~J., {Mao}, S., \& {White}, S.~D.~M. 1998, \mnras, 295, 319

\bibitem[{{Moore} {et~al.}(1996){Moore}, {Katz}, {Lake}, {Dressler}, \&
  {Oemler}}]{Moore-96}
{Moore}, B., {Katz}, N., {Lake}, G., {Dressler}, A., \& {Oemler}, A. 1996,
  \nat, 379, 613

\bibitem[{{Moran} {et~al.}(2012){Moran}, {Heckman}, {Kauffmann}, {Dav{\'e}},
  {Catinella}, {Brinchmann}, {Wang}, {Schiminovich}, \& {et al.}}]{Moran-12}
{Moran}, S.~M., {Heckman}, T.~M., {Kauffmann}, G., {et~al.} 2012, \apj, 745, 66

\bibitem[{{Mu{\~n}oz-Mateos} {et~al.}(2007){Mu{\~n}oz-Mateos}, {Gil de Paz},
  {Boissier}, {Zamorano}, {Jarrett}, {Gallego}, \& {Madore}}]{Munoz-Mateos-07}
{Mu{\~n}oz-Mateos}, J.~C., {Gil de Paz}, A., {Boissier}, S., {et~al.} 2007,
  \apj, 658, 1006

\bibitem[{{Muzzin} {et~al.}(2012){Muzzin}, {Wilson}, {Yee}, {Gilbank},
  {Hoekstra}, {Demarco}, {Balogh}, {van Dokkum}, \& {et al.}}]{Muzzin-12}
{Muzzin}, A., {Wilson}, G., {Yee}, H.~K.~C., {et~al.} 2012, \apj, 746, 188

\bibitem[{{Omand} {et~al.}(2014){Omand}, {Balogh}, \&
  {Poggianti}}]{Omand-Balogh-Poggianti-14}
{Omand}, C.~M.~B., {Balogh}, M.~L., \& {Poggianti}, B.~M. 2014, \mnras, 440,
  843

\bibitem[{{Papaderos} {et~al.}(2013){Papaderos}, {Gomes}, {V{\'{\i}}lchez},
  {Kehrig}, {Lehnert}, {Ziegler}, {S{\'a}nchez}, {Husemann}, \& {et
  al.}}]{Papaderos-13}
{Papaderos}, P., {Gomes}, J.~M., {V{\'{\i}}lchez}, J.~M., {et~al.} 2013, \aap,
  555, L1

\bibitem[{{Paturel} {et~al.}(2003){Paturel}, {Theureau}, {Bottinelli},
  {Gouguenheim}, {Coudreau-Durand}, {Hallet}, \& {Petit}}]{Paturel-03a}
{Paturel}, G., {Theureau}, G., {Bottinelli}, L., {et~al.} 2003, \aap, 412, 57

\bibitem[{{Peng} {et~al.}(2010){Peng}, {Lilly}, {Kova{\v c}}, {Bolzonella},
  {Pozzetti}, {Renzini}, {Zamorani}, {Ilbert}, \& {et al.}}]{Peng-10b}
{Peng}, Y.-J., {Lilly}, S.~J., {Kova{\v c}}, K., {et~al.} 2010, \apj, 721, 193

\bibitem[{{P{\'e}rez} {et~al.}(2013){P{\'e}rez}, {Cid Fernandes}, {Gonz{\'a}lez
  Delgado}, {Garc{\'{\i}}a-Benito}, {S{\'a}nchez}, {Husemann}, {Mast},
  {Rod{\'o}n}, \& {et al.}}]{Perez-13}
{P{\'e}rez}, E., {Cid Fernandes}, R., {Gonz{\'a}lez Delgado}, R.~M., {et~al.}
  2013, \apjl, 764, L1

\bibitem[{{Rees} \& {Ostriker}(1977)}]{Rees-Ostriker-77}
{Rees}, M.~J., \& {Ostriker}, J.~P. 1977, \mnras, 179, 541

\bibitem[{{Roche} {et~al.}(2010){Roche}, {Bernardi}, \&
  {Hyde}}]{Roche-Bernardi-Hyde-10}
{Roche}, N., {Bernardi}, M., \& {Hyde}, J. 2010, \mnras, 407, 1231

\bibitem[{{S{\'a}nchez} {et~al.}(2012{\natexlab{a}}){S{\'a}nchez}, {Kennicutt},
  {Gil de Paz}, {van de Ven}, {V{\'{\i}}lchez}, {Wisotzki}, {Walcher}, {Mast},
  \& {et al.}}]{Sanchez-12}
{S{\'a}nchez}, S.~F., {Kennicutt}, R.~C., {Gil de Paz}, A., {et~al.}
  2012{\natexlab{a}}, \aap, 538, A8

\bibitem[{{S{\'a}nchez} {et~al.}(2012{\natexlab{b}})}]{sanchez12}
{S{\'a}nchez}, S.~F., {et~al.} 2012{\natexlab{b}}, \aap, 538, A8

\bibitem[{{S{\'a}nchez} {et~al.}(2014){S{\'a}nchez}, {Rosales-Ortega},
  {Iglesias-P{\'a}ramo}, {Moll{\'a}}, {Barrera-Ballesteros}, {Marino},
  {P{\'e}rez}, {S{\'a}nchez-Blazquez}, \& {et al.}}]{Sanchez-14}
{S{\'a}nchez}, S.~F., {Rosales-Ortega}, F.~F., {Iglesias-P{\'a}ramo}, J.,
  {et~al.} 2014, \aap, 563, A49

\bibitem[{{S{\'a}nchez-Bl{\'a}zquez} {et~al.}(2014){S{\'a}nchez-Bl{\'a}zquez},
  {Rosales-Ortega}, {M{\'e}ndez-Abreu}, {P{\'e}rez}, {S{\'a}nchez}, {Zibetti},
  {Aguerri}, {Bland-Hawthorn}, {Catal{\'a}n-Torrecilla}, {Cid Fernandes}, {de
  Amorim}, {de Lorenzo-Caceres}, {Falc{\'o}n-Barroso}, {Galazzi},
  {Garc{\'{\i}}a Benito}, {Gil de Paz}, {Gonz{\'a}lez Delgado}, {Husemann},
  {Iglesias-P{\'a}ramo}, {Jungwiert}, {Marino}, {M{\'a}rquez}, {Mast},
  {Mendoza}, {Moll{\'a}}, {Papaderos}, {Ruiz-Lara}, {van de Ven}, {Walcher}, \&
  {Wisotzki}}]{Sanchez-Blazquez-14}
{S{\'a}nchez-Bl{\'a}zquez}, P., {Rosales-Ortega}, F.~F., {M{\'e}ndez-Abreu},
  J., {et~al.} 2014, \aap, 570, A6

\bibitem[{{Sarzi} {et~al.}(2013){Sarzi}, {Alatalo}, {Blitz}, {Bois},
  {Bournaud}, {Bureau}, {Cappellari}, {Crocker}, \& {et al.}}]{Sarzi-13}
{Sarzi}, M., {Alatalo}, K., {Blitz}, L., {et~al.} 2013, \mnras, 432, 1845

\bibitem[{{Schawinski} {et~al.}(2007){Schawinski}, {Thomas}, {Sarzi},
  {Maraston}, {Kaviraj}, {Joo}, {Yi}, \& {Silk}}]{Schawinski-07b}
{Schawinski}, K., {Thomas}, D., {Sarzi}, M., {et~al.} 2007, \mnras, 382, 1415

\bibitem[{{Schawinski} {et~al.}(2014){Schawinski}, {Urry}, {Simmons},
  {Fortson}, {Kaviraj}, {Keel}, {Lintott}, {Masters}, \& {et
  al.}}]{Schawinski-14}
{Schawinski}, K., {Urry}, C.~M., {Simmons}, B.~D., {et~al.} 2014, \mnras, 440,
  889

\bibitem[{{Schiminovich} {et~al.}(2007){Schiminovich}, {Wyder}, {Martin},
  {Johnson}, {Salim}, {Seibert}, {Treyer}, {Budav{\'a}ri}, \& {et
  al.}}]{Schiminovich-07}
{Schiminovich}, D., {Wyder}, T.~K., {Martin}, D.~C., {et~al.} 2007, \apjs, 173,
  315

\bibitem[{{Sharp} {et~al.}(2015){Sharp}, {Allen}, {Fogarty}, {Croom},
  {Cortese}, {Green}, {Nielsen}, {Richards}, {Scott}, {Taylor}, {Barnes},
  {Bauer}, {Birchall}, {Bland-Hawthorn}, {Bloom}, {Brough}, {Bryant}, {Cecil},
  {Colless}, {Couch}, {Drinkwater}, {Driver}, {Foster}, {Goodwin},
  {Gunawardhana}, {Ho}, {Hampton}, {Hopkins}, {Jones}, {Konstantopoulos},
  {Lawrence}, {Leslie}, {Lewis}, {Liske}, {L{\'o}pez-S{\'a}nchez}, {Lorente},
  {McElroy}, {Medling}, {Mahajan}, {Mould}, {Parker}, {Pracy}, {Obreschkow},
  {Owers}, {Schaefer}, {Sweet}, {Thomas}, {Tonini}, \& {Walcher}}]{Sharp-15}
{Sharp}, R., {Allen}, J.~T., {Fogarty}, L.~M.~R., {et~al.} 2015, \mnras, 446,
  1551

\bibitem[{{Silk}(1977)}]{Silk-77}
{Silk}, J. 1977, \apj, 211, 638

\bibitem[{{Skrutskie} {et~al.}(2006){Skrutskie}, {Cutri}, {Stiening},
  {Weinberg}, {Schneider}, {Carpenter}, {Beichman}, {Capps}, \& {et
  al.}}]{Skrutskie-06}
{Skrutskie}, M.~F., {Cutri}, R.~M., {Stiening}, R., {et~al.} 2006, \aj, 131,
  1163

\bibitem[{{Strateva} {et~al.}(2001){Strateva}, {Ivezi{\'c}}, {Knapp},
  {Narayanan}, {Strauss}, {Gunn}, {Lupton}, {Schlegel}, \& {et
  al.}}]{Strateva-01}
{Strateva}, I., {Ivezi{\'c}}, {\v Z}., {Knapp}, G.~R., {et~al.} 2001, \aj, 122,
  1861

\bibitem[{{Suh} {et~al.}(2010){Suh}, {Jeong}, {Oh}, {Yi}, {Ferreras}, \&
  {Schawinski}}]{Suh-10}
{Suh}, H., {Jeong}, H., {Oh}, K., {et~al.} 2010, \apjs, 187, 374

\bibitem[{{Szomoru} {et~al.}(2012){Szomoru}, {Franx}, \& {van
  Dokkum}}]{Szomoru-12}
{Szomoru}, D., {Franx}, M., \& {van Dokkum}, P.~G. 2012, \apj, 749, 121

\bibitem[{{Szomoru} {et~al.}(2013){Szomoru}, {Franx}, {van Dokkum}, {Trenti},
  {Illingworth}, {Labb{\'e}}, \& {Oesch}}]{Szomoru-13}
{Szomoru}, D., {Franx}, M., {van Dokkum}, P.~G., {et~al.} 2013, \apj, 763, 73

\bibitem[{{Taylor}(2005)}]{Taylor-05}
{Taylor}, M.~B. 2005, in Astronomical Data Analysis Software and Systems XIV,
  ed. P.~{Shopbell}, M.~{Britton}, \& R.~{Ebert}, Vol. 347, 29

\bibitem[{{Thomas} \& {Davies}(2006)}]{Thomas-Davies-06}
{Thomas}, D., \& {Davies}, R.~L. 2006, \mnras, 366, 510

\bibitem[{{Thomas} {et~al.}(2004){Thomas}, {Maraston}, \& {Korn}}]{Thomas-04}
{Thomas}, D., {Maraston}, C., \& {Korn}, A. 2004, \mnras, 351, L19

\bibitem[{{Thomas} {et~al.}(2010){Thomas}, {Maraston}, {Schawinski}, {Sarzi},
  \& {Silk}}]{Thomas-10}
{Thomas}, D., {Maraston}, C., {Schawinski}, K., {Sarzi}, M., \& {Silk}, J.
  2010, \mnras, 404, 1775

\bibitem[{{Tifft}(1963)}]{Tifft-63}
{Tifft}, W.~G. 1963, \aj, 68, 302

\bibitem[{{Toomre} \& {Toomre}(1972)}]{Toomre-Toomre-72}
{Toomre}, A., \& {Toomre}, J. 1972, \apj, 178, 623

\bibitem[{{Tortora} {et~al.}(2010){Tortora}, {Napolitano}, {Cardone},
  {Capaccioli}, {Jetzer}, \& {Molinaro}}]{Tortora-10a}
{Tortora}, C., {Napolitano}, N.~R., {Cardone}, V.~F., {et~al.} 2010, \mnras,
  407, 144

\bibitem[{{Tortora} {et~al.}(2011){Tortora}, {Napolitano}, {Romanowsky},
  {Jetzer}, {Cardone}, \& {Capaccioli}}]{Tortora-11}
{Tortora}, C., {Napolitano}, N.~R., {Romanowsky}, A.~J., {et~al.} 2011, \mnras,
  418, 1557

\bibitem[{{Vogt} {et~al.}(2004){Vogt}, {Haynes}, {Giovanelli}, \&
  {Herter}}]{Vogt-04}
{Vogt}, N.~P., {Haynes}, M.~P., {Giovanelli}, R., \& {Herter}, T. 2004, \aj,
  127, 3300

\bibitem[{{Wang} {et~al.}(2013){Wang}, {Kauffmann}, {J{\'o}zsa}, {Serra}, {van
  der Hulst}, {Bigiel}, {Brinchmann}, {Verheijen}, \& {et al.}}]{Wang-13b}
{Wang}, J., {Kauffmann}, G., {J{\'o}zsa}, G.~I.~G., {et~al.} 2013, \mnras, 433,
  270

\bibitem[{{Weinmann} {et~al.}(2009){Weinmann}, {Kauffmann}, {van den Bosch},
  {Pasquali}, {McIntosh}, {Mo}, {Yang}, \& {Guo}}]{Weinmann-09}
{Weinmann}, S.~M., {Kauffmann}, G., {van den Bosch}, F.~C., {et~al.} 2009,
  \mnras, 394, 1213

\bibitem[{{White} \& {Frenk}(1991)}]{White-Frenk-91}
{White}, S.~D.~M., \& {Frenk}, C.~S. 1991, \apj, 379, 52

\bibitem[{{Wilkinson} {et~al.}(2015)}]{Wilkinson-15}
{Wilkinson}, D.~M., {Maraston}, C., {Thomas}, D., {et~al.} 2015, \mnras\ accepted.

\bibitem[{{Williams} {et~al.}(2009){Williams}, {Quadri}, {Franx}, {van Dokkum},
  \& {Labb{\'e}}}]{Williams-09}
{Williams}, R.~J., {Quadri}, R.~F., {Franx}, M., {van Dokkum}, P., \&
  {Labb{\'e}}, I. 2009, \apj, 691, 1879

\bibitem[{{Wisnioski} {et~al.}(2015){Wisnioski}, {F{\"o}rster Schreiber},
  {Wuyts}, {Wuyts}, {Bandara}, {Wilman}, {Genzel}, {Bender}, {Davies},
  {Fossati}, {Lang}, {Mendel}, {Beifiori}, {Brammer}, {Chan}, {Fabricius},
  {Fudamoto}, {Kulkarni}, {Kurk}, {Lutz}, {Nelson}, {Momcheva}, {Rosario},
  {Saglia}, {Seitz}, {Tacconi}, \& {van Dokkum}}]{Wisnioski-15}
{Wisnioski}, E., {F{\"o}rster Schreiber}, N.~M., {Wuyts}, S., {et~al.} 2015,
  \apj, 799, 209

\bibitem[{{Wolf} {et~al.}(2005){Wolf}, {Gray}, \&
  {Meisenheimer}}]{Wolf-Gray-Meisenheimer-05}
{Wolf}, C., {Gray}, M.~E., \& {Meisenheimer}, K. 2005, \aap, 443, 435

\bibitem[{{Wolf} {et~al.}(2009){Wolf}, {Arag{\'o}n-Salamanca}, {Balogh},
  {Barden}, {Bell}, {Gray}, {Peng}, {Bacon}, \& {et al.}}]{Wolf-09}
{Wolf}, C., {Arag{\'o}n-Salamanca}, A., {Balogh}, M., {et~al.} 2009, \mnras,
  393, 1302

\bibitem[{{Worthey} \& {Ottaviani}(1997)}]{Worthey-Ottaviani-97}
{Worthey}, G., \& {Ottaviani}, D.~L. 1997, \apjs, 111, 377

\bibitem[{{Wuyts} {et~al.}(2012){Wuyts}, {F{\"o}rster Schreiber}, {Genzel},
  {Guo}, {Barro}, {Bell}, {Dekel}, {Faber}, \& {et al.}}]{Wuyts-12}
{Wuyts}, S., {F{\"o}rster Schreiber}, N.~M., {Genzel}, R., {et~al.} 2012, \apj,
  753, 114

\bibitem[{{Wuyts} {et~al.}(2013){Wuyts}, {F{\"o}rster Schreiber}, {Nelson},
  {van Dokkum}, {Brammer}, {Chang}, {Faber}, {Ferguson}, {Franx}, {Fumagalli},
  {Genzel}, {Grogin}, {Kocevski}, {Koekemoer}, {Lundgren}, {Lutz}, {McGrath},
  {Momcheva}, {Rosario}, {Skelton}, {Tacconi}, {van der Wel}, \&
  {Whitaker}}]{Wuyts-13}
{Wuyts}, S., {F{\"o}rster Schreiber}, N.~M., {Nelson}, E.~J., {et~al.} 2013,
  \apj, 779, 135

\bibitem[{{Wyder} {et~al.}(2007){Wyder}, {Martin}, {Schiminovich}, {Seibert},
  {Budav{\'a}ri}, {Treyer}, {Barlow}, {Forster}, \& {et al.}}]{Wyder-07}
{Wyder}, T.~K., {Martin}, D.~C., {Schiminovich}, D., {et~al.} 2007, \apjs, 173,
  293

\bibitem[{{Yan} {et~al.}(2015)}]{Yan-15}
{Yan}, R., {Tremonti}, C., {Bershady}, M.~A., {et~al.} 2015, to be submitted

\bibitem[{{Yang} {et~al.}(2007){Yang}, {Mo}, {van den Bosch}, {Pasquali}, {Li},
  \& {Barden}}]{Yang-07}
{Yang}, X., {Mo}, H.~J., {van den Bosch}, F.~C., {et~al.} 2007, \apj, 671, 153

\bibitem[{{Yi} {et~al.}(2005){Yi}, {Yoon}, {Kaviraj}, {Deharveng}, {Rich},
  {Salim}, {Boselli}, {Lee}, \& {et al.}}]{Yi-05}
{Yi}, S.~K., {Yoon}, S.-J., {Kaviraj}, S., {et~al.} 2005, \apjl, 619, L111

\bibitem[{{York} {et~al.}(2000){York}, {Adelman}, {Anderson}, {Anderson},
  {Annis}, {Bahcall}, {Bakken}, {Barkhouser}, \& {et al.}}]{York-00}
{York}, D.~G., {Adelman}, J., {Anderson}, Jr., J.~E., {et~al.} 2000, \aj, 120,
  1579

\bibitem[{{Zhang} {et~al.}(2013){Zhang}, {Li}, {Kauffmann}, \&
  {Xiao}}]{Zhang-13}
{Zhang}, W., {Li}, C., {Kauffmann}, G., \& {Xiao}, T. 2013, \mnras, 429, 2191

\bibitem[{{Zibetti} {et~al.}(2009){Zibetti}, {Charlot}, \&
  {Rix}}]{Zibetti-Charlot-Rix-09}
{Zibetti}, S., {Charlot}, S., \& {Rix}, H.-W. 2009, \mnras, 400, 1181

\bibitem[{{Zwaan} {et~al.}(2005){Zwaan}, {Meyer}, {Staveley-Smith}, \&
  {Webster}}]{Zwaan-05}
{Zwaan}, M.~A., {Meyer}, M.~J., {Staveley-Smith}, L., \& {Webster}, R.~L. 2005,
  \mnras, 359, L30

\end{thebibliography}
\end{document}